\documentclass{aa}
\pdfoutput=1
\usepackage[varg]{txfonts}

\usepackage{xcolor}
\usepackage{siunitx}
\usepackage[inline]{enumitem}
\usepackage{graphicx}
\usepackage{bm}
\usepackage{multirow}

\graphicspath{{graphics/}}

\usepackage{natbib}
\usepackage[breaklinks=true]{hyperref}
\bibpunct{(}{)}{;}{a}{}{,} %% natbib format for A&A and ApJ

\newcommand{\OII}{[\ion{O}{ii}]}
\newcommand{\OIII}{[\ion{O}{iii}]}
\newcommand{\Halpha}{H$\alpha$}
\newcommand{\Halphamath}{{\rm{H}}\alpha}
\newcommand{\Galfit}{\textsc{Galfit}}
\newcommand{\FAST}{\textsc{FAST}}
\newcommand{\MPFIT}{\textsc{MPFIT}}
\newcommand{\MPFITEXY}{\textsc{MPFITEXY}}
\newcommand{\Ltsfit}{\textsc{LtsFit}}
\newcommand{\Mocking}{\textsc{MocKinG}}
\newcommand{\Camel}{\textsc{Camel}}
\newcommand{\Multinest}{\textsc{MultiNest}}
\newcommand{\Lephare}{\textsc{LePhare}}
\newcommand{\matplotlib}{\textsc{matplotlib}}
\newcommand{\scipy}{\textsc{scipy}}
\newcommand{\numpy}{\textsc{numpy}}
\newcommand{\astropy}{\textsc{astropy}}

\newcommand{\NMAGIC}{2730}
\newcommand{\NOIISample}{1142}
\newcommand{\NMorphoSample}{890}
\newcommand{\NKinematicSample}{593}
\newcommand{\NMSSample}{447}
\newcommand{\NTFRSample}{146}

\begin{document}
	%%%%%%%%%%%%%%%%%%%%%%%%%%%%%%
	%         TITLE PAGE         %
	%%%%%%%%%%%%%%%%%%%%%%%%%%%%%%

	\title{Scaling relations of $z\sim 0.25 - 1.5$ galaxies in various environments from the morpho-kinematic analysis of the MAGIC sample \thanks{Based on observations made with ESO telescopes at the Paranal Observatory under programs 094.A-0247, 095.A-0118, 096.A-0596, 097.A-0254, 099.A-0246, 100.A-0607, 101.A-0282, 102.A-0327, and 103.A-0563.}\fnmsep\thanks{Tables \ref{tab:catalog_CDS} is only available in electronic form at the CDS via anonymous ftp to cdsarc.u-strasbg.fr (130.79.128.5) or via http://cdsweb.u-strasbg.fr/cgi-bin/qcat?J/A+A/}
	}
	\titlerunning{Scaling relations of $z\sim 0.25 - 1.5$ galaxies in various environments}
	\author{W. Mercier\inst{\ref{IRAP}}\fnmsep\thanks{\email{wilfried.mercier@irap.omp.eu}}\and B. Epinat\inst{\ref{LAM}, \ref{CFHT}}\and T. Contini\inst{\ref{IRAP}}\and V. Abril-Melgarejo\inst{\ref{LAM}, \ref{STScI}}\and L. Boogaard\inst{\ref{Leiden}}\and J. Brinchmann\inst{\ref{Leiden}, \ref{Porto}}\and H. Finley\inst{\ref{IRAP}}\and D. Krajnovi\'{c}\inst{\ref{Postdam}}\and L. Michel-Dansac\inst{\ref{CRAL}}\and E. Ventou\inst{\ref{IRAP}}\and N. Bouché\inst{\ref{CRAL}}\and J. Dumoulin\inst{\ref{LAM}}\and Juan C. B. Pineda\inst{\ref{Santander}}}

	\institute{Institut de Recherche en Astrophysique et Planétologie (IRAP), Université de Toulouse, CNRS, UPS, CNES, 31400 Toulouse, France\label{IRAP}
	\and
	Aix Marseille Univ, CNRS, CNES, LAM, Marseille, France\label{LAM}
	\and
	Leiden Observatory, Leiden University, PO Box 9513, 2300 RA Leiden, The Netherlands\label{Leiden}
	\and
	Instituto de Astrofísica e Ciências do Espaço, Universidade do Porto, CAUP, Rua das Estrelas, PT4150-762 Porto, Portugal\label{Porto}
	\and
	Univ Lyon, Univ Lyon1, Ens de Lyon, CNRS, Centre de Recherche Astrophysique de Lyon UMR5574, F-69230, Saint-Genis-Laval, France\label{CRAL}
	\and
	Escuela de Fisica, Universidad Industrial de Santander, A.A. 678, Bucaramanga, Colombia\label{Santander}
	\and
	Space Telescope Science Institute, 3700 San Martin Drive, Baltimore, MD 21218, USA\label{STScI}
	\and
	Leibniz-Institut für Astrophysik Potsdam (AIP), An der Sternwarte 16, D-14482 Potsdam, Germany\label{Postdam}
	\and
	Canada-France-Hawaii Telescope, CNRS, 96743 Kamuela, Hawaii, USA \label{CFHT}
	}
	
	\date{Received 14 January 2022; accepted 14 April 2022}

	\abstract{The evolution of galaxies is influenced by many physical processes which may vary depending on their environment.}%
	{We combine Hubble Space Telescope (HST) and Multi-Unit Spectroscopic Explorer (MUSE) data of galaxies at $0.25~\lesssim~z~\lesssim~1.5$ to probe the impact of environment on the size-mass relation, the Main Sequence (MS) and the Tully-Fisher relation (TFR).}%
	{We perform a morpho-kinematic modelling of \NKinematicSample{} \OII{} emitters in various environments in the COSMOS area from the MUSE-gAlaxy Groups In Cosmos (MAGIC) survey. The HST F814W images are modelled with a bulge-disk decomposition to estimate their bulge-disk ratio, effective radius, and disk inclination. We use the \OII{}$\lambda\lambda3727,3729$ doublet to extract the galaxies ionised gas kinematic maps from the MUSE cubes, and we model those maps for a sample of \NTFRSample{} \OII{} emitters, including bulge and disk components constrained from morphology and a dark matter halo.}%
	{We find an offset of $\SI{0.03}{dex}$ ($1\sigma$ significant) on the size-mass relation zero point between the field and the large structure subsamples, with a richness threshold of $N=10$ to separate between small and large structures, and of $\SI{0.06}{dex}$ ($2\sigma$) with $N=20$. Similarly, we find a $\SI{0.1}{dex}$ ($2\sigma$) difference on the MS with $N=10$ and 0.15\,dex ($3\sigma$) with $N=20$. These results suggest that galaxies in massive structures are smaller by 14\% and have star formation rates reduced by a factor of $1.3-1.5$ with respect to field galaxies at $z \approx 0.7$. Finally, we do not find any impact of the environment on the TFR, except when using $N=20$ with an offset of 0.04\,dex ($1\sigma$). We discard the effect of quenching for the largest structures that would lead to an offset in the opposite direction. We find that, at $z~\approx~0.7$, if quenching impacts the mass budget of galaxies in structures, these galaxies would have been affected quite recently, for roughly $0.7-\SI{1.5}{Gyr}$. This result holds when including the gas mass, but vanishes once we include the asymmetric drift correction.}{}
	

	\maketitle
	
	%%%%%%%%%%%%%%%%%%%%%%%%%%%%%%%%
	%          1st SECTION         %
	%%%%%%%%%%%%%%%%%%%%%%%%%%%%%%%%
	
	\section{Introduction}
	
	The evolution of galaxies is not a trivial process as numerous physical mechanisms are at play, acting on different physical and time scales, and with different amplitudes. From an observational point of view, our understanding of galaxy evolution has greatly improved throughout roughly the last 25 years. First, from extended multi-band imaging and spectroscopic surveys of the local Universe (e.g. SDSS, 2dFGRS). Then, followed the advent of the Huble Space Telescope (HST) associated with 8-10m class telescopes (e.g. VLT, Keck) which allowed to probe and study galaxies in the more distant Universe by combining extremely deep images (e.g. HUDF, COSMOS) with large spectroscopic surveys (e.g. VVDS, zCOSMOS). And finally, to the development and continuous improvement of 3D spectrographs (e.g. SINFONI, KMOS, MUSE), whose data have allowed to study distant galaxies in even more detail. The current paradigm for galaxy evolution is that galaxies must have first formed their Dark Matter (DM) haloes in the early stages of the Universe, and only later started assembling their baryonic mass either by continuous accretion via the Circum Galactic Medium (CGM) of mainly cold gas from filaments located in the cosmic web \citep{kere_how_2005, ocvirk_bimodal_2008, bouche_signatures_2013, zabl_muse_2019}, by galactic wind recycling \citep{dave_missing_2009, hopkins_stellar_2012, schroetter_muse_2019}, and through galaxy mergers \citep{lopez-sanjuan_dominant_2012, ventou_muse_2017, mantha_major_2018, duncan_observational_2019, ventou_new_2019}. In particular, this scenario is favoured in order to explain the large star formation rates (SFR) measured in the past billion years which would have rapidly depleted the galaxies gas content and would have led the galaxies to an early quenching phase unless their gas reservoir was continuously replenished throughout cosmic time. Thus, the mass assembly of the galaxies baryonic components must be tightly linked to the evolution of their DM content.
	
	This picture is further supported by the fact that high redshift galaxies appear to be quite different from their local counterparts, indicative that they must have radically evolved in order to populate the Hubble sequence that we see today. Studies comparing the global properties between high and low redshift galaxies have indeed shown that the former tend to be on average smaller \citep{trujillo_strong_2007, van_der_wel_3d-hstcandels_2014, mowla_cosmos-dash:_2018} and less massive \citep{ilbert_galaxy_2010, muzzin_evolution_2013} than the latter. At the same time, galaxies have shown a rise of their mean SFR throughout cosmic time up to a peak of star formation at a redshift $z \sim 2$ before declining to the typical value of roughly $\SI{0.01}{M_{\odot}~yr^{-1}~Mpc^{-3}}$ measured today \citep{hopkins_normalization_2006}, and their molecular gas fraction is also found to be larger at high redshift \citep{tacconi_phibss_2018, freundlich_phibss2_2019, walter_evolution_2020}. In addition to their global properties, galaxies also show clear signs of morphological and kinematic evolution. Several studies have indeed highlighted the fact that the proportion of triaxial systems and thick disks increases as we go to higher redshifts, with low mass galaxies having a larger tendency to be triaxial \citep{van_der_wel_geometry_2014, zhang_evolution_2019}. This would suggest a trend for star-forming galaxies to flatten as they evolve, going from prolate to oblate shapes. At the same time, intermediate to high redshift galaxies are found to have on average more complex and perturbed gas kinematics with a larger velocity dispersion than their local counterparts \citep{flores_3d_2006, yang_images._2008, Epinat2010}. While understanding the evolution of the different galaxy populations in their intricate details is a particularly tedious task to accomplish, it has become clear that there must exist a finite set of physical mechanisms at play which drives the bulk of the evolution in order to explain the various scaling relations first discovered in the local Universe, but which have been shown to hold at intermediate and high redshift. Among these we can cite the Schmidt-Kennicut relation \citep[e.g.][]{schmidt_rate_1959, kennicutt_star_1998}, mass-size relation \citep[e.g.][]{shen_size_2003, mowla_cosmos-dash:_2018}, Main Sequence, hereafter MS, \citep[e.g.][]{noeske_star_2007, whitaker_constraining_2014}, Tully-Fisher relation, hereafter TFR, \citep[e.g.][]{tully_new_1977, Contini2016, tiley_krosssami_2019, Abril-Melgarejo2021} or mass-metalicity relation \citep[e.g.][]{tremonti_origin_2004, erb_massmetallicity_2006}.
	
	As to now, one of the key questions is whether the transition seen from high to low redshift between morphologically disturbed, particularly active galaxies to mostly relaxed, low SF massive systems is mainly driven by in-situ physical phenomena such as supernovae-driven galactic super winds and Active Galactic Nucleii (AGN) feedback or, on the contrary, is driven by the environment within which these galaxies lie. This question has led discussions about the impact of galaxy clusters onto the physical properties, morphology, and kinematic of their constituent galaxies. The two main mechanisms which can affect star formation in galaxies located in clusters with respect to those in the lowest density environments (hereafter field) are bursts of star formation and quenching \citep[e.g. see][for an analysis of environment and mass quenching in the local Universe]{peng_mass_2010}. While the latter is not specifically inherent to galaxy clusters, these massive structures tend to accelerate its effect either through hydrodynamical mechanisms such as ram-pressure stripping \citep[e.g.][]{gunn_infall_1972, boselli_evidence_2019} or thermal evaporation \citep[e.g.][]{cowie_evaporation_1977, cowie_thermal_1977}, or through gravitational mechanisms such as galaxy harassment \citep[e.g.][]{cortese_dawes_2021}.
	
	Until quite recently, few studies tried to investigate the well known scaling relations as a function of the galaxies environment, except for the MS. Indeed, the MS is probably one of the most studied scaling relations as a function of environment as it can be used to directly probe the impact of quenching on the evolution of galaxies. Following the recent data release announcement of the GOGREEN and GCLASS surveys \citep{balogh_gogreen_2020}, aimed at probing the impact of dense environments on intermediate redshift ($0.8 < z < 1.5$) galaxies properties, \citet{old_erratum_2020, old_gogreen_2020} explored the environmental dependence of the star forming MS between massive clusters and field galaxies. Using the \OII{} doublet flux as a proxy for the SFR, they found the cluster galaxies SFR to be on average 1.4 times lower than that of their field sample, the difference being more pronounced for low stellar masses. Alternatively, \citet{erfanianfar_non-linearity_2016}, using data from the COSMOS, AEGIS, ECDFS, and CDFN fields, could not find any difference in the MS between field galaxies and those in structures in the redshift range $0.5 < z < 1.1$, but a similar trend to that of \citet{old_gogreen_2020} in the lowest redshift regime ($0.15 < z < 0.5$). On the other hand, \citet{nantais_h_2020} could not find any significant difference between field and SpARCS \citep{muzzin_spectroscopic_2009} cluster galaxies at a redshift $z \sim 1.6$, which they explained either by the fact that galaxies might have been accreted too recently to show signs of quenching, or that the clusters might be not mature enough yet at this redshift to produce measurable environmental effects on these galaxies.
	
	The environmental impact on the size-mass relation began to be studied only in the last decade by \citet{maltby_environmental_2010}. Using galaxies from the STAGES survey \citep{gray_stages_2009}, they found no difference in the size-mass relation for massive galaxies ($M_{\star} > 10^{10} {\rm{M}}_{\odot}$) and a significant offset for intermediate to low mass galaxies, consistent with field spiral galaxies being about 15\% larger than those in clusters at $z \sim 0.16$. Alternatively, \citet{kuchner_effects_2017} found a similar relation at high mass 
rather than at low mass for late-type galaxies at $z = 0.44$ where cluster galaxies were smaller than their field counterparts, and \citet{matharu_hst_2019} also found the same trend when comparing the size-mass relation between field and cluster galaxies at $z \sim 1$. However, \citet{kelkar_galaxy_2015}, using data from the ESO Distant Cluster Survey, could not find any difference between field and cluster galaxies in the redshift range $0.4 < z <0.8$. 
	
	Finally, regarding the TFR, \citet{Pelliccia2019} searched for differences between two samples of galaxies in groups and clusters from the ORELSE sample \citep{ORELSE}, using long-slit spectroscopy data to derive the galaxies kinematic. Their conclusion was that they could not find any significant difference between the two TFR and therefore claimed for no impact of the environment. More recently, \citet{Abril-Melgarejo2021} analysed a sample of $z \sim 0.7$ galaxies located in galaxy groups from the MAGIC survey (Epinat et al., in prep.) using MUSE and HST data. By comparing their TFR with that from the KMOS3D \citep{ubler_evolution_2017}, KROSS \citep{tiley_krosssami_2019}, and ORELSE \citep{Pelliccia2019} samples, they found a significant offset in the TFR zero point which they attributed to a possible impact of the environment since these samples targetted different populations of galaxies (galaxies in groups and clusters versus galaxies in clusters and in the field). This result led them to two different interpretations of this offset: 
	\begin{enumerate*}[label=(\roman*)]
		\item a quenching of star formation visible in the massive structures which led to a decrease in stellar mass with respect to the field,
		\item a baryon contraction phase for the galaxies in groups and clusters which led to an increase in circular velocity for these galaxies.
	\end{enumerate*}
	However, they also indicated that comparing samples from different datasets, with physical quantities derived from different tools, methods, and models, and with different selection functions leads to many uncertainties which might compromise the interpretation. Thus, they argued that, in order to study in a robust way the impact of the environment on the TFR, one would need to apply in a self-consistent manner the same methodology and models on galaxies located in various environments (field, groups, and clusters), which is the goal of this paper.
	
	Indeed, in this paper, we push beyond the previous analysis performed by \citet{Abril-Melgarejo2021} and investigate differences in three main scaling relations (size-mass, MS, and TFR) when using samples targetting different environments, with HST and MUSE data from the MAGIC (MUSE gAlaxy Groups In Cosmos) survey. Because this survey targets galaxies located in galaxy groups and clusters, as well as foreground and background galaxies in a similar redshift range without prior selection, by applying the same procedure to model the morphology with HST images and the kinematic with MUSE cubes using the \OII{} doublet, we expect to probe in detail and with reduced uncertainties the impact of the environment on these relations. 
	
	This paper is structured as follows. In Sect.\,\ref{sec:MUSE_HST}, we present the HST and MUSE data. In Sect.\,\ref{sec:Sample properties}, we introduce the initial MAGIC sample, the structure identification, and we explain how we derived the galaxies global properties (stellar mass and SFR). In Sect.\,\ref{sec:morpho}, we present the morphological modelling performed with \textsc{Galfit} on the entire \OII{} emitter sample with reliable redshifts, the aperture correction applied for the stellar mass, and the prescription we applied to derive an average disk thickness as a function of redshift. In Sect.\,\ref{sec:kin_modelling}, we describe the kinematic modelling using the \OII{} doublet as a kinematic tracer, as well as the mass models used to constrain the kinematic from the stellar distribution. In Sect.\,\ref{sec:selection}, we discuss the selection criteria applied to select samples to study the size-mass, MS, and TFR. Finally, we focus in Sect.\,\ref{Sec:analysis} on the analysis of the three scaling relations as a function of environment. Throughout the paper, we assume a $\Lambda$CDM cosmology with H$_0 = \SI{70}{km~s^{-1}~Mpc^{-1}}$, $\Omega_{\rm{M}} = 0.3$ and $\Omega_\Lambda = 0.7$. 
	
	\section{MUSE and HST data}
	\label{sec:MUSE_HST}
	
	\subsection{MUSE observations and data reduction}
	\label{sec:MUSE_data}
	
	Galaxies studied in this paper are part of the MAGIC survey. This survey targeted 14 galaxy groups located in the COSMOS area \citep{Scoville2007} selected from the COSMOS group catalogue of \citet{Knobel2012} in the redshift range $0.5<z<0.8$, and observed during Guaranteed Time Observations (GTO) as part of an observing program studying the effect of the environment on $\SI{8}{Gyr}$ of galaxy evolution (PI: T.Contini). Though more details will be given in the MAGIC survey paper (Epinat et al. in prep), we provide in what follows a summary of the data acquisition and reduction.
	
	In total, 17 different MUSE fields were observed over seven periods. For each target, Observing Blocks (OB) of four 900 seconds exposures were combined, including a small dithering pattern, as well as a rotation of the field of $\SI{90}{\degree}$ between each exposure. The final combined data cubes have total exposure times ranging between 1 and 10 hours. Because kinematic studies are quite sensitive to spatial resolution, we required observations to be carried out under good seeing conditions with a Point Spread Function (PSF) Full Width at Half Maximum (FWHM) lower than $\SI{0.8}{\arcsec}$, except in cases where the Adaptive Optics (AO) system was used.
	
	The MUSE standard pipeline \citep{Weilbacher2020} was used for the data reduction on each OB individually. Observations with AO used the v2.4 version, whereas the others used v1.6, except for the MUSE observations of COSMOS group CGr30 which used v1.2. Default sky subtraction was applied on each science exposure before aligning and combining them using stars located in the field. To improve sky subtraction, the Zurich Atmosphere Purge software \citep[ZAP;][]{ZAP} was then applied onto the final combined data cube. The reduction leads to data and variance cubes with spatial and spectral sampling of $\SI{0.2}{\arcsec}$ and $\SI{1.25}{\angstrom}$, respectively, in the spectral range $4750 - \SI{9350}{\angstrom}$.
	
	As shall be discussed in more detail in Sect.\,\ref{sec:kin_modelling}, the kinematic maps, which are extracted from the MUSE data cubes, serve as a basis for the kinematic modelling. Among those kinematic maps are the ionized gas velocity field and velocity dispersion maps which are both highly affected by the limited spectral (Line Spread Function - LSF) as well as spatial (PSF)  resolutions of MUSE data through beam smearing. Because extracting reliable kinematic parameters depends on correctly taking into account the impact of the beam smearing in the kinematic models of the galaxies, it is therefore important to know the values of the MUSE PSF and LSF FWHM at the wavelength of observation. The MUSE LSF is modelled using the prescription from \citet{bacon_muse_2017} and \citet{guerou_muse_2017} who derived the wavelength dependence of the MUSE LSF FWHM in the Hubble Ultra Deep Field (HUDF) and Hubble Deep Field South (HDFS) as
	
	\begin{equation}
		\rm{FWHM_{LSF}} = \lambda^2 \times 5.866 \times 10^{-8} - \lambda \times 9.187 \times 10^{-4} + 6.040,
	\end{equation}
	where FWHM$_{\rm{LSF}}$ and $\lambda$ are both in \AA. 
	
	Because of the atmospheric turbulence, we expect the PSF FWHM to be reduced with increasing wavelength. As was shown in \citet{bacon_muse_2017}, the change of the PSF with wavelength can be quite accurately modelled with a declining linear relation. To derive the slope and zero point of this relation in each MUSE field, we extracted as many stars as possible, only keeping those with a reliable MUSE redshift measurement of $z \sim 0$. For each star, 100 sub-cubes of spatial dimension $10 \times \SI{10}{\rm{pixels}}$ were extracted at regular intervals along the MUSE wavelength range and later collapsed into narrow band images using a fixed redshift slice depth of $\Delta z = 0.01$, scaling with wavelength as $\Delta \lambda = \Delta z \times \lambda$. Each narrow band image was modelled with \Galfit{} \citep{GALFIT} using a symmetric
	\begin{enumerate*}[label=(\roman*)]
		\item 2D Gaussian profile,
		\item Moffat profile with a free $\beta$ index.
	\end{enumerate*}
	We found consistent results between these two models, and therefore decided to use the Gaussian values in the following analysis. In order to remove small scale variations while keeping the global declining trend of interest in the wavelength dependence of the PSF FWHM, we applied a rolling average with a window of 5 data points for all the stars. For each MUSE field, the median wavelength dependence of the PSF FWHM of the stars in the field was fitted with a linear relation. We find a median value of $\SI{0.65}{\arcsec}$ for the MUSE PSF FWHM and $\SI{2.55}{\AA}$ for the LSF FWHM (roughly $\SI{50}{km s^{-1}}$). The values of the slope and zero point retrieved from the best-fit models were later used in the kinematic modelling (see Sect.\,\ref{sec:kin_modelling}).

	\subsection{HST data}
	
	In addition to using MUSE observations to extract the ionized gas kinematic, we also made use of Hubble Space Telescope Advanced Camera for Surveys (HST-ACS) images and photometry to model the morphology of the galaxies (see Sect.\,\ref{sec:morpho modelling}). For each galaxy we extracted stamps of dimension $\SI{4}{\arcsec} \times \SI{4}{\arcsec}$ in the F814W filter from the third public data release of the HST-ACS COSMOS observations \citep{Koekemoer2007, Massey2010}. These images have the best spatial resolution available ($\lesssim \SI{0.1}{\arcsec}$, that is $\sim \SI{600}{pc}$ at $z \sim 0.7$) for HST data in the COSMOS field with a spatial sampling of $\SI{0.03}{\arcsec/pixel}$, as is required to extract precise morphological parameters, with an exposure time of $\SI{2028}{s}$ per HST tile. At the same time, this filter corresponds to the reddest band available (I-band) and therefore to the oldest stellar populations probed by HST data, being less affected by star forming clumps and with smoother stellar distributions.
	
	As for MUSE data, a precise knowledge of the HST PSF in this filter is required to extract reliable morphological parameters. To model the HST PSF FWHM, a circular Moffat profile was fitted onto 27 non saturated stars located in our MUSE fields. The theoretical values of the HST PSF parameters, retrieved from the best-fit Moffat profile, used in the morphological modelling (see Sect.\,\ref{sec:morpho modelling}) correspond to the median values of the 27 best fit models parameters and are $\rm{FWHM_{HST}} = \SI{0.0852}{\arcsec}$ and $\beta = 1.9$ respectively \citep{Abril-Melgarejo2021}.

	\section{Galaxies samples properties}
	\label{sec:Sample properties}
	
	\subsection{Initial MAGIC sample}
		
	\begin{figure}[hbt!]
		\includegraphics[scale=0.7]{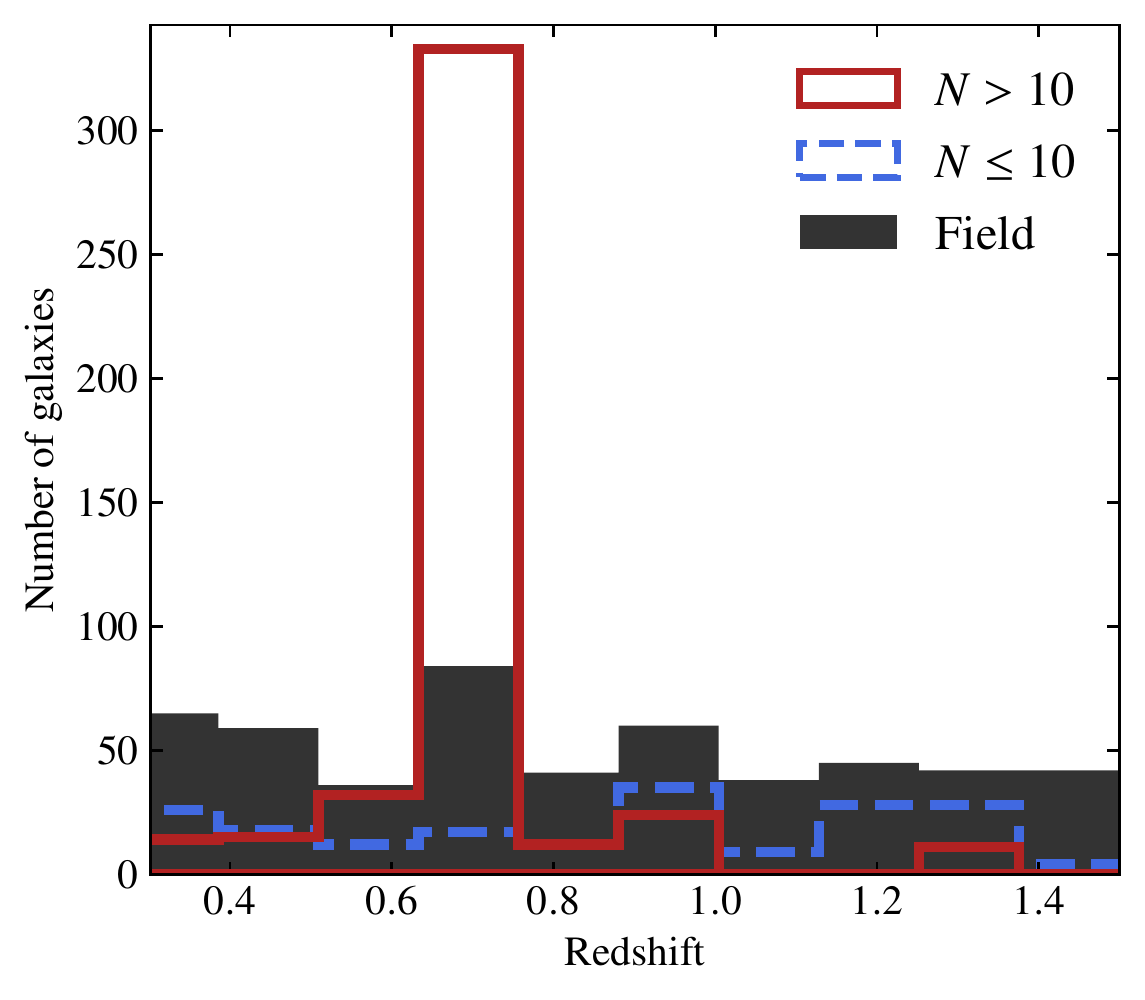}
		\caption{Redshift distribution for the three initial sub-samples defined in Sect.\,\ref{sec:FoF}. The field galaxies (grey area) and galaxies in small structures (dashed blue line) samples have relatively flat distributions. The peak of the distribution for galaxies in large structures (red line) is located at a redshift $z \sim 0.7$ and is driven by the largest structures ($40~\lesssim~N~\lesssim~100$) found in the COSMOS area of the MAGIC sample.}
		\label{fig:redshift_distribution}
	\end{figure}
	
	Observations carried out for the MAGIC survey were targetting already known galaxy groups in the COSMOS field such that all the galaxies in these fields up to $z \sim 1.5$  were already detected from previous broad band photometry and listed in the COSMOS2015 catalogue of \citet{cosmos2015} up to a $3\sigma$ limiting magnitude of 27 in z++ band. The spectroscopic redshift of the objects in the COSMOS2015 catalogue located in the observed MUSE fields were estimated with the redshift finding algorithm \textsc{MARZ} \citep{Hanton2016} using both absorption and emission features. At the redshift of the targetted groups ($z \sim 0.7$) the strongest emission lines are \OII$\lambda\lambda3727,3729$, \OIII$\lambda 5007$, and H$\beta$, and the main absorption lines are Ca\textsc{ii} H$\lambda3968.47$, Ca\textsc{ii} K$\lambda3933.68$, G~band from CH molecules, and Balmer absorption lines. Following \citet{inami2017}, a PSF weighted spectrum was extracted for each source and a robust redshift determination was obtained using the strongest absorption and emission lines. In each case, a redshift confidence flag was assigned ranging from $\rm{CONFID} = 1$ (tentative redshift) to $\rm{CONFID} = 3$ (high confidence).
	
	Initially, the catalogue contained \NMAGIC{} objects, including stars in our Galaxy, intermediate, and high redshift ($z \geq 1.5$) galaxies, 51\% of which having reliable spectroscopic redshifts (CONFID $> 1$). As described in Sect.\,\ref{sec:kin_modelling}, the kinematic of the galaxies is extracted from the \OII{} doublet. Therefore, as a starting point, we decided to restrict the sample of galaxies to \OII{} emitters with reliable redshifts only, that is galaxies in the redshift range $0.25 \lesssim z \lesssim 1.5$ with CONFID $> 1$. The main reason for considering \OII{} emitters only is that the bulk of galaxies located in the targeted groups is located at a redshift $z \sim 0.7$ where the \OII{} doublet is redshifted into the MUSE wavelength range and happens to be among the brightest emission lines. Thus, using this emission line combines the advantages of having a high signal-to-noise ratio (S/N) extended ionised gas emission, while probing galaxies within a quite large redshift range roughly corresponding to $\SI{8}{Gyr}$ of galaxy evolution. Using the aforementioned criteria onto the initial MAGIC sample and without applying any further selection, the \OII{} emitters sample contains \NOIISample{} galaxies. The main physical properties of this sample, along with other samples defined later in the text are shown in Table\,\ref{tab:samples}.
	
	\begin{table}
		\centering
		\caption{Median properties for the different samples of galaxies defined in Sect.\,\ref{sec:summary_samples}.}
		\resizebox{0.48\textwidth}{!}{%
		\begin{tabular}{lcccccc}
		\hline
		\hline
		Sample          & Selection  & Number              & $\log_{10} M_{\star}$ & $R_{\textit{eff}, \rm{d}}$ & B/D ($R_\textit{eff}$) & $\log_{10}$ SFR$_z$ \\ 
		                &            &                     & [M$_{\odot}$]         & kpc                        &                        & [M$_{\odot}$ yr$^{-1}$] \\
		(1)             & (2)        & (3)                 & (4)                   & (5)                        & (6)                    & (7) \\
		\hline \\[-9pt]
		\OII{} emitters &            & \NOIISample{}       & $9.2^{+1.2}_{-1.1}$   &                            &                        & \\[3pt]
		Morphological   &            & \NMorphoSample{}    & $9.4^{+1.1}_{-1.0}$   & $2.5^{+2.6}_{-1.3}$        & $0.2^{+1.5}_{-0.2}$    &  \\[3pt]
		Kinematic       &            & \NKinematicSample{} & $9.3^{+0.9}_{-0.9}$   & $2.6^{+2.6}_{-1.4}$        & $0.1^{+1.0}_{-0.1}$    & $-0.2^{+0.5}_{-0.5}$ \\[3pt]
		MS              & (i)        & \NMSSample{}        & $9.3^{+0.7}_{-0.7}$   & $2.8^{+2.4}_{-1.5}$        & $0.0^{+0.3}_{-0.0}$    & $-0.2^{+0.5}_{-0.5}$ \\[3pt]
		TFR             & (i) to (v) & \NTFRSample{}       & $9.6^{+0.6}_{-0.6}$   & $3.9^{+2.1}_{-1.2}$        & $0.0^{+0.2}_{-0.0}$    & $\,0.0^{+0.4}_{-0.4}$ \\[3pt]
		\hline\\
		\end{tabular}}
		\label{tab:samples}
			{\small\raggedright {\bf Notes:} (1) Sample name, (2) selection criteria applied from Sect.\,\ref{sec:samples_selection}, (3) number of galaxies, (4) SED-based stellar mass, (5) disk effective radius, (6) bulge-to-disk flux ratio at radius $R_{\textit{eff}}$, (7) \OII{}-based SFR corrected for redshift evolution by normalising at a redshift $z_0 \approx 0.7$. In this table, each sample is a sub-sample of the one located just above. Stellar masses and SFR values are given in an aperture of 3\,\arcsec. Uncertainties correspond to the 16th and 84th percentiles.\par}
	\end{table}
	
	\subsection{Structures identification and characterisation}
	\label{sec:FoF}
	
	A crucial point when one wants to look at the effect of the environment on galaxies properties and evolution is to efficiently characterise the environment where galaxies lie. Galaxies are usually split into three main categories depending on their environment
	\begin{enumerate*}[label=(\roman*)]
		\item field galaxies which do not belong to any structure,
		\item galaxies in groups which are gravitationally bound to a small number of other galaxies
		\item galaxies in clusters which are gravitationally bound to a large number of galaxies.
	\end{enumerate*}
	Because there is no sharp transition between a galaxy group and a galaxy cluster, and also because it is not particularly relevant for this discussion to disentangle between these two cases, we will refer to both in the following parts as structures.
	
	The characterisation of the galaxies environment and their potential membership to a structure was performed with a 3D Friends-of-Friends (FoF) algorithm. Structure membership assignment was performed galaxy per galaxy given that the sky projected and the line of sight velocity separations were both below two thresholds set to $\SI{450}{kpc}$ and $\SI{500}{km \per s}$, respectively, as was suggested by \citet{Knobel2009}. We checked that varying the thresholds around the aforementioned values by small amounts did not change significantly the structure memberships (see MAGIC survey paper, Epinat et al. in prep for more details). As shown in Fig.\,\ref{fig:redshift_distribution}, the bulk of the structures is located in the redshift range $0.6 < z < 0.8$ since most of them belong to the COSMOS wall \citep{scoville_large_2007, iovino_high_2016}, a large scale filamentary structure located at a redshift $z \approx 0.72$. Among these structures, those with at least 10 members were studied in a previous paper \citep{Abril-Melgarejo2021}. In order to probe in detail the environmental dependence on galaxies properties, we will use throughout the following sections three subsamples, 
	\begin{enumerate*}[label=(\roman*)]
		\item the field galaxies subsample which contains galaxies not assigned to any structure as well as galaxies belonging to structures with up to three members,
		\item the small structures subsample which is made of galaxies belonging to structures having between three and ten members,
		\item the large structures subsample containing galaxies in structures with more than ten members.
	\end{enumerate*}		
	Within the \OII{} emitters sample, 45\% belong to the field, 20\% are in small structures, and 35\% are in the large structures subsample.
	
	\subsection{Stellar mass and star formation rates}
	\label{sec:gal_properties}
	
	\begin{figure}
		\includegraphics[scale=0.7]{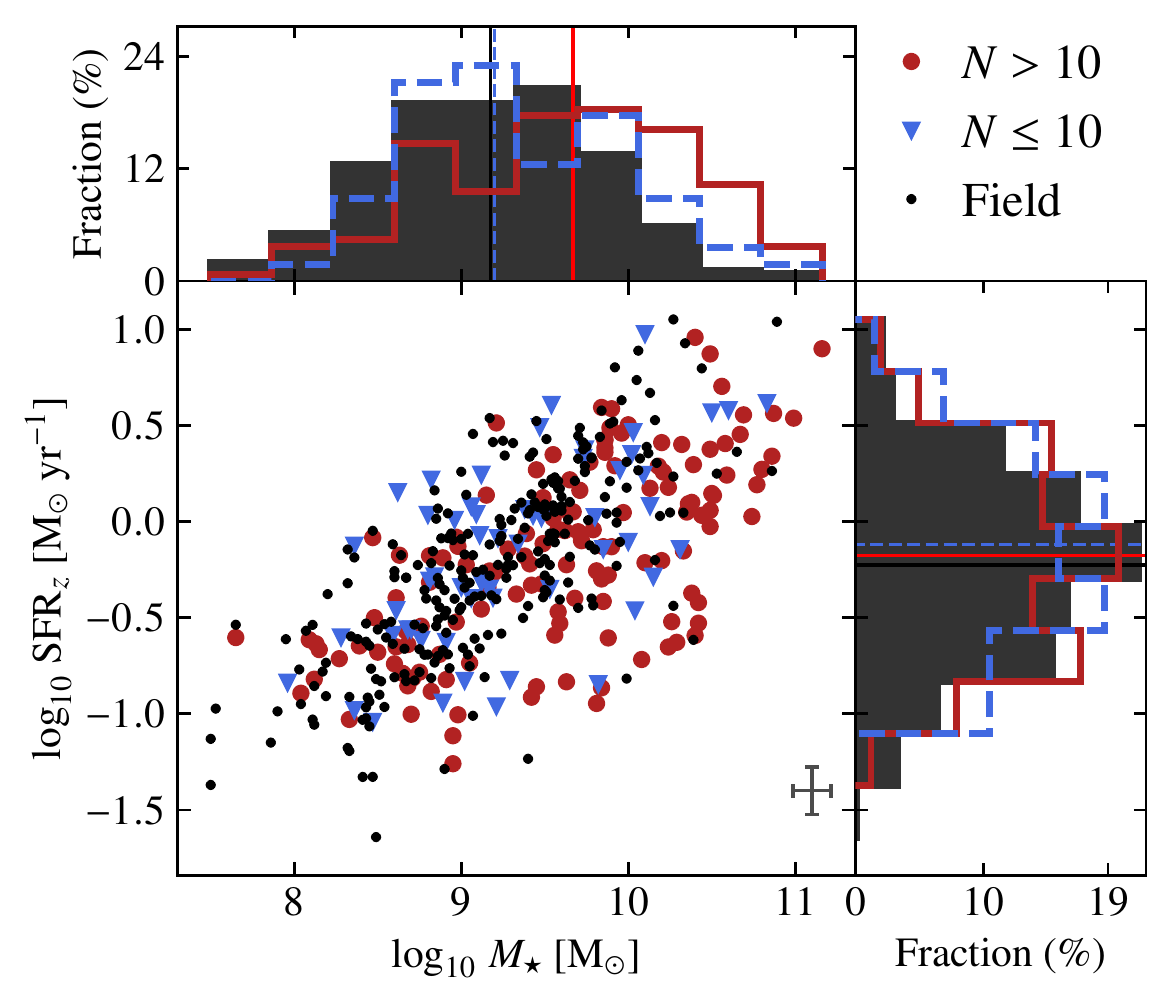}
		\caption{SFR-M$_{\star}$ diagram for galaxies from the kinematic sample (see Sect.\,\ref{sec:samples_selection}). Galaxies are separated between the field (black points), small structures (blue triangles), and large structures (red circles). The typical stellar mass and SFR error is shown on the bottom right. The SFR was normalised to a redshift $z_0 = 0.7$. The SFR and mass distributions are shown as top and right histograms, respectively, with the median values for each sample represented as lines with similar colours.}
		\label{fig:MS_full}
	\end{figure}
	
	Since galaxies are located in the COSMOS area, we used the same 32 photometric bands as in \citet{epinat_ionised_2018} and \cite{Abril-Melgarejo2021} found in \citet{Laigle2016} (COSMOS2015) catalogue to derive additional physical parameters such as stellar masses and Star Formation Rates (SFR). We used the Spectral Energy Distribution (SED) fitting code \FAST{} \citep{FAST} with a synthetic library generated from the Stellar Population Synthesis (SPS) models of \citet{Conroy2010} using a \citet{Chabrier2003} Initial Mass Function (IMF), an exponentially declining SFR, a \citet{Calzetti2000} extinction law, and fixing the redshift of the galaxy to the spectroscopic redshift derived from the MUSE spectrum. The SED output parameters, including the stellar mass, SFR, and stellar metallicity, as well as their $1\sigma$ error, correspond to the values retrieved from the best-fit model of the SED, using the photometric bands values from \citet{Laigle2016} catalogue, and integrated within a circular aperture of diameter $\SI{3}{\arcsec}$.
	
	After performing a careful comparison between the stellar masses and SFR values computed with FAST and those given in the COSMOS2015 catalogue (computed using \Lephare{} SED fitting code), we found consistent results for the stellar masses with, on average, a scatter of $0.2-\SI{0.3}{dex}$. On the other hand, we found larger discrepancies between the SFR values, around $0.7-\SI{0.8}{dex}$. Given that the origin of this discrepancy is unclear, and that SED-based SFR estimates usually have quite large uncertainties \citep[e.g.][]{wuyts_star_2011, leja_hot_2018}, we decided to use emission lines instead to compute the SFR. Ultimately, one would want to use \Halpha{} as tracer of star formation, but given the MUSE wavelength range, this would restrict the sample to $z \lesssim 0.4$ galaxies. Instead, following \citet{kennicutt_jr_global_1998}, we can use the \OII{} doublet to compute the SFR in the entire \OII{} emitters sample, as long as we can correct for Galactic and intrinsic extinctions, that is
	
	\begin{equation}
		{\rm{SFR}}~[{\rm{M_{\odot}~yr^{-1}}}] = (1.4 \pm 0.4) \times 10^{-41}~L_{\rm{\OII{}}}~[{\rm{erg~s^{-1}}}],
	\end{equation}
	where SFR has not been normalised yet to account for the redshift evolution of the MS, $L_{\rm{\OII{}}} = 4 \pi D_L^2 F_{\rm{\OII{}, corr}}$ is the \OII{} luminosity, with $D_L$ the luminosity distance, and $F_{\rm{\OII{}, corr}}$ the extinction corrected \OII{} flux, which must be corrected for intrinsic extinction at the rest-frame \Halpha{} wavelength \citep{kennicutt_integrated_1992, kennicutt_jr_global_1998}, computed as
	
	\begin{equation}
		F_{\rm{\OII{}, corr}} = F_{\rm{\OII{}}} \times 10^{0.4 \left ( A_{\Halphamath{}} + A_{\rm{\OII{}, MW}} \right )},
	\end{equation}
	with $F_{\rm{\OII{}}}$ the uncorrected \OII{} flux integrated in an aperture of $\SI{3}{\arcsec}$, $A_{\Halphamath{}}$ the intrinsic extinction computed at the rest-frame \Halpha{} wavelength, and $A_{\rm{\OII{}, MW}}$ the Galactic extinction computed at the observed \OII{} wavelength assuming a \citet{cardelli_relationship_1989} extinction law and $R_V = 3.1$. In order to compute the intrinsic extinction, one needs to know the extinction in a given band or at a given wavelength, for instance in the V band. This value is provided by \FAST{} but, similarly to the SFR, it usually comes with large uncertainties. Given that the extinction plays an important role when deriving the SFR, we decided not to rely on the values from \FAST{}. Instead, we used the prescription from \citet{gilbank_local_2010, gilbank_erratum_2011}, which parametrises the extinction for \Halpha{} using the galaxies stellar mass as
	
	\begin{equation}
		A_{\Halphamath{}} = 51.201 - 11.199 \log_{10} \left ( \frac{M_{\star}}{\rm{M}_{\odot}} \right ) + 0.615 \log_{10}^2 \left ( \frac{M_{\star}}{\rm{M}_{\odot}} \right ),
	\end{equation}
	for stellar masses $M_{\star} > \SI{e9}{M_{\odot}}$, and as a constant value below. When using the \OII{}-based SFR in the analysis (Sect.\ref{Sec:analysis}), we checked that using the SED-based extinction rather than the prescription from \citet{gilbank_local_2010} to correct for intrinsic extinction did not change our conclusions.
	
	The SFR-stellar mass plane for the kinematic sample (see Sect.\,\ref{sec:kin_modelling_kin_sample}), as well as the stellar mass and SFR distributions are shown in Fig.\,\ref{fig:MS_full}. In this figure and in what follows, we have taken out the zero point evolution of the MS by normalising the individual SFR values to a redshift $z_0 = 0.7$ using the prescription

	\begin{equation}
		\log_{10} {\rm{SFR}}_z = \log_{10} {\rm{SFR}} - \alpha \log_{10} \left ( \frac{1+z}{1+z_0} \right ),
	\end{equation}
	where SFR and SFR$_z$ are the unnormalised and normalised SFR, respectively, and $\alpha$ is a scale factor. We used a value of $\alpha~=~2.8$ from \citet{speagle_highly_2014}, which is larger than the value of $\alpha = 1.74$ derived and used in \citet{boogaard_muse_2018} and \citet{Abril-Melgarejo2021}. The main reason for normalising the redshift evolution with a larger slope is that the prescription from \citet{boogaard_muse_2018} was derived on the low mass end ($\log_{10} M_{\star}/{\rm{M_{\odot}}} \lessapprox 9$) of the MS. However, most of our galaxies have stellar masses larger than this threshold where the redshift evolution of the MS is much steeper \citep[e.g.][]{whitaker_constraining_2014}.
	
	\section{Galaxies morphology}
	\label{sec:morpho}
	
	\subsection{Morphological modelling}
	\label{sec:morpho modelling}
	
	To recover the galaxies morphological parameters, we performed a multi-component decomposition using the modelling tool \Galfit{} on HST-ACS images observed with the F814W filter. In order to have a fair comparison with previous findings from \citet{Abril-Melgarejo2021}, we used the same methodology to model the morphology of galaxies. Therefore, we performed a multi-component decomposition with
	\begin{enumerate*}[label=(\roman*)]
		\item a spherically symmetric de Vaucouleurs profile\footnote{Sérsic profile with fixed Sérsic index $n=4$, axis-ratio $b/a=1$, and $\rm{PA}=\SI{0}{\degree}$} aimed at modelling the central parts of the galaxies (hereafter bulge),
		\item a razor-thin exponential disk \footnote{Sérsic profile with a fixed Sérsic index $n=1$} describing an extended disk (hereafter disk).
	\end{enumerate*}
	In most cases, we expect the disk component to dominate the overall flux budget, except within the central parts where the bulge is usually concentrated. In very rare cases where the galaxies do not show any bulge component, \Galfit{} always converged towards a disk component only model. On the opposite, in the case of elliptically shaped galaxies, \Galfit{} usually converges towards a single de Vaucouleurs component. We do not systematically try to model additional features which may appear in very few cases such as clumps, central bars or spiral arms. When clumps do appear, the multi-component decomposition is usually carried out without masking the clumps first. If the clumps seem to bias the morphological parameters of the main galaxy, a second run is done by either masking the clumps or adding other Sérsic profiles at their location. Unless there is no significant improvement in the robustness of the fitting process, the masked model is usually kept. Other cases may be galaxies in pairs or with small sky projected distances, which are modelled with an additional Sérsic profile at the second galaxy location, or out-of-stamps bright stars which can contaminate the light distribution of some galaxies, in which case it is usually modelled with an additional sky gradient.
	
	The aforementioned procedure was applied on the \OII{} emitters sample. Among the \NOIISample{} galaxies, a few of them could not be reliably modelled with neither a bulge-disk decomposition, nor with a single disk or single bulge profile. Such galaxies turned out to be
	\begin{enumerate*}[label=(\roman*)]
		\item low, or very low S/N objects for which the noise is contributing too much to the light distribution to extract reliable morphological parameters,
		\item very small galaxies for which the disk is barely resolved and the bulge not resolved at all.
	\end{enumerate*}
	After removing those cases, we get a morphological sample of \NMorphoSample{} galaxies (i.e. 77\% of the \OII{} sample) which can be reliably modelled using this decomposition. 
	
	\subsection{Morphological properties}
	\label{sec:morpho_properties}
	
	\begin{figure}
		\includegraphics[scale=0.7]{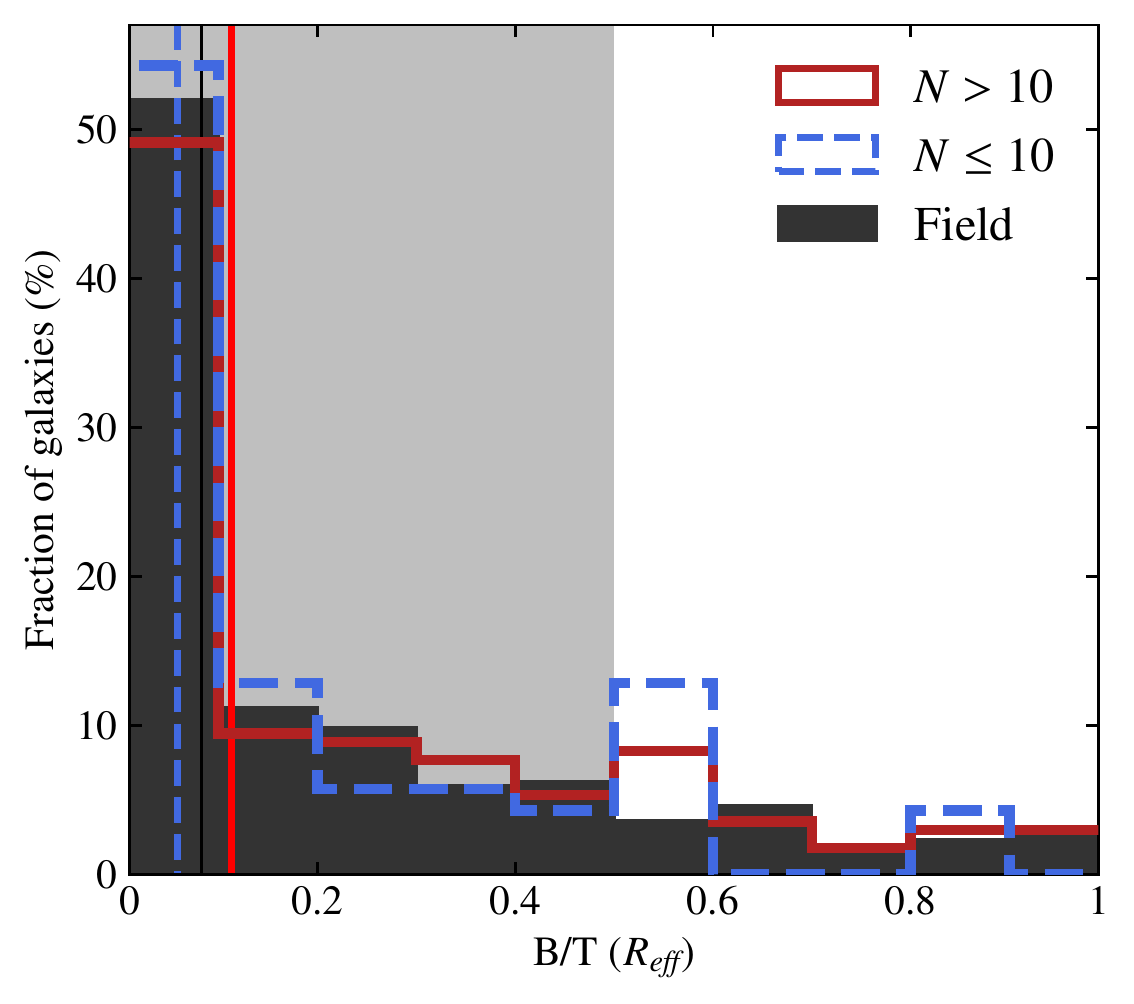}
		\caption{Bulge to total flux ratio distribution computed at one effective radius for galaxies in the morphological sample located in various environments. The legend is similar to Fig.\,\ref{fig:redshift_distribution}. The vertical lines correspond to the median B/T values for each sample. The grey area in the background indicates which galaxies were selected in the kinematic sample (see Sect.\,\ref{sec:kin_modelling_kin_sample}).}
		\label{fig:B/T_distribution}
	\end{figure}
	
	The multi-component decomposition provides two scale parameters, the effective radius of the disk $R_{\rm{\textit{eff}, d}}$, and that of the bulge $R_{\rm{\textit{eff}}, b}$, but, in practice, we are more interested in the effective radius of the total distribution of light in the plane of the disk $R_{\rm{\textit{eff}}}$. Even though there is no analytical formula linking $R_{\textit{eff}}$, $R_{\textit{eff}, \rm{d}}$ and $R_{\textit{eff}, \rm{b}}$, it can be shown from the definition of these three parameters that finding $R_{\textit{eff}}$ amounts to solving the following equation (see Appendix\,\ref{Appendix:bulge_disk_decomposition} for the derivation)
	
	\begin{equation}
		\label{eq:find_re}
		\begin{split}
			& 10^{-{\rm{mag_d}} /2.5} \left [ \gamma \left (2, b_1 \frac{R_{\textit{eff}}}{R_{\textit{eff}, \rm{d}}}  \right ) - 0.5 \right ]  + \\
			& 10^{-{\rm{mag_b}} /2.5} \left [ \gamma \left (8, b_4 \left (\frac{R_{\textit{eff}}}{R_{\textit{eff}, \rm{b}}} \right )^{1/4} \right ) / \Gamma(8) - 0.5 \right ] = 0
		\end{split},
	\end{equation}
	where $\rm{mag_d}$ and $\rm{mag_b}$ stand for the disk and bulge apparent total magnitudes as provided by \Galfit{}, respectively, $b_1 \approx 1.6783$, $b_4 \approx 7.6692$, $\Gamma$ is the complete gamma function, and $\gamma$ the lower incomplete gamma function. Equation\,\ref{eq:find_re} is solved for each galaxy using a zero search algorithm considering the two following additional arguments
	\begin{enumerate*}[label=(\roman*)]
		\item it always admits a single solution,
		\item $R_{\textit{eff}}$ must be located between $R_{\textit{eff}, \rm{d}}$ and $R_{\textit{eff}, \rm{b}}$.
	\end{enumerate*}
	To get an estimate of the error on the effective radius, we generate for each galaxy 1\,000 realisations by perturbing the bulge and disk magnitudes and effective radii using the errors returned by \Galfit{} and assuming Gaussian distributions. For each realisation, we solve Eq.\,\ref{eq:find_re} and then compute the error as the $1\sigma$ dispersion around the median value. The majority of the galaxies in the morphological sample are disk dominated, 80\% of them having a bulge-to-total flux ratio B/T $(R_{\textit{eff}}) < 0.5$, with B/T as defined in Appendix\,\ref{Appendix:bulge_disk_decomposition}. As can be seen in Fig.\,\ref{fig:B/T_distribution}, B/T distributions for galaxies from the morphological sample in the field, small, and large structure subsamples are mostly similar, with very few bulge dominated objects. There appears to be an excess of galaxies located in small structures with respect to field galaxies in the range $0.5 \lesssim {\rm{B/T}} \lesssim 0.6$ but, given the small number of galaxies in this bin (9), this excess may not be significant.
	
	\subsection{Stellar mass correction}
	\label{sec:mass_correction}
	
	\begin{figure}
		\includegraphics[scale=0.7]{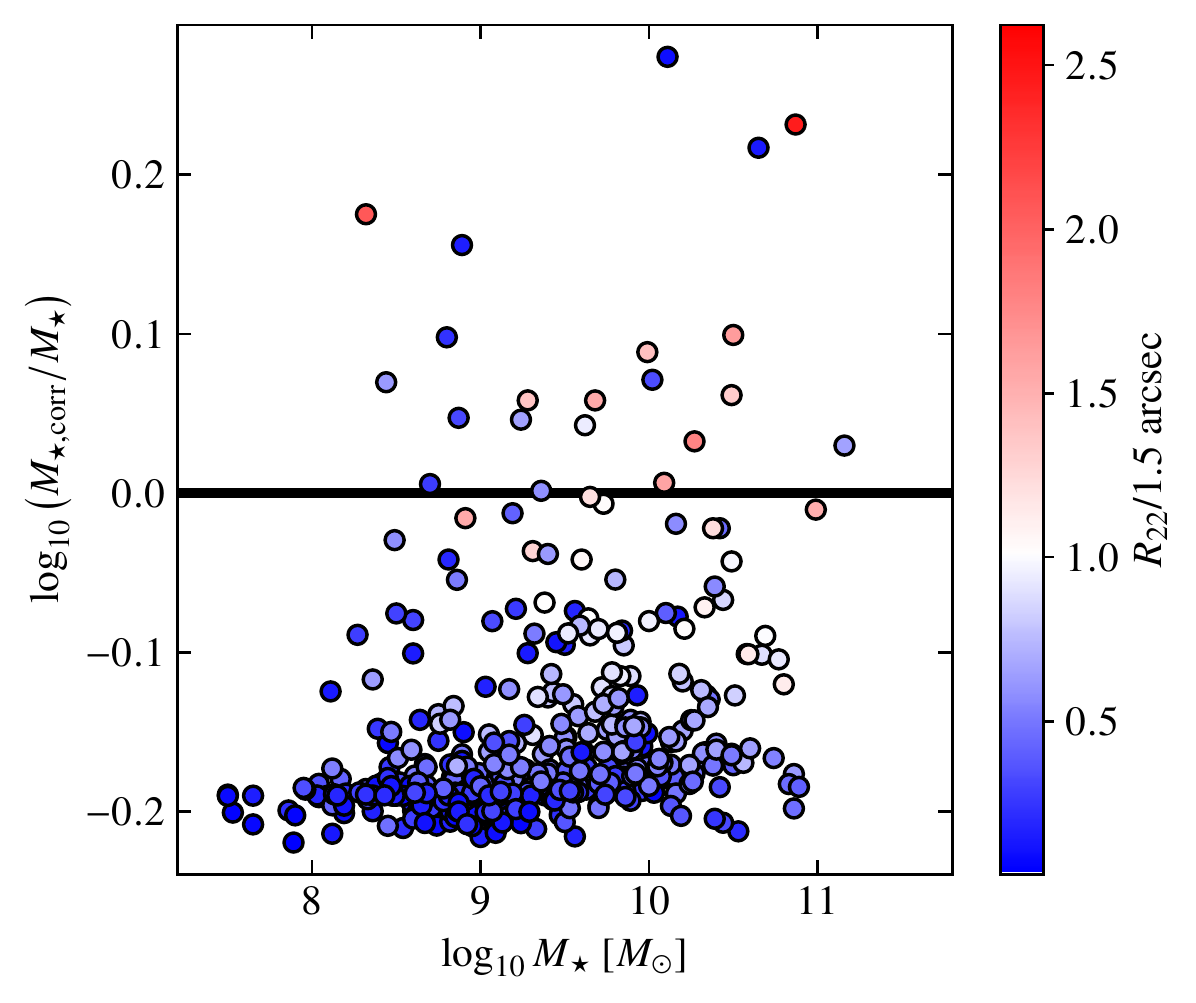}
		\caption{Impact of stellar mass correction as a function of the SED-based stellar mass for galaxies from the morphological sample. Overall, the correction lowers the stellar mass, reducing as much as by a factor of 1.5. We see that the smaller the disk radius $R_{\textit{eff}, \rm{d}}$ (or equivalently $R_{22}$), the larger the stellar mass reduction, consistent with the fact that the SED-based stellar mass computed in an aperture of $\SI{3}{\arcsec}$ usually overestimates the real value, though in practice this effect can be compensated by sky projection and PSF effects.}
		\label{fig:mass_correction}
	\end{figure}
	
	As mentioned in Sect.\,\ref{sec:gal_properties}, the galaxies stellar mass is retrieved from the SED fitting on the photometric bands in a circular aperture of $\SI{3}{\arcsec}$ on the plane of the sky. On the other hand, the gas rotation velocity $V_{22}$ (see Sect.\,\ref{Sec:analysis}), is usually derived at $R_{22} = 2.2 \times R_{\rm{d}}$, where $R_{\rm{d}} = R_{\textit{eff}, \rm{d}} / b_1$ is the disk scale length defined as the e-folding length with respect to the central value. This means that the SED-based stellar mass corresponds to the integrated mass within a cylinder of diameter $\SI{3}{\arcsec}$ orthogonal to the plane of the sky, whereas the kinematic is derived from the contribution of the mass located within a sphere of radius $R_{22}$. Therefore, directly comparing the kinematics with the SED-based stellar mass in scaling relations such as the TFR adds additional uncertainties due to projection effects (inclination), different sizes ($R_{\textit{eff}, \rm{d}}$, $R_{\textit{eff}, \rm{b}}$), and different bulge and disk contributions (B/D). Thus, we decided to apply a correction to the SED-based stellar mass estimate in the following way, assuming a constant mass to light ratio across the galaxy
	
	\begin{equation}
		M_{\star, \rm{corr}} = \frac{F_{\rm{sph}}~(R_{22})}{F_{\rm{circ}}~(\SI{1.5}{\arcsec})} M_{\star},
		\label{eq:stellar_mass_correction}
	\end{equation}
	where $M_{\star}$ and $M_{\star, \rm{corr}}$ are the uncorrected and corrected stellar masses measured in a $\SI{3}{\arcsec}$ circular aperture on the plane of the sky and in a sphere of radius $R_{22}$ around the galaxy centre, respectively. In Eq.\,\ref{eq:stellar_mass_correction}, $F_{\rm{sph}}$ corresponds to the integrated flux in a sphere of radius $R_{22}$, while $F_{\rm{circ}}$ corresponds to the integrated flux in a $\SI{3}{\arcsec}$ circular aperture on the plane of the sky.
	
	In order to compute the mass correction, a high resolution 2D model was generated for each galaxy, projected on the sky given the axis ratio returned by \Galfit{}, and taking into account the impact of the MUSE PSF, whereas the flux in a sphere of radius $R_{22}$ was integrated without taking into account the impact of the inclination, nor convoluting the surface brightness profile with the PSF. Taking into account the impact of the inclination and the PSF is important for the sky-projected model since the flux is integrated in a fixed aperture. Indeed, a higher inclination will result in integrating the flux to larger distances along the minor axis, whereas higher PSF FWHM values will result in loosing flux since it will be spread further out. On the other hand, because the dynamical mass is derived in Sect.\,\ref{sec:kin_modelling} from a forward model of the ionised gas kinematics taking into account the geometry of the galaxy and the impact of the PSF, the flux model integrated within a sphere of radius $R_{22}$ must be pristine of projection and instrumental effects (i.e. inclination and PSF).
	
	The impact of the stellar mass correction is shown in Fig.\,\ref{fig:mass_correction}. For most galaxies the correction reduces the stellar mass, reaching at its maximum a factor of roughly 1.5. The main reason is that for $R_{22} < \SI{1.5}{\arcsec}$, the lower the disk effective radius, the more overestimated the SED-based stellar mass should be, though this argument must be mitigated by the fact that the inclination, the bulge contribution, and the PSF convolution can also play an important role in some cases, explaining why some galaxies have positive stellar mass corrections even with small disk effective radii.

	\subsection{Stellar disk inclination and thickness}
	\label{sec:disk_thickness}
	
	\begin{figure}[hbt!]
		\includegraphics[scale=0.7]{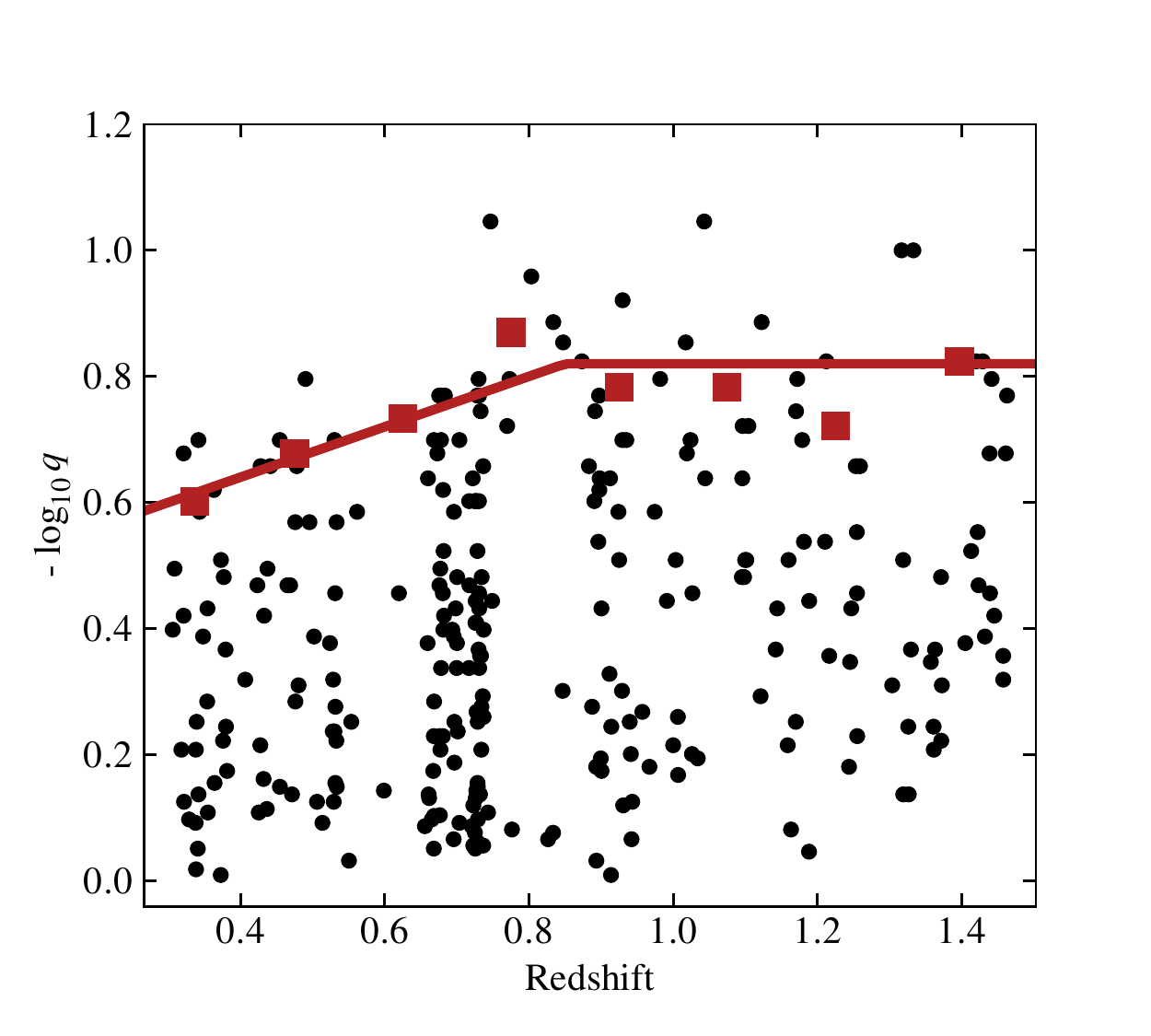}
		\caption{Observed axis ratio $q$ as a function of redshift for galaxies from the morphological sample (black points) after removing bulge dominated galaxies and those with small disk sizes. The median values for the six most edge-on galaxies in redshift bins of width $\Delta z = 0.15$ are shown as red squares. The red line represents the thickness prescription which was applied. Independently of mass, galaxies tend to have thinner disks at larger redshifts which may be due to the fact that we probe younger stellar populations at higher redshifts when observing in a single band.}
		\label{fig:inc_vs_z}
	\end{figure}
	
	\begin{figure}[hbt!]
		\includegraphics[scale=0.7]{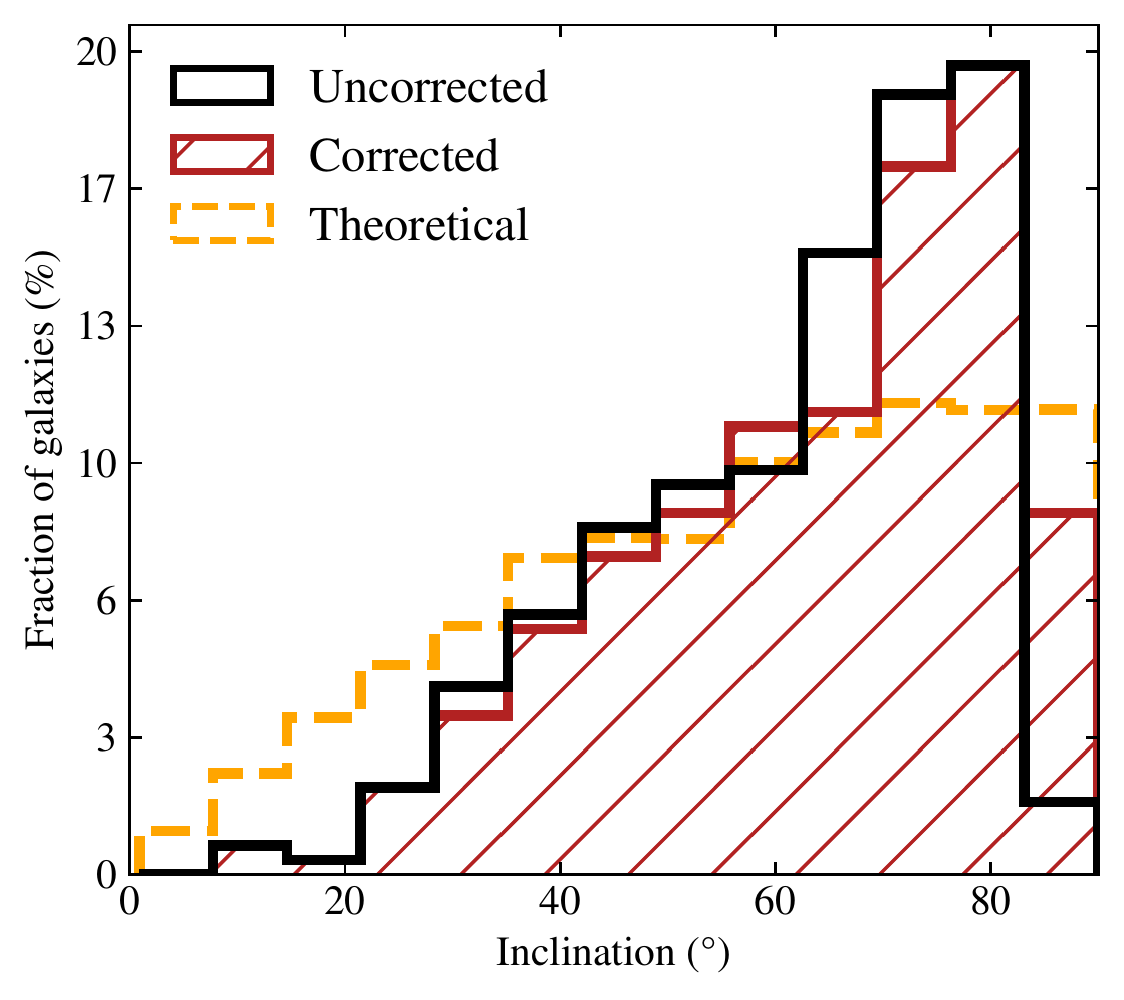}
		\caption{Distribution of disk inclination for galaxies from the morphological sample, after removing bulge dominated galaxies and those with small disk sizes. We show the distribution before correcting for the finite thickness of the disk (black line) and after the correction (red hatched area). The orange dashed line represents the binned theoretical distribution expected for randomly orientated disk galaxies. The correction tends to increase the fraction of edge-on galaxies. While being closer to the theoretical distribution at large inclinations, the corrected inclinations still do not match the distribution of randomly inclined galaxies.}
		\label{fig:inc_distribution}
	\end{figure}
	
	In Sect.\,\ref{sec:morpho modelling}, we have assumed that the surface brightness of the stellar disk can be represented by a razor-thin exponential profile, but in practice we expect most disk components to have non-zero thickness. Not taking into account this finite thickness can bias morphological and kinematic measurements, especially in the central parts, and the circular velocity. In turn, this can bias the derived dynamical parameters such as the baryon fraction. This effect becomes even more relevant when considering that the stellar disks thickness is expected to evolve with redshift and mass. By modelling the $q = b/a$ distribution, with $a$ and $b$ the apparent major and minor axes of the disk, respectively, for star-forming $z \leq 2.5$ galaxies in the CANDELS field and from the SDSS catalogue, \citet{van_der_wel_geometry_2014} found that galaxy disks become thicker with increasing stellar mass and at larger redshift. Similarly, \citet{zhang_evolution_2019}, by looking at the $q - \log a$ plane, reached a fairly similar conclusion. On top of that, galaxies exhibiting a combination of a blue thin and a red thick stellar disks are expected to have an observed thickness which varies with rest-frame wavelength. This effect can be observed in the catalogue of edge-on SDSS galaxies of \citet{bizyaev_catalog_2014}, where the disk thickness of $z \lesssim 0.05$ galaxies tends to almost systematically increase when measured in the $g$, $r$, and $i$ bands, respectively. In order to get an estimate of the disk thickness in our sample of galaxies, we used the methodology described in \citet{heidmann_inclination_1972} and \citet{bottinelli_h_1983}. If galaxies located at a given redshift $z$, with a fixed stellar mass $M_\star$, and emitting at a fixed rest-frame wavelength $\lambda$ have a typical non-zero thickness $q_0(\lambda, z, M_\star)$, then the observed axis ratio $q$ for the majority of the galaxies should reach a minimum value equal to $q_0$ for edge-on galaxies. In our case, because the morphology is derived at a fixed observed wavelength $\lambda_{\rm{obs}} \approx \SI{8140}{\angstrom}$ (F814W HST filter), this condition can be written as
	
	\begin{equation}
		\label{eq:thickness_criterion}
		q_0 \left (\lambda_{\rm{obs}}/(1+z), z, M_\star \right ) \lesssim q,
	\end{equation}
	where $\lambda_{\rm{obs}}$ is the observed wavelength. The distribution of the observed axis ratio as a function of redshift is shown in Fig.\,\ref{fig:inc_vs_z}. We see that the minimum observed axis ratio (i.e. highest $-\log_{10} q$) seems to decrease with redshift up to $z \approx 0.8-0.9$ and remains roughly constant afterwards. This trend, which seems inconsistent with the fact that the disk thickness has been previously observed to increase with redshift, can be explained by the fact that higher redshift galaxies are seen at a bluer rest-frame wavelength which probes younger stellar populations, and probably thinner disks. Due to the lack of edge-on galaxies in various mass bins, we do not observe a clear dependence of $q$ on stellar mass, and therefore decided to model only the redshift dependence of $q$. In order to avoid placing too much weight on outliers which may have thinner disks than the typical thickness expected at a given redshift, we separated galaxies in eight redshift bins and computed the median thickness of the six most edge-on galaxies in each bin. The dependence of the stellar disk thickness with redshift is given by
	
	\begin{equation}
		- \log_{10} q_0 =
		\begin{cases}
			0.48 + 0.4 z & \mbox{if }z \leq 0.85 \\
			0.48 & \mbox{otherwise}.
		\end{cases}
		\label{Eq:finite_thickness}
	\end{equation}
	
	In the case of a razor-thin disk, the inclination $i$ is related to the observed axis-ratio $q$ through the relation $\cos i = q$. However, for a disk with non-zero thickness, the relation between $i$ and $q$ will depend on the exact geometry of the disk. Assuming our disk galaxies can be well approximated by oblate spheroidal systems, we have \citep{bottinelli_h_1983}
	
	\begin{equation}
		\label{Eq:correction_inclination1}
		\cos^2 i = (q^2 - q_0^2) / (1 - q_0^2).
	\end{equation}
	
	In Fig.\,\ref{fig:inc_distribution}, we show the distribution of the disk inclination for galaxies from the morphological sample (see Sect.\,\ref{sec:kin_modelling_kin_sample}) assuming razor-thin disks (black line), and after applying the thickness correction using Eqs.\,\ref{Eq:finite_thickness} and \ref{Eq:correction_inclination1} (red hatched area). As expected, correcting for the disk thickness significantly increases the number of edge-on galaxies. Nevertheless, compared to the theoretical distribution (orange line), none of the distributions are consistent with randomly inclined galaxies. We find that we have an excess of galaxies in the range $\SI{60}{\degree} \lesssim i \lesssim \SI{80}{\degree}$. The reason why we are still missing some edge-on galaxies ($i > \SI{80}{\degree}$) might be that we did not try to model the impact of the dust which is known to affect more severely edge-on galaxies. Nevertheless, the inclination distribution we get is quite similar to the distributions found in other studies where they also lack edge-on galaxies \citep{padilla_stellar_2009, foster_sami_2017}.
	
	\section{Galaxies kinematic}
	\label{sec:kin_modelling}
	
	\begin{figure}[hbt!]
		\includegraphics[scale=0.7]{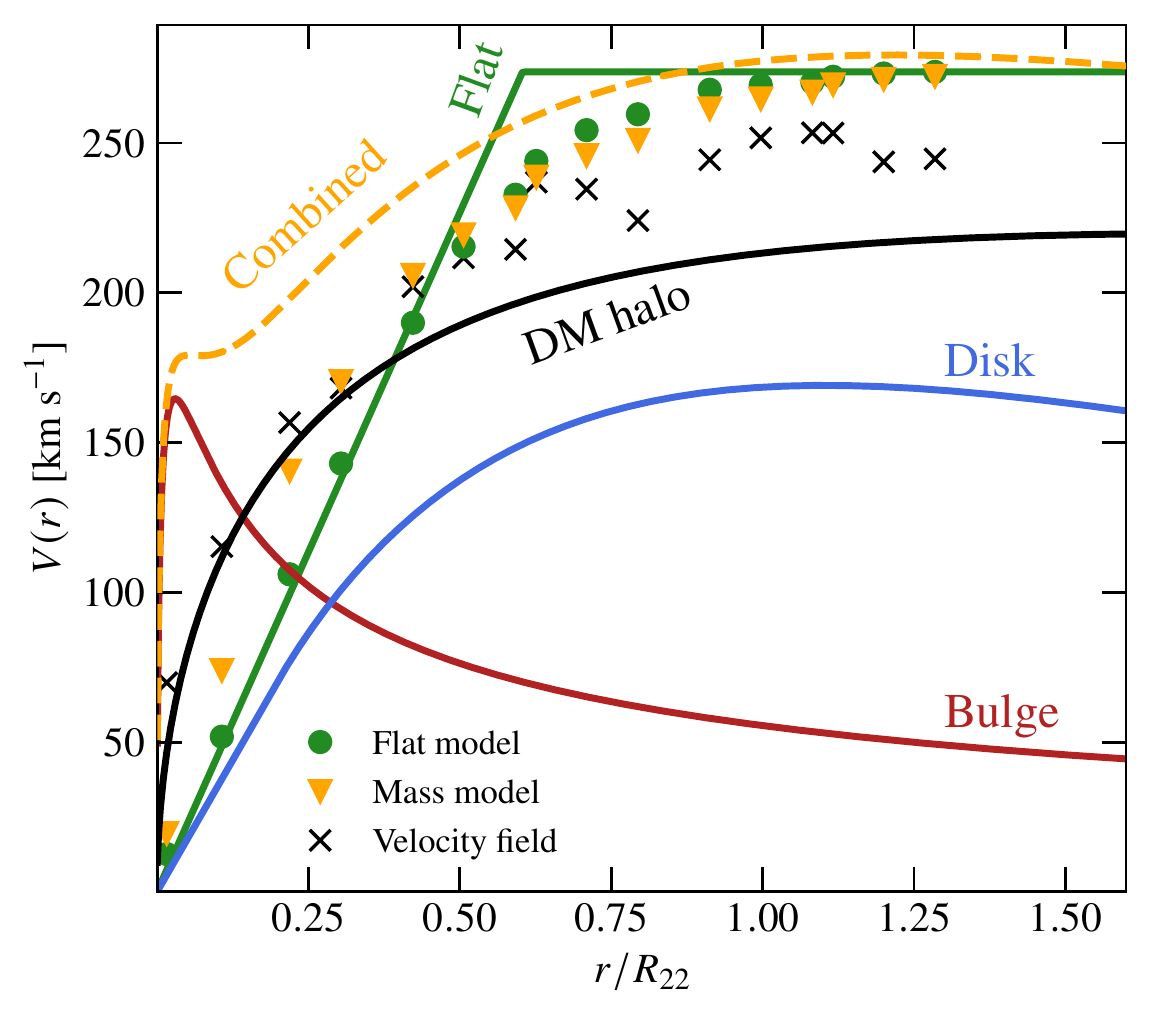}
		\caption{Rotation curves for the flat (green line) and mass models (orange dashed) of galaxy \textsc{104-CGr79} at redshift $z = 0.53$. The components are the bulge (red), the thin disk (blue), and the DM halo (black). We also show the de-projected (but beam-smeared) observed rotation curves extracted along the major axis from the observed velocity field map (black crosses), from the best-fit velocity field flat model (green circles), and from the best-fit velocity field mass model (orange triangles). The largest difference between the flat and mass models is found in the inner parts where the beam smearing is the strongest. The total dynamical mass differs slightly between models, with the flat one being 4\% higher than the mass model one.}
		\label{fig:comp_mass_models}
	\end{figure}
	
	\begin{figure*}
		\centering
		\includegraphics[scale=0.5]{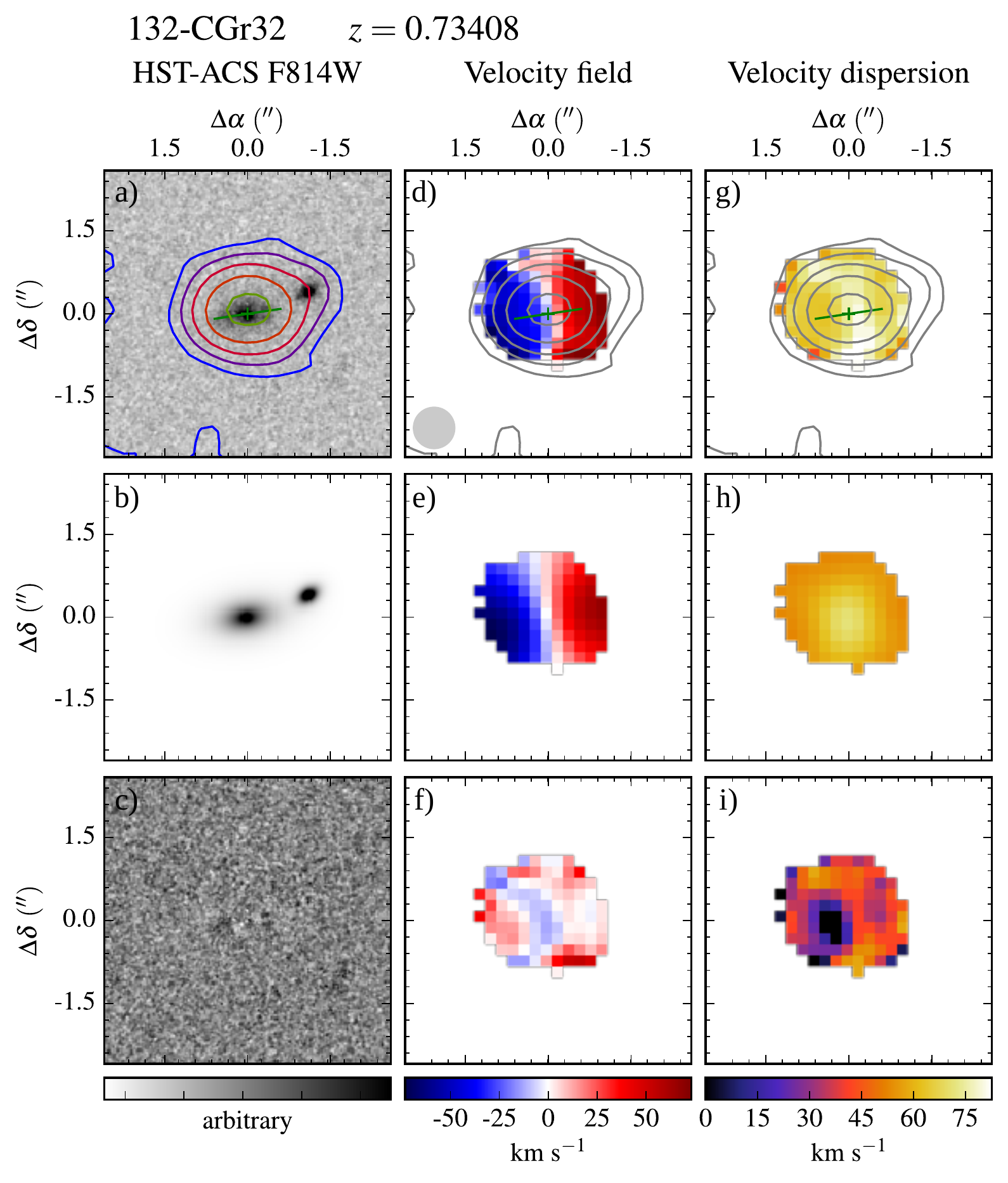}
		\includegraphics[scale=0.5]{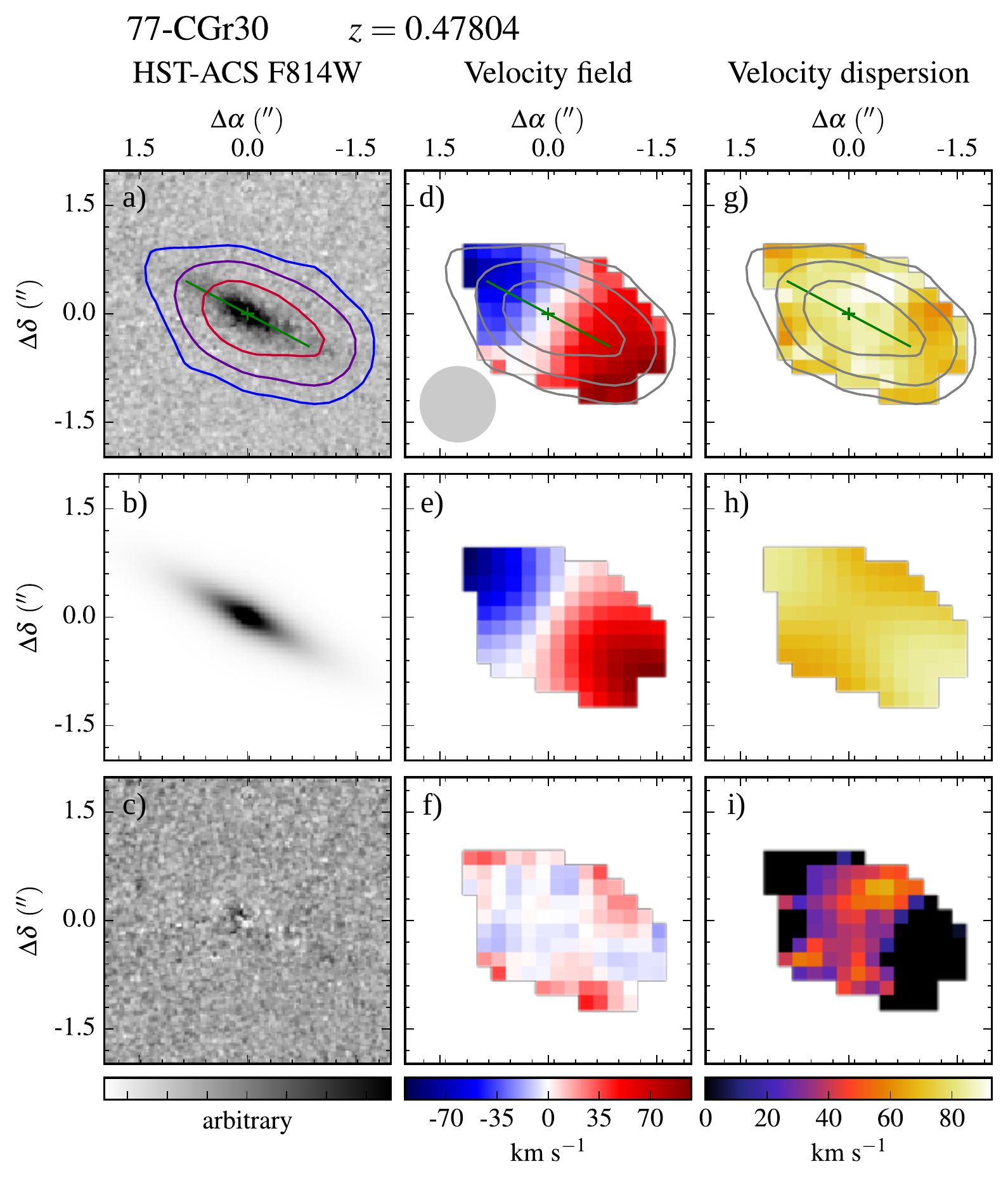}
		\includegraphics[scale=0.5]{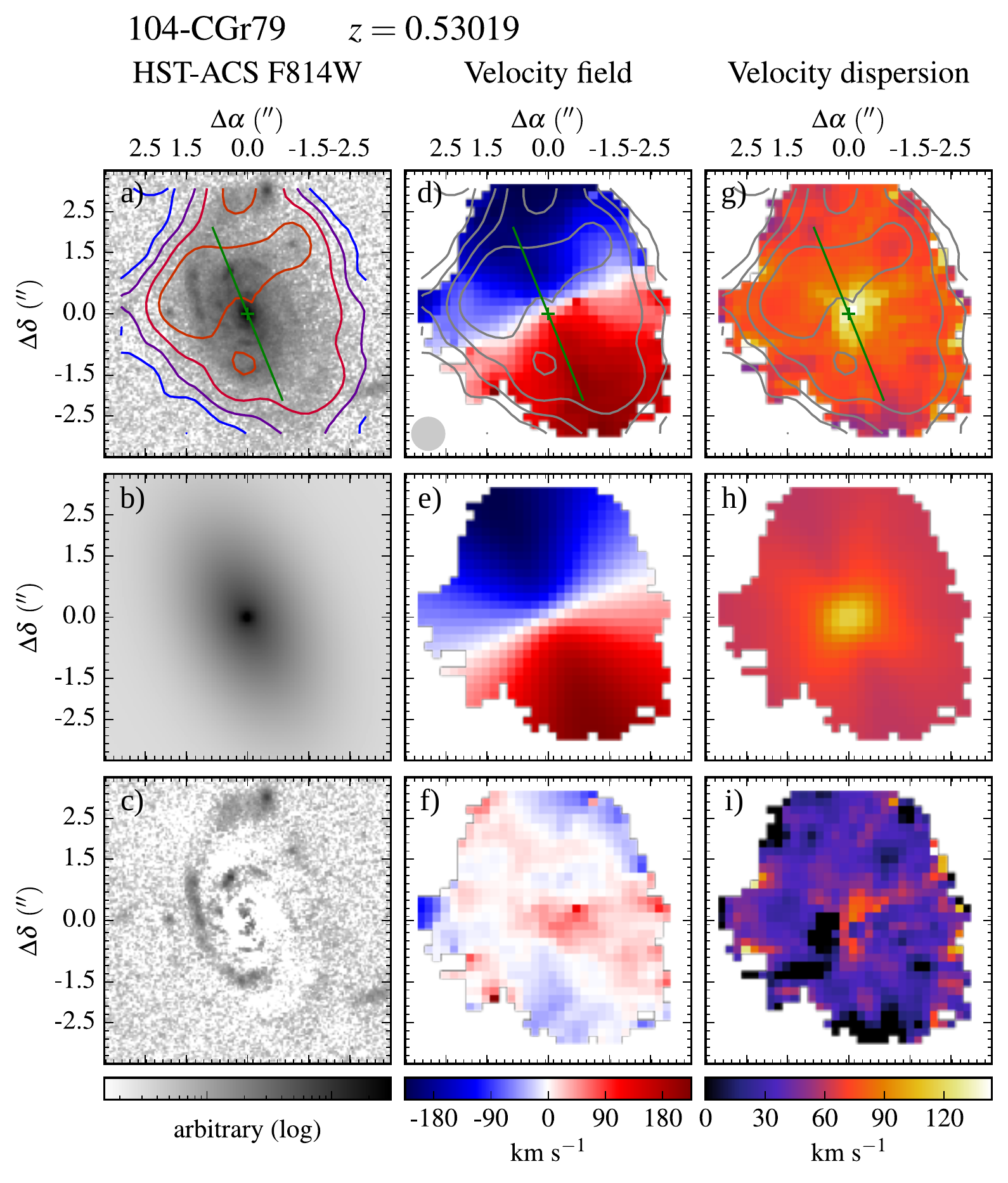}
		\includegraphics[scale=0.5]{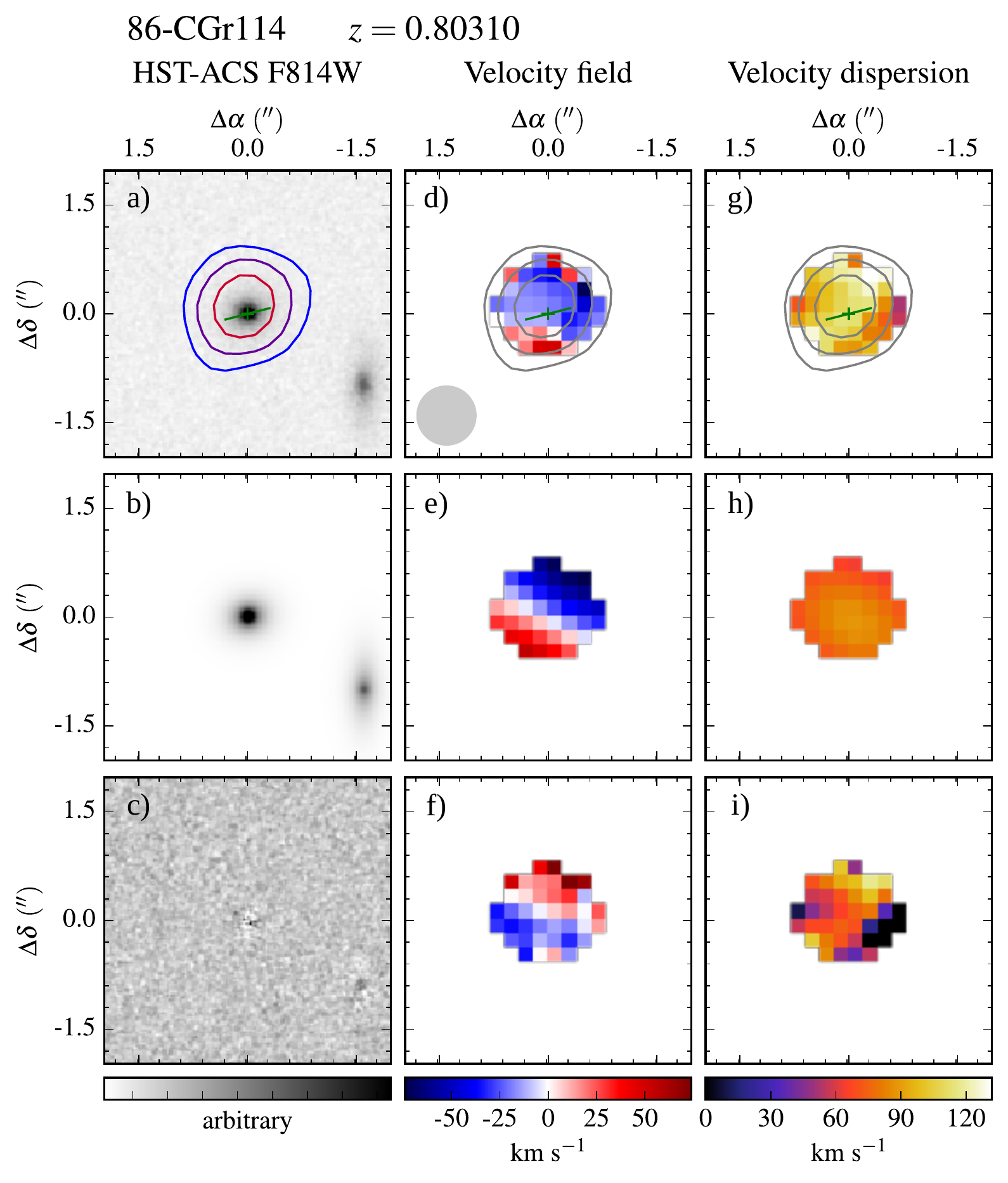}
		\caption{Examples of morpho-kinematic modelling for galaxies with IDs \textsc{132-CGr32} and \textsc{104-CGr79} (both in large structures), \textsc{77-CGr30} and \textsc{86-CGr114} (both in the field). In each panel, from top to bottom and left to right: (a) HST-ACS image, (b) \Galfit{} model, (c) HST residuals, (d) \Camel{} velocity field, (e) \Mocking{} velocity field model, (f) Velocity field residuals, (g) \Camel{} velocity dispersion map, (h) \Mocking{} beam smearing model including spectral resolution broadening, (i) beam smearing and LSF corrected velocity dispersion map. The morpho-kinematic centre and the morphological position angle are shown in the HST image and the \Camel{} maps as a green cross and a green line whose length corresponds to $R_{22}$, respectively. The PSF FWHM is indicated as the grey disk in the velocity field. The \OII{} surface brightness distribution is overlaid on top of the HST and MUSE \OII{} flux maps with contours at levels $\Sigma_{\OII{}} = 2.5, 5, 10, 20, 40~{\rm{and}}~\SI{80e-18}{erg~s^{-1}~cm^{-2}~arcsec^{-2}}$.}
		\label{fig:examples_models}
	\end{figure*}
	
	\subsection{Kinematic modelling}
	\label{sec:kin_modelling_kin_sample}
	
	Following the analysis in \citet{Abril-Melgarejo2021}, we derived the ionised gas kinematics from the \OII{} doublet only. For each galaxy, we extracted a sub-datacube with spatial dimensions $30\times 30$ pixels around their centre and then performed a sub-resolution spatial smoothing using a 2D Gaussian kernel with a FWHM of 2 pixels in order to increase the S/N per pixel without worsening the datacube spatial resolution. From this smoothed version of the datacube, the \OII{} doublet was fitted spaxel by spaxel by two Gaussian profiles with rest-frame wavelengths of $\SI{3727}{\angstrom}$ and $\SI{3729}{\angstrom}$ respectively, assuming identical intrinsic velocity and velocity dispersion. Additionally, given the assumed photo-ionization mechanisms producing the \OII{} doublet \citep{Osterbrock2006}, we further constrained the flux ratio between the two lines as $0.35 \leq F_{\OII{}\lambda 3727} / F_{\OII{}\lambda 3729} \leq 1.5$. The aforementioned steps were performed with the emission line fitting python code \Camel{}\footnote{\url{https://gitlab.lam.fr/bepinat/CAMEL}}, using a constant value to fit the continuum, and the MUSE variance cubes to weight the fit and estimate the noise. From this procedure, we recovered 2D maps for the following quantities: \OII{} fluxes, S/N, velocity field, and velocity dispersion, as well as their corresponding spaxel per spaxel error estimation from the fit. To avoid fitting any noise or sky residuals which might appear in the flux and kinematic maps, especially in the outer parts of the galaxies, we cleaned the 2D maps in two successive steps
	
	\begin{enumerate}[label=(\roman*)]
		\item through an automatic procedure, only keeping spaxels with $\rm{S/N} \geq 5$ and $\rm{FWHM}_{\OII{}} \geq 0.8 \times \rm{FWHM_{LSF}} (z)$, where $\rm{FWHM}_{\OII{}}$ and $\rm{FWHM_{LSF}}$ are the \OII{} spatial PSF and spectral LSF FWHM, respectively,
		\item by visually inspecting the automatically cleaned velocity fields and manually removing remaining isolated spaxels or those with large velocity discontinuities with respect to their neighbours.
	\end{enumerate}
	
	This led to the removal of 293 galaxies from the morphological sample (around 30\%), mainly because they did not show any velocity field in their cleaned maps due to too low S/N per pixel. Because this cleaning process is mainly driven by S/N considerations, it is roughly similar to applying an \OII{} integrated flux selection criterion of $F_{\OII{}} \gtrsim 2 \times \SI{e18}{erg~s^{-1}~cm^{-2}}$.
	
	The kinematics of the ionised gas in the remaining galaxies was modelled as a razor-thin rotating disk, using the method of line moments as described in \citet{Epinat2010}. This method can quickly derive velocities maps by combining rotation curves\footnote{We use the term rotation curve to refer to the circular velocity as a function of radius of models, and explicitly write that it is observed otherwise.} from various components, taking into account the impact of spatial resolution on the derived velocity field, and the combined effect of the limited spatial and spectral resolutions on the velocity dispersion map. To derive the kinematics (circular velocity and velocity dispersion), we performed a mass modelling, taking into account prior knowledge from the morphological modelling. By using the best-fit \Galfit{} bulge and disk parameters from Sect.\,\ref{sec:morpho modelling}, and the disk thickness derived in Sect.\,\ref{sec:disk_thickness}, we fixed the rotation curves of the stellar disk and bulge components. Below we provide the main characteristics of the mass models used in the modelling, and we refer the reader to Appendix\,\ref{Appendix:mass_modelling} for a detailed description of the models and rotation curves, as well as how implementing a finite thickness for the stellar disk impacts the estimate of the rotation of the gas. We assumed a double exponential density profile for the disk, which provides a surface brightness profile which is fairly similar to a single exponential distribution once projected onto the sky. In order to derive a density profile for the bulge component, one would need to numerically solve the inverse Abel transform for a de Vaucouleurs profile. But, because we required to have an analytical form for the bulge density, we decided to use instead a Hernquist profile \citep{hernquist_analytical_1990}. As shown in Fig.\,\ref{fig:sersic_hernquist} and \ref{fig:comparison_sersic_Hernquist}, for the typical bulge parameters found in our sample, this functional form gives fairly reasonable surface brightness profiles once projected onto the sky. Finally, the dark matter halo was modelled using a Navarro–Frenk–White (NEW) profile \citep{navarro_structure_1996} with free parameters to account for constant or slowly declining observed rotation curves at large radii. This choice of DM parametrisation may not be entirely suitable with respect to observations which favour cored DM distributions. However, the core-cusp problem mainly affects the inner parts of the profiles. On the other hand, our goal is not to study the shape of DM haloes as a function of radius but rather to derive the ionised gas kinematics, the baryon, and the DM fractions where the inner shape of the DM halo has little impact on these quantities \citep{korsaga_ghasp_2019}. In addition, because beam smearing strongly affects ground based observations of intermediate redshift galaxies, constraining robustly the inner DM halo distribution in detail remains a challenging problem but within reach \citep[e.g.][]{genzel_rotation_2020, bouche_muse_2021}. The effect of beam smearing can be seen in Fig.\,\ref{fig:comp_mass_models} where we compare the best-fit rotation curves between a mass model and a simpler flat model for galaxy {\textsc{104-CGr79}}. Even though the intrinsic rotation curves in the inner parts differ (dashed orange line versus green full line), the deprojected (but beam-smeared) rotation curves are almost the same.
	
	For each galaxy, a 2D velocity field model is generated and fitted onto the velocity field extracted from the cube. Since beam smearing artificially increases the value of the velocity dispersion recovered from the cube, especially near the central parts, modelling it and quadratically removing it from the velocity dispersion map allows us to extract a much more reliable estimator of the overall velocity dispersion in each galaxy.
	Given the above description, the kinematic model requires the following parameters:
	\begin{enumerate*}[label=(\roman*)]
		\item centre coordinates,
		\item inclination,
		\item kinematic PA,
		\item systemic redshift $z_{\rm{s}}$,
		\item disk rotation curve parameters $V_{\rm{RT, max}}$, $V_{\rm{corr, max}}$, $R_{\rm{d}}$ (see Appendix\,\ref{Appendix:thin_disk}),
		\item bulge rotation curve parameters $V_{\rm{b, max}}$, $a$ (see Appendix\,\ref{Appendix:bulge_model}),
		\item DM halo rotation curve parameters $V_{\rm{h, max}}$ and $r_{\rm{s}}$ (see Appendix\,\ref{Appendix:DM_halo}),
		\item PSF size.
	\end{enumerate*}
	However, there exists a strong degeneracy between the kinematic centre and $z_{\rm{s}}$ on one side, and the inclination of the disk and $V_{\rm{h, max}}$ on the other side, which is even stronger when the data is highly impacted by beam smearing. Therefore, to remove this degeneracy we fixed the kinematic centre and inclination assuming they are identical to their morphological counterparts. As previously stated, we also fix the parameters of the disk and bulge components since we assume they are entirely constrained from the morphology. Thus, the centre coordinates, the inclination, the disk and bulge rotation curve parameters ($V_{\rm{RT, max}}$, $V_{\rm{corr, max}}$, $R_{\rm{d}}$, $V_{\rm{b, max}}$ and $a$) and the PSF model are fixed, whereas the kinematic PA, the systemic redshift and the DM halo rotation curve parameters ($V_{\rm{h, max}}$ and $r_{\rm{s}}$) are free.
	
	The kinematic modelling described above was performed with the new kinematic fitting code \Mocking{}\footnote{\url{https://gitlab.lam.fr/bepinat/MocKinG}} using the python implementation of \Multinest{} \citep{MULTINEST, PyMULTINEST}. \Multinest{} is a bayesian tool using a multinodal nested sampling algorithm to explore parameter space and extract inferences, as well as posterior distributions and parameter error estimation. To check our results, we ran \Mocking{} a second time but using this time the Levenberg–Marquardt algorithm, with the python implementation \textsc{cat\_mpfit}\footnote{\url{https://www-astro.physics.ox.ac.uk/~mxc/software/}} of \MPFIT{} \citep{MPFIT}. Kinematic parameters were compared between these two methods as well as with earlier results obtained with an IDL code used in several previous studies \citep{Epinat2009, Epinat2010, Epinat2012, Vergani2012, Contini2016, Abril-Melgarejo2021}. A comparison of circular velocities obtained with \Multinest{} and \MPFIT{} can be found in Fig.\,\ref{fig:rotation_comparison}. We find consistent results between the methods, with \Multinest{} providing more robust results. Thus, we use values from \Multinest{} in the following parts. In addition, we performed a similar kinematic modelling but using an ad-hoc flat model for the rotation curve as described in \citet{Abril-Melgarejo2021}, in order to check the mass modelling and assess its reliability. After checking the morphological, kinematic, and mass models on the remaining galaxies, we decided to remove four additional objects: 
	\begin{enumerate*}[label=(\roman*)]
		\item {\textsc{106-CGr84}}, {\textsc{21-CGr114}} and {\textsc{101-CGr79}} because they show signs of mergers in their morphology and kinematics, which may bias the measure of their dynamics, as well as their stellar mass estimate and thus their mass modelling,
		\item {\textsc{13-CGr87}} because it lies on the edge of the MUSE field with only half of its \OII{} flux map visible.
	\end{enumerate*}
	Once these objects are removed, we get a kinematic sample of \NKinematicSample{} galaxies with morphological and kinematic mass and flat models.
		
	An example of a mass model with its corresponding flat model is shown in Fig.\,\ref{fig:comp_mass_models} for a disk-like galaxy with a non-zero (but weak) bulge contribution. The mass model rotation curve (orange dashed line) for the galaxy, which appears to be dark matter dominated, is consistent with the simpler flat model (green line), especially at $R_{22}$ where the rotation velocity is inferred. Examples of full morpho-kinematic models for four types of galaxies are shown in Fig.\,\ref{fig:examples_models} with, on the top left corner, a galaxy with a close companion in its HST image and with a velocity field similar to that of a large fraction of galaxies in our sample, on the top right corner an edge-on galaxy, on the bottom left corner a large disk dominated galaxy with visible arms and clumps, and on the bottom right corner a small galaxy with a prominent bulge and a highly disturbed velocity field. These four examples give a decent overview of the types of galaxies, morphologies and kinematics we have to deal with in the MAGIC survey.
	
	\section{Samples selection}
	\label{sec:selection}
	
	\subsection{Selection criteria}
	\label{sec:samples_selection}
	
	\begin{table*}
		\centering
		\caption{Median properties for various subsamples of galaxies from the MS sample.}
		\begin{tabular}{cccccccc}
		\hline
		\hline
		Subsample & Number & Proportion (\%) & $\log_{10} M_{\star}$ & $\log_{10} M_{\rm{g}}$ & $\log_{10} M_{\rm{dyn}}$ & $\log_{10}$ SFR$_z$ & $R_{\textit{eff}, \rm{d}}$ \\ 
		          &     &     & [M$_{\odot}$] & [M$_{\odot}$]     & [M$_{\odot}$] & [M$_{\odot}$ yr$^{-1}$] & [kpc] \\
		(1)       & (2) & (3) & (4)           & (5)               & (6)           & (7)                     & (8) \\
		\hline \\[-9pt]
		Field     & 256 & 57  & $9.0^{+0.7}_{-0.7}$ & $8.6^{+0.6}_{-0.6}$ & $\ \ 9.8^{+0.8}_{-0.9}$ & $-0.1^{+0.6}_{-0.7}$ & $2.6^{+2.4}_{-1.4}$ \\[3pt]
		Small     & 56  & 13  & $9.0^{+0.9}_{-0.5}$ & $8.7^{+0.6}_{-0.4}$ & $\ \ 9.8^{+0.6}_{-0.8}$ & $-0.2^{+0.8}_{-0.5}$ & $2.8^{+2.2}_{-1.3}$ \\[3pt]
		Large     & 135 & 30  & $9.5^{+0.7}_{-0.9}$ & $8.7^{+0.6}_{-0.7}$ & $10.2^{+0.7}_{-1.0}$    & $-0.2^{+0.6}_{-0.5}$ & $3.3^{+2.5}_{-1.6}$ \\[3pt]
		\hline \\[-9pt]
		Small-5   & 293 & 66  & $9.0^{+0.7}_{-0.6}$ & $8.6^{+0.6}_{-0.6}$ & $\ \ 9.8^{+0.8}_{-0.9}$ & $-0.1^{+0.6}_{-0.7}$ & $2.6^{+2.4}_{-1.4}$ \\[3pt]
		Large-5   & 154 & 34  & $9.5^{+0.7}_{-0.9}$ & $8.7^{+0.6}_{-0.6}$ & $10.1^{+0.8}_{-0.9}$    & $-0.2^{+0.6}_{-0.5}$ & $3.2^{+2.5}_{-1.6}$ \\[3pt]
		\hline \\[-9pt]
		Small-10  & 312 & 70  & $9.0^{+0.7}_{-0.6}$ & $8.6^{+0.6}_{-0.6}$ & $\ \ 9.8^{+0.8}_{-0.9}$ & $-0.1^{+0.7}_{-0.7}$ & $2.6^{+2.3}_{-1.4}$ \\[3pt]
		\hline \\[-9pt]
		Small-15  & 345 & 77  & $9.1^{+0.7}_{-0.7}$ & $8.7^{+0.6}_{-0.6}$ & $\ \ 9.9^{+0.8}_{-1.0}$ & $-0.1^{+0.8}_{-0.7}$ & $2.7^{+2.4}_{-1.4}$ \\[3pt]
		Large-15  & 102 & 23  & $9.5^{+0.8}_{-0.8}$ & $8.6^{+0.7}_{-0.6}$ & $10.1^{+0.9}_{-0.9}$    & $-0.2^{+0.5}_{-0.5}$ & $3.2^{+2.6}_{-1.5}$ \\[3pt] 
		\hline \\[-9pt]
		Small-20  & 370 & 83  & $9.1^{+0.7}_{-0.7}$ & $8.7^{+0.6}_{-0.6}$ & $\ \ 9.9^{+0.8}_{-1.0}$ & $-0.1^{+0.6}_{-0.7}$ & $2.8^{+2.4}_{-1.5}$ \\[3pt]
		Large-20  & 77  & 17  & $9.5^{+0.7}_{-0.9}$ & $8.5^{+0.7}_{-0.7}$ & $10.0^{+0.7}_{-0.9}$    & $-0.3^{+0.5}_{-0.5}$ & $3.2^{+2.1}_{-1.8}$ \\[3pt]
		\hline
		          &     &     &                     &                     &                         &                      & \\
		
		\end{tabular}
		\label{tab:subsamples}
		
		{\small\raggedright {\bf Notes:} (1) Subsample name, we do not show the Large-10 subsample since it is identical to the Large one, (2) Number of galaxies in each subsample, (3) Proportion of galaxies in each subsample, (4) SED-based stellar mass, (5) gas mass derived from the extinction corrected \OII{} flux, (6) dynamical mass from the mass models including the stellar disk, stellar bulge, and DM halo, (7) \OII{}-based SFR normalised at a redshift $z_0 \approx 0.7$, (8) disk effective radius. Masses are computed within $R_{22} = 2.2 R_{\rm{d}}$, with $R_{\rm{d}}$ the disk scale length. Uncertainties correspond to the 16th and 84th percentiles.\par}
	\end{table*}
	
	\begin{figure}[hbt!]
		\includegraphics[scale=0.7]{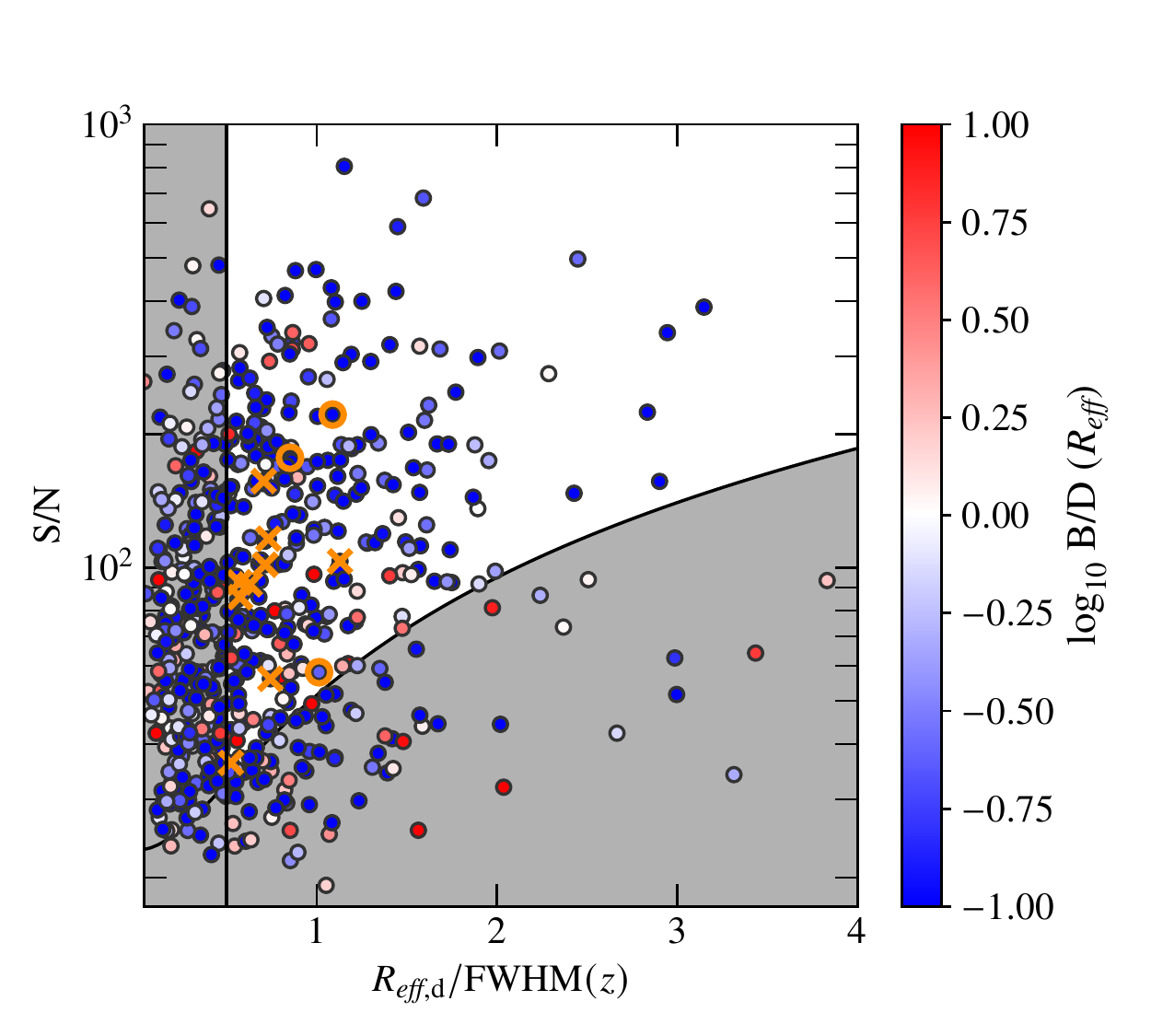}
		\caption{S/N-disk effective radius-B/D selection plot for galaxies from the kinematic sample. The disk size selection criterion is represented as the vertical line. The S/N selection criterion depends on the FWHM which varies with redshift and MUSE field. As an example, we show the S/N limit used for a typical FWHM of $\SI{0.65}{\arcsec}$. Points are colour coded according to their bulge to disk ratio computed at one effective radius. The grey areas give an idea of the galaxies eliminated by the size and S/N selection criteria. We also show the ten galaxies eliminated by selection criterion v (orange crosses) and the three we decided to keep (orange circles).}
		\label{fig:sample_selection}
	\end{figure}

	Before analysing morpho-kinematics scaling relations as a function of environment, and following the discussion in \citet{Abril-Melgarejo2021} (Sect.\,3.6), we must first apply a few selection criteria on the kinematic sample depending on the scaling relation studied. The three relations analysed in this paper are the size-mass relation, MS, and TFR. Among the three, the TFR is the one which requires the most stringent criteria since we must ensure that we have good constraints on both the stellar mass and the kinematic measurements, which translates as having reliable constraints on the disk parameters (size, inclination), on the \OII{} S/N, and on the dynamical modelling. On the other hand, we only require to have disk-dominated MS galaxies to analyse the size-mass and MS relations. Thus, we define a common sample for both the size-mass and MS relations, named the MS sample, by applying the following selection criterion
	
	\begin{enumerate}[label=(\roman*)]
		\item ${\rm{B/D}}~(R_{\textit{eff}}) \leq 1$,
	\end{enumerate}
	where ${\rm{B/D}}~(R_{\textit{eff}})$ is the bulge-to-disk flux ratio computed at one effective radius. This criterion ensures that we only have disk-dominated galaxies in the sample. In \citet{Abril-Melgarejo2021}, we used a second selection criterion to remove red sequence galaxies located below the MS since we were only interested in star forming galaxies. For the kinematic sample, applying this criterion would only remove two additional galaxies, since most of the red sequence galaxies also tend to be bulge dominated. Thus, we decided not to apply this criterion in the next parts. When applying the B/D selection, we end up with a MS sample of \NMSSample{} galaxies.
	
	Concerning the TFR, we must ensure that we have good constraints on the disk size, inclination, and \OII{} S/N, as well as on the dynamical modelling, since they can all have significant impact on the kinematics and the derived dynamical masses. To ensure the TFR is not impacted by poor constraints on any of these parameters, we apply the following additional criteria on top of the B/D selection
	
	\begin{enumerate}[label=(\roman*)]
		\setcounter{enumi}{1}
		\item $R_{\textit{eff}, \rm{d}} \geq 0.5 \times {\rm{FWHM}} (z)$,
		\item $(\rm{S/N})_{\rm{tot}} \geq 40 \times \sqrt{\pi \left [R_{\textit{eff}, \rm{d}}^2 + \left ({\rm{FWHM}}(z) / 2 \right )^2 \right ]}$,
		\item $\SI{25}{\degree} \leq i \leq \SI{75}{\degree}$,
		\item $f_{\star} \leq 1 - \Delta f_{\star}$,
	\end{enumerate}
	where $R_{\textit{eff}, \rm{d}}$ is the disk effective radius and FWHM$(z)$ the MUSE PSF FWHM computed at the \OII{} doublet wavelength at the redshift of the galaxy (see Sect.\,\ref{sec:MUSE_data}), both in arcsec. In criterion (iv), $i$ is the inclination after correcting for the finite thickness of the stellar disk, and in (v), $f_{\star} = M_{\star, \rm{corr}} / (M_{\star, \rm{corr}} + M_{\rm{DM}})$ is the stellar fraction, with $M_{\star, \rm{corr}}$ and $M_{\rm{DM}}$ the stellar and dark matter halo mass, respectively, both computed at $R_{22}$. The uncertainty on the stellar fraction $\Delta f_{\star}$ is computed by propagating measurement and fit errors on both the stellar mass and the circular velocity. In (iii), the total S/N is computed as 
	 
	\begin{equation}
		({\rm{S/N}})_{\rm{tot}} = \sum_{x, y} F_{\OII{}} (x, y) \bigg / \sqrt{\sum_{x, y} \left [ \frac{F_{\OII{}} (x, y)}{{\rm{S/N}} (x,y)} \right ]^2},
		\label{eq:SNR}
	\end{equation}		
	where $F_{\OII{}} (x,y)$ and S/N$(x,y)$ correspond to the \OII{} flux and S/N cleaned maps, respectively \citep[see][]{Abril-Melgarejo2021}. Criterion (ii) is used to remove unresolved galaxies, that is for which the stellar disk is smaller than the PSF, and criterion (iii) takes into account the dependence of the S/N on the effective radius, and is derived by assuming a constant surface brightness map, as well as a constant S/N map with a S/N per pixel of at least eight across one observed effective radius ($R^2_{\rm{obs}} = R^2_{\textit{eff}} + ({\rm{FWHM}}(z)/2)^2$). As a consistency check, we also looked at how using a different threshold $\rm{(S/N)_{tot}}~\geq~30$ would impact the selection. This threshold adds 40 new galaxies, but the majority are either small with respect to their MUSE PSF FWHM or do not show clear velocity gradients. Thus, we decided to use the former criterion in the next parts. We show in Fig.\,\ref{fig:sample_selection}, the galaxies distribution and selection in terms of S/N, $R_{\textit{eff}, \rm{d}} / \rm{FWHM}$, and B/D for galaxies from the kinematic sample.  Criterion (iv) removes face-on and edge-on galaxies because, for the former, uncertainties are too large to reliably constrain the rotation of the ionised gas and, for the latter, the mass models used in the kinematic modelling are much more loosely constrained. 
	
	Finally, criterion (v) identifies galaxies whose dynamical modelling failed, that is for which we overestimated the contribution of baryons to the total rotation curve. This corresponds to 13 galaxies in the kinematic sample. Among them, we decided to remove ten galaxies, namely {\textsc{85-CGr35}}, {\textsc{28-CGr26}}, {\textsc{257-CGr84}}, {\textsc{113-CGr23}}, {\textsc{83-CGr23}}, {\textsc{38-CGr172}}, {\textsc{130-CGr35}}, {\textsc{110-CGr30}}, {\textsc{105-CGr114}} and {\textsc{100-CGr172}}. These objects are shown as orange crosses in Fig.\,\ref{fig:sample_selection}. Most of them tend to be quite small or with low S/N values even though they pass criteria (i) and (ii), but also have velocity fields with a quite low amplitude ($\sim 30-\SI{40}{km~s^{-1}}$). This means that any uncertainty on their morphological modelling and mass-to-light ratio will have a stronger impact on their dynamical modelling. In addition, galaxies {\textsc{85-CGr35}} and {\textsc{28-CGr26}} have disturbed morphologies and/or kinematics which may be due to past merger events or to a more complex morphology than the bulge-disk decomposition performed in Sect.\,\ref{sec:morpho modelling}. On the contrary, after carefully investigating their morphology and kinematics, we decided to keep galaxies {\textsc{378-CGr32}}, {\textsc{20-CGr84}} and {\textsc{19-CGr84}} since they seemed to be intrinsically "baryon dominated". After applying criteria (i) to (v), we end up with a TFR sample of \NTFRSample{} galaxies.
	
	In Sect.\,\ref{Sec:analysis}, we may apply two additional selection criteria when it is necessary to have comparable parameter distributions between different environments:
	
	\begin{enumerate}[label=(\roman*)]
		\setcounter{enumi}{5}
		\item $\log_{10} M_{\star}~[{\rm{M}}_{\odot}] \leq 10$,
		\item $0.5 < z < 0.9$.
	\end{enumerate}
	
	 Criterion (vi) is used to have comparable samples in terms of stellar mass (see stellar mass distributions in Fig.\,\ref{fig:selection_impact_distributions}), whereas (vii) only keeps galaxies in a $\SI{1}{Gyr}$ interval around a redshift $z~\approx~0.7$ where most of the galaxies in the large structures are located. Thus, this criterion allows us to check that our results may not be impacted by a potential redshift evolution. 
	 
	 \subsection{Summary of the different samples and subsamples}
	 \label{sec:summary_samples}
	 
	 To clarify the difference between the various samples used in this paper, we provide below a summary of their characteristics. We also show in Table\,\ref{tab:samples} the distribution of their main physical parameters represented by their median value, 16th and 84th percentiles.
	 
	 \begin{enumerate}[label=(\arabic*)]
	 	\item {\bf{\OII{} emitters sample}}: \NOIISample{} \OII{} emitters with reliable spectroscopic redshift in the range $0.25~\lesssim~z~\lesssim~1.5$
	 	\item {\bf{Morphological sample}}: \NMorphoSample{} galaxies from the \OII{} emitters sample with reliable bulge-disk decomposition
	 	\item {\bf{Kinematic sample}}: \NKinematicSample{} galaxies from the morphological sample with reliable kinematics
	 	\item {\bf{MS sample}}: \NMSSample{} disk-dominated galaxies from the kinematic sample selected in B/D only. This sample is used to study the size-mass and MS relations.
	 	\item {\bf{TFR sample}}: \NTFRSample{} disk-dominated galaxies from the MS sample with selection criteria from (i) to (v) applied to only keep galaxies with well constrained kinematics. This sample is used to study the TFR.
	 \end{enumerate}
	 
	 We show in Table\,\ref{tab:subsamples} the median properties for each environment-based subsample of galaxies from the MS sample later used in the analysis. Among these, we show the field, small, and large structure ones defined in Sect.\,\ref{sec:FoF}. Alternatively, when analysing the TFR in Sect.\,\ref{sec:TFR}, we will also split the entire sample into two subsamples only: a field/small structures subsample on the one hand, and a large structure subsample on the other hand. This separation is performed because using the previously defined subsamples would lead to too few galaxies in the small structures to reliably constrain their TFR. In the following and in Table\,\ref{tab:subsamples}, we will refer to these subsamples as Small-{\textit{N}} and Large-{\textit{N}}, where {\textit{N}} corresponds to the richness threshold used to classify galaxies in either the field/small structure or large structure subsamples. We note that the terms {\textit{small}} and {\textit{large}} used to name the subsamples never refer to neither the size, nor the mass of the structures, but only to the number of galaxy members. 
	 
	 The main properties shown in Table\,\ref{tab:subsamples} are the total number and the proportion of galaxies in each subsample, the stellar, gas, and dynamical masses computed within $R_{22} = 2.2 R_{\rm{d}}$, with $R_{\rm{d}}$ the disk scale length, the extinction corrected \OII{}-based SFR, and the median disk effective radius $R_{\textit{eff}, \rm{d}}$. All the subsamples have mostly similar gas mass and SFR distributions. However, the subsamples targetting the largest structures tend to have on average larger disk sizes and stellar masses. Their dynamical masses are slightly larger as well, though the difference between small and large structures at a fixed threshold is roughly 0.3-0.4\,dex, similar to the difference seen in stellar masses, indicative that these massive structures do not host, on average, more massive DM haloes. Interestingly, when using the largest threshold values $N = 15, 20$, the large structure subsamples have larger stellar masses ($\Delta \log_{10} M_{\star} \approx \SI{0.5}{dex}$), but similar dynamical masses with respect to the small structure subsamples. One of the key difference visible in Fig.\,\ref{fig:MS_full} is the stellar mass distribution. The large structure subsample is more extended than the field and the small structure subsamples towards larger stellar masses, so that almost all the galaxies beyond $M_{\star} > \SI{e10}{M_{\odot}}$ are located in the large structures. These massive galaxies also tend to have the largest SFR values, though their impact on the SFR distribution is not as clearly visible as in the stellar mass distribution.
	
	The MAGIC catalogue containing the main morpho-kinematics and physical properties for galaxies from the MS sample is available at the CDS. We provide in Table\,\ref{tab:catalog_CDS} a description of the columns appearing in the catalogue. Appendix\,\ref{Sec:example_morpho_kin_maps} contains the morpho-kinematics maps as shown in Fig.\,\ref{fig:examples_models} for all galaxies in the TFR sample.

	\section{Analysis}
	\label{Sec:analysis}
	
	We focus the analysis on the size-mass relation, MS, and TFR. We consider the MS and TFR samples, and separate galaxies in three different subsamples targetting different environments. For the size-mass relation, we use the corrected stellar mass $M_{\star, \rm{corr}}$ which better traces the disk and bulge masses within a sphere of radius $R_{22}$ (see Sect.\,\ref{sec:mass_correction}), and the disk scale length $R_{\rm{d}} = R_{\textit{eff}, \rm{d}} / b_1$ for the size of our galaxies, where $R_{\textit{eff}, \rm{d}}$ is the disk effective radius and $b_1 \approx 1.6783$. We also use $M_{\star, \rm{corr}}$ for the TFR, as well as the total circular velocity $V_{22}$ derived at $R_{22}$ from the best-fit mass and flat models for the velocity. This $R_{22}$ value corresponds to where the peak of rotation for the disk component is reached and is typically used in similar studies \citep{Pelliccia2019, Abril-Melgarejo2021}. Lastly, for the MS, we use the SED-based stellar mass $M_{\star}$ derived in an aperture of $\SI{3}{\arcsec}$ and the extinction corrected and normalised \OII{} SFR as described in Sect.\,\ref{sec:gal_properties}. Each scaling relation is fitted with the form
	
	\begin{equation}
		\log_{10} y = \beta + \alpha (\log_{10} x - p),
		\label{eq:general_fit}
	\end{equation}
	where $y$ is the dependent variable, $x$ is the independent variable, and $p$ is a pivot point equal to the median value of $\log_{10} x$ when using the full samples (MS or TFR). For each relation, we decided to always use stellar mass as the independent variable, so that the pivot point is $p = 9.2$. As pointed out in \citet{MPFITEXY, Pelliccia2017}, this is justified for the TFR as fitting the opposite relation yields a slope biased towards lower values, while for the size-mass and MS relations we find more robust fits and smaller dispersion.
	
	In order to have fits not biased by points with underestimated errors in $x$ and $y$, we quadratically added an uncertainty on the error of both independent and dependent variables in each scaling relation. Based on \citet{Abril-Melgarejo2021}, we decided to quadratically add an uncertainty of $\SI{0.2}{dex}$ on the stellar mass and the SFR, and of $\SI{20}{km~s^{-1}}$ on the velocity, consistent with typical uncertainties and systematics found in the literature. For the size estimate, we added a slightly lower uncertainty of $\SI{0.065}{dex}$, which corresponds to a relative error of roughly 15\%, slightly below the more or less 30\% scatter  \citet{kuchner_effects_2017} found when comparing size measurements between Subaru and HST data.
	
	We used two different tools to perform the fits. The first one is \Ltsfit{} \citep{ltsfit}, a python implementation of the Least Trimmed Square regression technique from \citet{rousseeuw_computing_2006}, and the second one is \MPFITEXY{} \citep{MPFITEXY} IDL wrapper of \MPFIT{}. Both methods take into account uncertainties on $x$ and $y$, as well as the intrinsic scatter of each relation, but \Ltsfit{} implements a robust method to identify and remove outliers from the fit. However, it currently does not have an option to fix the slope. Therefore, whenever we needed to fix the slope, we used \MPFITEXY{}, removing beforehand outliers found by \Ltsfit{}.	
	
	\subsection{Impact of selection}
	\label{sec:selection_impact}
	
	\begin{table}
		\centering
		\caption{Comparison of fit parameters for the scaling relations with various selection criteria. The B/D selection, that is limiting the sample to disk-dominated galaxies, is always applied.}
		\resizebox{0.48\textwidth}{!}{%
		\begin{tabular}{cccccc}
			\hline
			\hline
			Relation                           & Selection          & Number                & $\alpha$        & $\beta$ \\
			(1)                                & (2)                & (3)                   & (4)             & (5) \\
			\hline \\[-9pt]
			\multirow{4}{*}{Size-Mass}         &                    & \ \ \NMSSample (10)   & $0.33 \pm 0.02$ & $\ \ 0.22 \pm 0.01$ \\
			                                   & (ii)               & \ \ 270 (11)          & $0.16 \pm 0.01$ & $\ \ 0.34 \pm 0.01$ \\
			                                   & (iii)              & 389 (4)               & $0.35 \pm 0.02$ & $\ \ 0.21 \pm 0.01$ \\
			                                   & (ii), (iii)        & 223 (8)               & $0.18 \pm 0.02$ & $\ \ 0.33 \pm 0.01$ \\[3pt]
			\hline \\[-9pt]
			\multirow{4}{*}{MS}                &                    & \ \ \NMSSample (14)   & $0.61 \pm 0.02$ & $-0.25 \pm 0.02$ \\
			                                   & (ii)               & \ \ 270 (13)          & $0.66 \pm 0.03$ & $-0.32 \pm 0.02$ \\
									           & (iii)              & 389 (4)               & $0.61 \pm 0.02$ & $-0.21 \pm 0.02$ \\
									           & (ii), (iii)        & 224 (6)               & $0.66 \pm 0.03$ & $-0.28 \pm 0.02$ \\[3pt]
			\hline \\[-9pt]
			\multirow{5}{*}{TFR}               &                    & \ \ \NMSSample (23)   & $0.34 \pm 0.01$ & $\ \ 2.03 \pm 0.01$ \\
			                                   & (ii)               & 270 (5)               & $0.27 \pm 0.02$ & $\ \ 2.07 \pm 0.01$ \\
									           & (iii)              & \ \ 389 (21)          & $0.36 \pm 0.01$ & $\ \ 2.01 \pm 0.01$ \\
									           & (ii), (iii)        & 223 (7)               & $0.31 \pm 0.02$ & $\ \ 2.03 \pm 0.01$ \\
			                                   & (ii) to (v)        & \NTFRSample{} (1)     & $0.29 \pm 0.02$ & $\ \ 2.01 \pm 0.01$ \\[3pt]
			\hline
			& & & & &
			
		\end{tabular}}
		\label{tab:comp}
		
		{\small\raggedright {\bf Notes:} (1) Scaling relation fitted, (2) Selection criteria used (see Sect.\,\ref{sec:samples_selection}), (3) Number of galaxies with outliers shown in parentheses, (4) Best-fit slope, (5) Best-fit zero point. Errors on fit parameters correspond to $1\sigma$ uncertainties.\par}
	\end{table}
	
	\begin{figure}[htb!]
		\centering
		\includegraphics[scale=0.7]{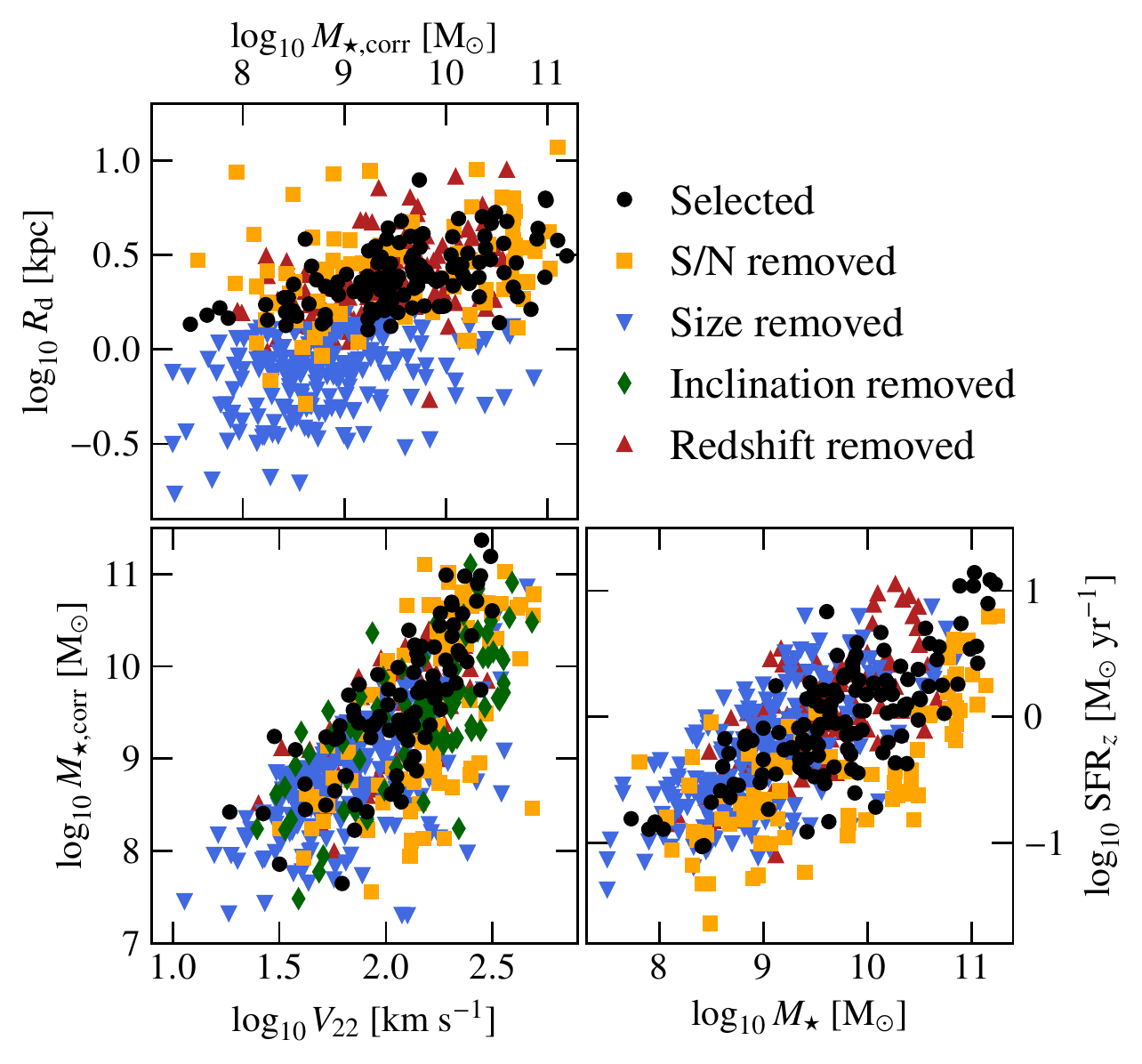}
		\caption{Impact of the different selection criteria from Sect.\,\ref{sec:samples_selection} applied on the size-mass relation (top left), TFR (bottom left), and MS (bottom right). Black circles represent galaxies which remain when all the selection criteria are applied. Given that some selections remove similar galaxies, we show those removed by the S/N (orange square), size (blue lower triangles), and redshift criteria (red upper triangles), in this order. Additionally, we also show in the TFR galaxies removed by the inclination selection (green diamonds) before applying the redshift selection.}
		\label{fig:selection_criteria_impact}
	\end{figure}
	
	We start by looking at how the aforementioned scaling relations are impacted by the different selection criteria used to select the MS and TFR samples. To do so, we fitted each scaling relation using the MS sample with \Ltsfit{}, letting the slope free, and we looked at the impact of the size (ii) and/or S/N (iii) criteria on the best-fit results. Additionally, since we also apply the inclination (iv) and the mass modelling uncertainty (v) selections on the TFR, we also consider their impact on the slope and zero point of this relation. The results for each scaling relation are shown in Table\,\ref{tab:comp}. We also show in Fig.\,\ref{fig:selection_criteria_impact} the population of galaxies removed by each selection criterion, as well as the galaxies removed when applying a redshift cut $0.5 < z < 0.9$ (red upper triangles), and the remaining galaxies (black points). We find that the size-mass relation is mainly impacted by the size selection for both the slope and zero point, while the S/N criterion has a weaker effect. When removing small galaxies, the slope is biased towards lower values, and this effect is more important for field galaxies than for galaxies in other subsamples. Similarly, the MS is mainly affected by the size selection while the S/N selection has almost no impact. This result may seem surprising given that, as can be seen in Fig.\,\ref{fig:selection_criteria_impact}, size-removed (blue lower triangles), and S/N-removed (oranges squares) galaxies tend to lie along the MS, but on opposite parts. However, the size selection has a stronger impact since it mainly removes low mass galaxies, biasing the slope to larger values driven by more massive galaxies.
	
	Finally, similarly to the size-mass and MS relations, the TFR is also mainly impacted by the size selection. Removing small galaxies changes the slope to lower values, driven by more massive galaxies. However, when applying the size and S/N selections, both the slope and zero point values become close to the original ones. Because of the mass models used, the TFR is quite tight and those criteria tend to remove almost symmetrically galaxies with low and high circular velocity as can be seen in Fig.\,\ref{fig:selection_criteria_impact}, so that the remaining galaxies fall along the original TFR without any bias. Important selection criteria for the TFR are the inclination and mass modelling uncertainty (iv and v). Among the two, criterion (v) has the weakest impact since it only removes a handful of galaxies, whereas the inclination selection (iv) tends to remove a significant fraction of galaxies with larger circular velocities than the bulk of galaxies with stellar masses beyond $\SI{e9}{M_{\odot}}$. These galaxies probably have overestimated circular velocities, so that including them in the fit of the TFR would lead to a slope biased towards larger values.
	
	Because the size and S/N selection criteria were defined to select galaxies with reliable morphology and kinematics for the mass modelling, and because they can bias the slope and zero point of the size-mass and MS relations, we decided not to apply them to select the MS sample, as described in Sect.\,\ref{sec:samples_selection}. However, these criteria, in combination with the inclination (iv) and mass modelling uncertainty (v) selections, are important to have an unbiased fit of the TFR. Thus, we decided to apply selection criteria from (i) to (v) to select the TFR sample in Sect.\,\ref{sec:samples_selection}.
	
	\subsection{Impact of the environment on the size-mass relation}
	\label{sec:mass-size}
	
	\begin{figure}[htb!]
		\centering
		\includegraphics[scale=0.7]{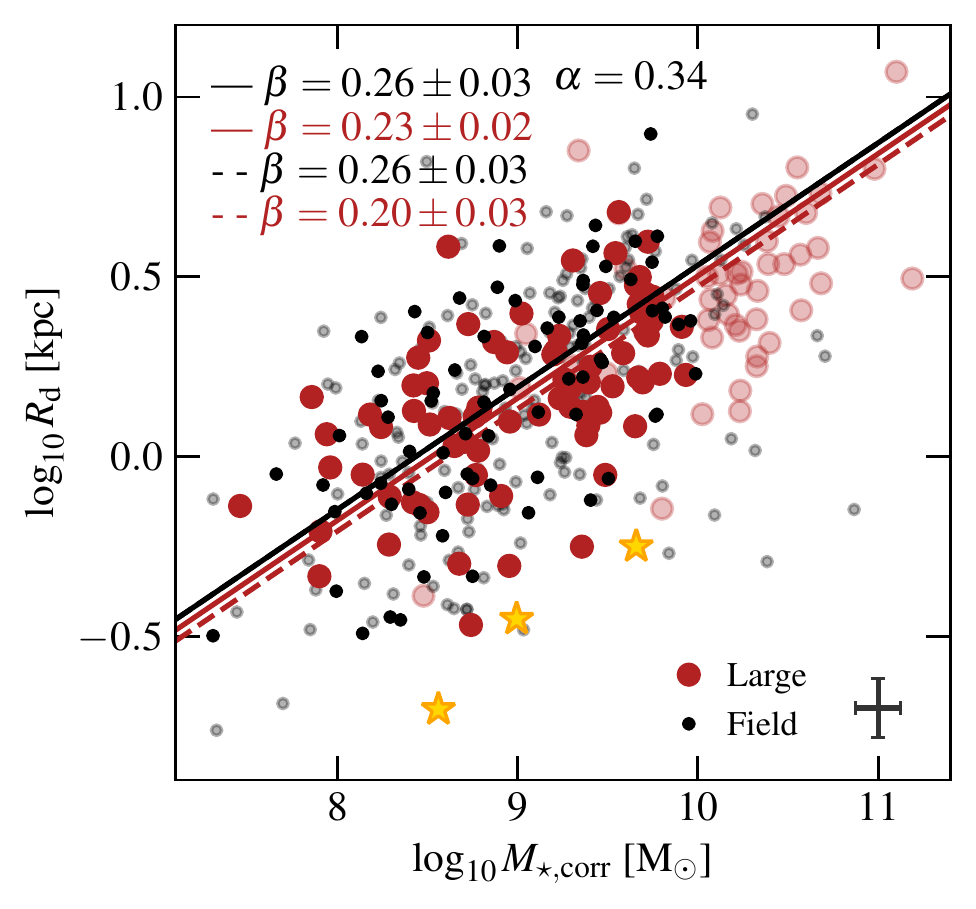}
		\caption{Size-mass relation for galaxies from the MS sample with additional mass and redshift cuts applied (vi and vii). Symbols and colours are similar to Fig.\,\ref{fig:MS_full}, and orange stars represent galaxies identified as outliers from the fit done with \Ltsfit{}. As an indication, we also show as semi-transparent symbols galaxies removed by the mass and redshift cuts. Best-fit lines are shown when using a richness threshold $N=10$ (full lines) and $N=20$ (dashed lines). The black dashed line is not visible since field galaxies have the same best-fit line for $N=10$ and $N=20$. We do not show galaxies in the small structure subsample since there remain too few galaxies after applying selection criteria (vi) and (vii). We also provide on the top left the slope and best-fit zero point for each subsample (see Eq.\,\ref{eq:general_fit} with $y = R_{\rm{d}}$ and $x = M_{\star}$). On the bottom right is shown the typical uncertainty on stellar mass and disk size as a grey errorbar. After controlling for differences in mass and redshift, we find a $1\sigma$ significant difference of $\SI{0.03}{dex}$ between subsamples with $N=10$, and a $2\sigma$ significant difference of $\SI{0.06}{dex}$ with $N=20$.}
		\label{fig:fit_size-mass}
	\end{figure}
	
	We fit the subsamples targetting different environments fixing the best-fit slope to the value from \Ltsfit{} when considering the entire MS sample with the same selection criteria. We further apply two additional selection criteria: a mass cut $M_{\star} < \SI{e10}{M_{\odot}}$ and a redshift cut $0.5 < z < 0.9$, in order to reduce the impact of different mass and redshift distributions between subsamples on the best-fit zero points. We show in Table\,\ref{tab:fit_MS_size-mass} the best-fit zero points as well as the slopes used for each fit, and in Fig.\,\ref{fig:fit_size-mass} the size-mass relation and its best-fit line when applying the mass (vi), and redshift cuts (vii). We also provide in Fig.\,\ref{fig:impact_on_mass-size_relation} the size-mass relation and its best-fit line when only applying the mass cut, and when applying neither mass nor redshift cuts.s
	
	We find a small offset in the zero point between subsamples. When applying both mass and redshift cuts, the difference amounts to $\SI{0.03}{dex}$, which is at most $1\sigma$ significant\footnote{The term $\sigma$ significant will always refer to the uncertainty on the zero point of the best-fit line.}. Similarly, when applying only the mass cut, we get a $1\sigma$ significant difference between the field subsample and the small and large structure subsamples. However, if we apply neither cuts, we get a slightly larger offset of $\SI{0.04}{dex}$ between the field and the large structure one, and almost similar zero points between the field and the small structure one. In Fig.\,\ref{fig:fit_size-mass} and in Table\,\ref{tab:fit_MS_size-mass}, we used the disk size to fit the size-mass relation, whereas other studies \citep[e.g.][]{maltby_environmental_2010} usually use a global radius. To check whether the choice of radius might have an impact on our results we fitted the size-mass relation, but using the global effective radius derived in Sect.\,\ref{sec:morpho_properties}. Even when using the global radius, we get the same trend as before, with an offset of $\SI{0.02}{dex}$ ($1\sigma$ significant). If we use instead a more stringent richness threshold of $N=20$ to separate galaxies between small and large structures, we do find a larger offset of $\SI{0.06}{dex}$ ($2\sigma$ significant) between the field and the large structure subsamples when using the disk radius as a size proxy, and a similar offset of $\SI{0.02}{dex}$ when using the global effective radius.
	
	\begin{table}[hbtp!]
		\centering
		\caption{Best-fit values for the size-mass and MS relations fitted on the MS sample. Optionally, we apply a mass cut $M_{\star} \leq \SI{e10}{M_{\odot}}$ (vi) and a redshift cut $0.5 \leq z \leq 0.9$ (vii). For each fit, the slope is fixed to the one from \Ltsfit{} on the entire MS sample using the same selection criteria. We do not show the small structures subsample when applying the redshift cut since there remain too few galaxies to reliably constrain its zero point. Bold values correspond to those shown in Fig.\,\ref{fig:fit_size-mass} and \ref{fig:fit_MS} (full lines).}
		\resizebox{0.48\textwidth}{!}{%
		\begin{tabular}{ccccccc}
			\hline
			\hline
			Subsample & Scaling relation & Selection & Nb. & Prop. (\%) & $\alpha$ & $\beta$ \\
			(1)       & (2)              & (3)       & (4) & (5)        & (6)      & (7) \\	
			\hline \\[-9pt]
			Field     & \multirow{3}{*}{\shortstack{Size-Mass}} & & 250 (6) & 57 & \multirow{3}{*}{\shortstack{$0.33$}} & $\ \ 0.23 \pm 0.02$ \\
			Small     & & & \ \ 54 (2) & 13 & & $\ \ 0.22 \pm 0.03$ \\
			Large     & & &    133 (2) & 30 & & $\ \ 0.19 \pm 0.02$ \\[0.3pt]
			\hline \\[-9pt]
			Field     & \multirow{3}{*}{\shortstack{Size-Mass}} & \multirow{3}{*}{\shortstack{(vi)}} & 237 (3) & 63 & \multirow{3}{*}{\shortstack{$0.41$}} & $\ \ 0.26 \pm 0.02$ \\
			Small     & & & \ \ 48 (2) & 13 & & $\ \ 0.24 \pm 0.04$ \\
			Large     & & & \ \ 93 (3) & 24 & & $\ \ 0.24 \pm 0.02$ \\[0.3pt]
			\hline \\[-9pt]
			{\bf{Field}} & \multirow{2}{*}{\shortstack{\bf{Size-Mass}}} & \multirow{2}{*}{\shortstack{\bf{(vi), (vii)}}} & \ \ {\bf{77 (1)}} & {\bf{45}} & \multirow{2}{*}{\shortstack{\bm{$0.34$}}} & {\bm{$\ \ 0.26 \pm 0.03$}} \\
			{\bf{Large}} & & & \ \ {\bf{84 (2)}} & {\bf{49}} & & {\bm{$\ \ 0.23 \pm 0.02$}} \\[0.3pt]
			\hline
			\hline \\[-9pt]
			Field     & \multirow{3}{*}{\shortstack{MS}} & & 251 (5) & 58 & \multirow{3}{*}{\shortstack{$0.61$}} & $-0.19 \pm 0.02$ \\
			Small     & & & \ \ 55 (1) & 13 & & $-0.22 \pm 0.05$ \\
			Large     & & &    126 (9) & 29 & & $-0.36 \pm 0.03$ \\[0.3pt]
			\hline \\[-9pt]
			Field     & \multirow{3}{*}{\shortstack{MS}} & \multirow{3}{*}{\shortstack{(vi)}} & 239 (1) & 63 & \multirow{3}{*}{\shortstack{$0.78$}} & $-0.18 \pm 0.02$ \\
			Small     & & & \ \ 47 (3) & 13 & & $-0.15 \pm 0.06$ \\
			Large     & & & \ \ 91 (5) & 24 & & $-0.29 \pm 0.04$ \\[0.3pt]
			\hline \\[-9pt]
			{\bf{Field}} & \multirow{2}{*}{\shortstack{\bf{MS}}} & \multirow{2}{*}{\shortstack{\bf{(vi), (vii)}}} & \ \ {\bf{78 (0)}} & {\bf{45}} & \multirow{2}{*}{\shortstack{\bm{$0.72$}}} & {\bm{$-0.22 \pm 0.04$}} \\
			{\bf{Large}} & & & \ \ {\bf{83 (3)}} & {\bf{48}} & & {\bm{$-0.32 \pm 0.04$}} \\[0.3pt]
			\hline
			          & & & & & & \\
		\end{tabular}}
		\label{tab:fit_MS_size-mass}
		
		{\small\raggedright {\bf Notes:} (1) Subsample name, (2) Scaling relation fitted, (3) Selection criteria applied, (4) Number of galaxies in each subsample with outliers shown in parentheses, (5) Proportion of galaxies in each subsample (after removing outliers), (6) Fixed slope, (7) Best-fit zero point. Errors on fit parameters correspond to $1\sigma$ uncertainties.\par}
	\end{table}
	
	Overall, if significant, the difference between the field and the largest structures when using $N=10$ is quite small. We note that this result is different from what was found in previous studies such as \citet{maltby_environmental_2010} or \citet{matharu_hst_2019}. Indeed, such studies always found a weak but significant dependence of the size-mass relation with environment. For instance, \citet{maltby_environmental_2010} found spiral galaxies in the field to be about 15\% larger than their cluster counterparts but, in our case, it would only amount to a size difference of roughly 7\%. Instead, using the offset value with $N=20$, we get a size difference of roughly 14\%, consistent with previous findings from \citet{maltby_environmental_2010} that galaxies in the most massive structures are more compact than those in the field. Given the models used in the morphological analysis and because the bulge-to-disk ratio is fairly similar between subsamples, the zero point of the size-mass relation directly translates in terms of the galaxies central surface mass density of the disk component (i.e. extrapolated from the Sérsic profile at $R=0$). Assuming the flux of the disk component dominates at $R_{22}$, using a slope $\alpha~=~0.34$ and a zero point $\beta_{\rm{sm}}$, we get the following scaling relation for the disk component central surface mass density $\Sigma_{\rm{M, d}} (0)$ as a function of stellar mass:
	
	\begin{equation}
		\log_{10} \Sigma_{\rm{M, d}} (0)~[{\rm{M_{\odot}/kpc}}^2] \approx 0.32 \log_{10} M_{\star, \rm{corr}}~[{\rm{M}}_{\odot}] + 5.65 - 2\beta_{\rm{sm}},
		\label{eq:central_surface_mass_density_vs_mass_simplified}
	\end{equation}
	where $\beta_{\rm{sm}}~=~0.26 \pm 0.03$ for the field subsample and $\beta_{\rm{sm}}~=~0.20 \pm 0.03$ for the large structure subsample when using a richness threshold of $N=20$. A change in the zero point of the size mass relation does not impact the slope of Eq.\,\ref{eq:central_surface_mass_density_vs_mass_simplified} but only its zero point. Thus, the $\SI{0.06}{dex}$ offset measured between the field and the most massive structures results in a negative offset of $\SI{-1.2}{dex}$ in Eq.\,\ref{eq:central_surface_mass_density_vs_mass_simplified}. We note that this interpretation remains true as long as we can neglect the flux of the bulge at $R_{22}$. However, when we cannot neglect it any more, then Eq.\,\ref{eq:central_surface_mass_density_vs_mass_simplified} would have an additional non-linear term which would be a function of the bulge central surface mass density and effective radius. In this case, the interpretation would be more complex as galaxies could have different bulge or disk physical properties as a function of environment but still align on the same size-mass relation. However, as is visible in Fig.\,\ref{fig:med_B/T}, the bulge contribution at $R_{22}$ is on average and independently of environment around 10\% of the total flux, which amounts to a scatter in the size-mass relation of about $\SI{0.1}{dex}$, which is sufficiently small to neglect at first order the bulge contribution in this relation.
	
	\subsection{Impact of the environment on the MS relation}
	\label{sec:MS}
	
	\begin{figure}[htb!]
		\centering
		\includegraphics[scale=0.7]{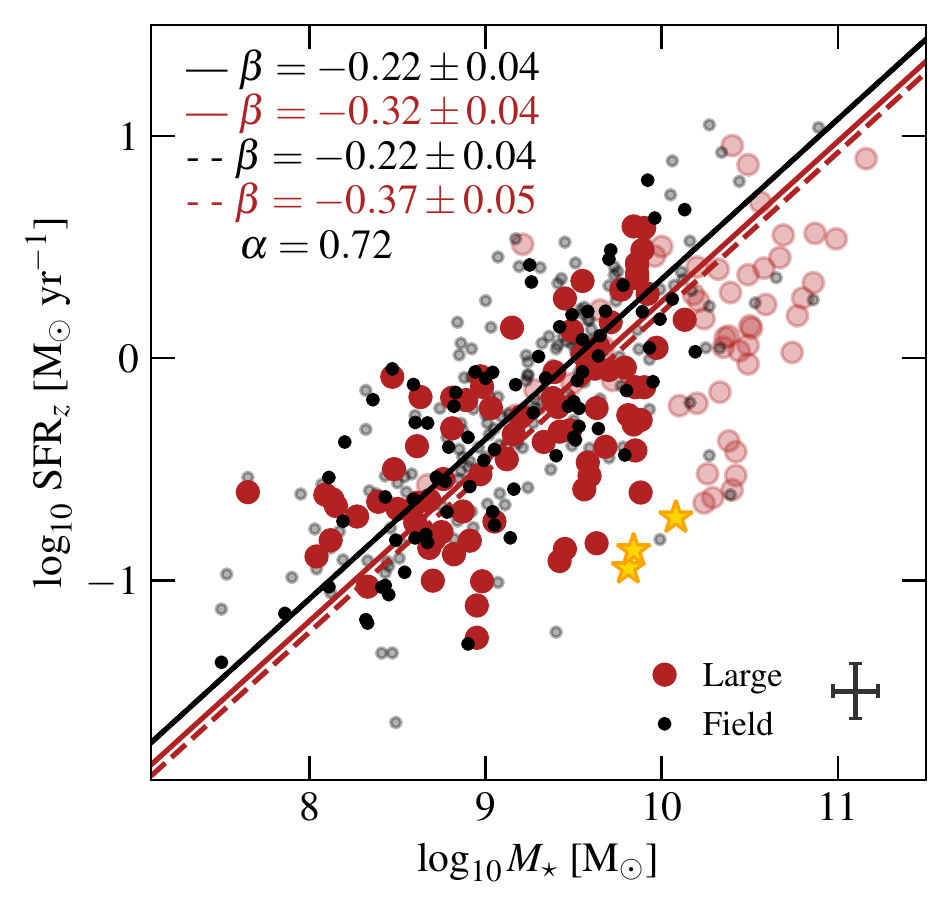}
		\caption{SFR-stellar mass relation for galaxies from the MS sample with additional mass and redshift cuts applied (vi and vii). Symbols and colours are similar to Fig.\,\ref{fig:MS_full}, and orange stars represent galaxies identified as outliers from the fit done with \Ltsfit{}. As an indication, we also show as semi-transparent symbols galaxies removed by the mass and redshift cuts. Best-fit lines are shown when using a richness threshold $N=10$ (full lines) and $N=20$ (dashed lines). We do not show galaxies in the small structure subsample since there remain too few galaxies after applying selection criteria (vi) and (vii). The SFR is normalised to a redshift $z_0~=~0.7$ (see Sect.\,\ref{sec:gal_properties}). We also provide on the top left the slope and best-fit zero point for each subsample (see Eq.\,\ref{eq:general_fit} with $y = \rm{SFR}$ and $x = M_{\star}$). On the bottom right, the typical uncertainty on stellar mass and SFR is shown as a grey errorbar. Even after controlling for differences in mass and redshift, we find a $2\sigma$ significant difference of $\SI{0.10}{dex}$ between subsamples with $N=10$, and a $3\sigma$ significant difference of 0.15\,dex with $N=20$.}
		\label{fig:fit_MS}
	\end{figure}

	To study the MS, we use the SED-based stellar mass and the \OII{} SFR corrected for extinction and normalised to a redshift $z_0~=~0.7$ as described in Sect.\,\ref{sec:gal_properties}. For this relation, applying both a mass and a redshift cut is important. Indeed, as can be seen in Fig.\,\ref{fig:MS_redshift}, the MS can be quite sensitive to redshift since there is still a small dichotomy between low and high redshift galaxies even after normalisation. The main reason for this effect is that the MAGIC survey is designed to blindly detect sources in a cone. The blind detection makes the survey flux-limited which means we are missing faint, low SFR galaxies in the highest redshift bin. Besides, we expect to see an excess of massive galaxies in the most massive structures with respect to the field which, in our sample, are all located at a redshift $z~\approx~0.7$. Thus, the survey design tends to create a dichotomy in mass, which is visible in SFR as well since we are focussing on star forming galaxies. Nevertheless, as can be seen in Table\,\ref{tab:fit_MS_size-mass}, the redshift cut has a much smaller effect than the mass cut, especially on the slope value from the best-fit line.
	
	We show in Fig.\,\ref{fig:fit_MS} the MS with both cuts applied for the field and large structure subsamples, as well as their best-fit lines and zero point values. We also provide in Fig.\,\ref{fig:impact_on_MS} the MS and its best-fit line when only applying the mass cut, and when applying neither mass nor redshift cuts. Independently of whether we apply a mass and/or redshift cut or not, we find a more than $2\sigma$ significant difference in the zero point ($\sim~\SI{0.1}{dex}$) between the field and large structure subsamples. However, there is almost no difference in the zero point between the field and the small structure subsamples. Independently of the cut applied, the field galaxies always have a larger zero point than the galaxies in the large structures. If we interpret this difference in terms of a SFR offset between the field and the largest structures, this would lead to an average SFR for the galaxies in the large structures which is about 1.3 times lower than that in the field. This factor is quite close to the recent value found by \citet{old_erratum_2020, old_gogreen_2020} using the GOGREEN and GCLASS surveys at redshift $z~\sim~1$. On the other hand, the reason why other studies such as \citet{nantais_h_2020} do not find any impact of the environment on the MS is still unclear. The effect of the redshift evolution of the MS might play a role, since \citet{nantais_h_2020} probe clusters at $z~\sim~1.6$ which is beyond the $0.5~<~z~<~0.9$ redshift range we restricted our fit to. Similarly, the impact of the environment on the MS may be segregated between low and high mass galaxies. As was reported in \citet{old_erratum_2020, old_gogreen_2020}, the MS seems to be more impacted in the lowest mass regime. This explanation would be compatible with our result where we mainly probe low to intermediate mass galaxies since we remove massive galaxies not to bias the fit.
	
	Similarly to Sect.\,\ref{sec:mass-size}, we performed the same fits but using a more stringent richness threshold of $N=20$ to separate between structures. When using this threshold combined with both mass and redshift cuts, we find a roughly $3\sigma$ significant difference of $\SI{0.15}{dex}$ ($\beta_{\rm{MS}}~=~-0.22~\pm~0.04$ for field galaxies, $\beta_{\rm{MS}}~=~-0.37~\pm~0.05$ for galaxies in the largest structures), consistent with our previous finding that galaxies in the largest structures have reduced SFR with respect to the field. With this offset, we get an average SFR in the most massive structures which is about 1.5 times lower than that in the field, still quite close to the value from \citet{old_erratum_2020, old_gogreen_2020}
	
	\subsection{Impact of the environment on the TFR}
	\label{sec:TFR}
	
	\begin{figure}[htb!]
		\centering
		\includegraphics[scale=0.7]{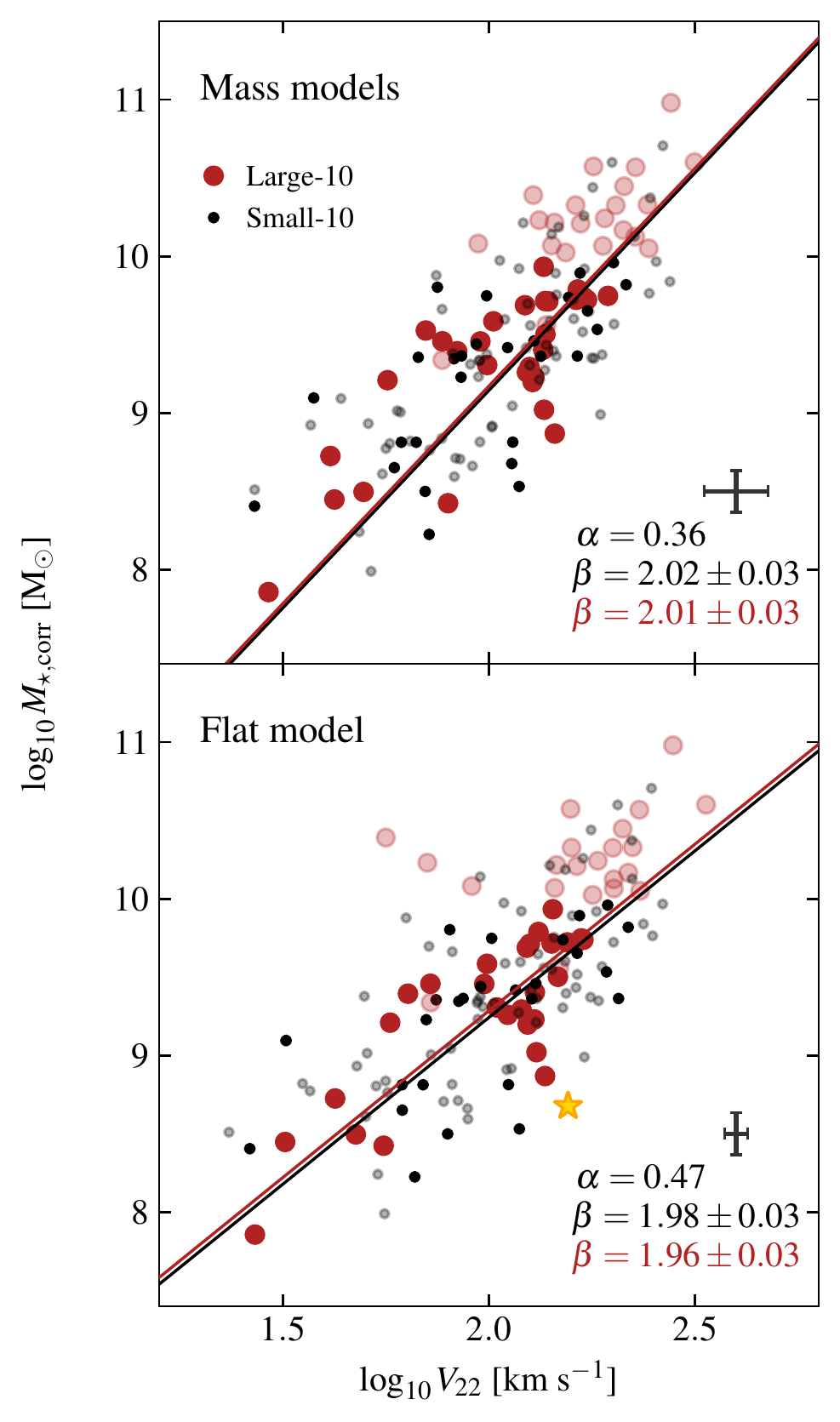}
		\caption{Stellar mass TFR at $R_{22}$ for galaxies from the TFR sample with mass and redshift cuts applied (vi and vii). The top panel shows the TFR using the velocity computed from the mass models, and the bottom one shows the TFR using the velocity from a flat model. Galaxies are split between field+small structure (black points) and large structure (red circles) subsamples using a richness threshold of $N = 10$. Orange stars represent galaxies identified as outliers from the fit done with \Ltsfit{}. As an indication, we also show as semi-transparent symbols galaxies removed by the mass and redshift cuts. Best-fit linear relations for both subsamples are shown as full lines. We provide in the bottom part of each panel the slope and best-fit zero points (see Eq.\,\ref{eq:general_fit} with $y = V_{22}$ and $x = M_{\star, \rm{corr}}$). The typical uncertainty on stellar mass and velocity is shown as a grey errorbar. After controlling for differences in mass and redshift, we do not find any impact of the environment on the zero point of both TFR.}
		\label{fig:TFR}
	\end{figure}
	
	\begin{table}[hbtp!]
		\centering
		\caption{Best-fit values for the TFR fitted on the TFR sample. Optionally, we also apply a mass cut $M_{\star} \leq \SI{e10}{M_{\odot}}$ (vi) and a redshift cut $0.5 \leq z \leq 0.9$ (vii). For each fit, the slope is fixed to the one from \Ltsfit{} on the entire kinematic sample using the same selection criteria. Bold values correspond to those shown in Fig.\,\ref{fig:TFR}.}
		\resizebox{0.48\textwidth}{!}{%
		\begin{tabular}{cccccccc}
			\hline
			\hline
			Subsample & Selec. & Number & Prop. (\%) & $\alpha_{\rm{TFR}}$ & $\beta_{\rm{TFR}}$ & $\alpha_{\rm{TFR}}^{\rm{flat}}$ & $\beta_{\rm{TFR}}^{\rm{flat}}$ \\
			(1)       & (2)    & (3)    & (4)        & (5)      & (6)     & (7)                  & (8)                 \\	
			\hline \\[-9pt]
			Small-5   & & \ \ 87 (1) & 60 & \multirow{8}{*}{\shortstack{$0.29$}} & $2.04 \pm 0.01$ & \multirow{8}{*}{\shortstack{$0.32$}} & $2.00 \pm 0.02$ \\
			Large-5   & & \ \ 58 (0) & 40 & & $1.99 \pm 0.01$ & & $1.97 \pm 0.01$ \\ 
			Small-10  & & \ \ 94 (1) & 65 & & $2.03 \pm 0.01$ & & $2.00 \pm 0.02$ \\
			Large-10  & & \ \ 51 (0) & 35 & & $2.00 \pm 0.02$ & & $1.97 \pm 0.01$ \\ 
			Small-15  & & 106 (1)    & 73 & & $2.02 \pm 0.01$ & & $1.99 \pm 0.01$ \\
			Large-15  & & \ \ 39 (0) & 27 & & $2.00 \pm 0.02$ & & $1.98 \pm 0.02$ \\ 
			Small-20  & & 117 (1)    & 81 & & $2.02 \pm 0.01$ & & $1.99 \pm 0.01$ \\
			Large-20  & & \ \ 28 (0) & 19 & & $1.98 \pm 0.02$ & & $1.97 \pm 0.02$ \\[3pt]
			\hline \\[-9pt]
			Small-5   & \multirow{8}{*}{(vi)} & \ \ 80 (1) & 69 & \multirow{8}{*}{\shortstack{$0.41$}} & $2.02 \pm 0.02$ & \multirow{8}{*}{\shortstack{$0.49$}} & $1.98 \pm 0.02$ \\
			Large-5   & & \ \ 36 (0) & 31 & & $1.98 \pm 0.03$ & & $1.95 \pm 0.03$ \\ 
			Small-10  & & \ \ 85 (1) & 73 & & $2.01 \pm 0.02$ & & $1.98 \pm 0.02$ \\
			Large-10  & & \ \ 31 (0) & 27 & & $1.99 \pm 0.03$ & & $1.95 \pm 0.03$ \\ 
			Small-15  & & \ \ 89 (1) & 77 & & $2.01 \pm 0.02$ & & $1.98 \pm 0.02$ \\
			Large-15  & & \ \ 27 (0) & 23 & & $2.00 \pm 0.03$ & & $1.95 \pm 0.03$ \\ 
			Small-20  & & \ \ 98 (1) & 84 & & $2.01 \pm 0.02$ & & $1.98 \pm 0.02$ \\
			Large-20  & & \ \ 18 (0) & 16 & & $1.97 \pm 0.04$ & & $1.92 \pm 0.04$ \\ [3pt]
			\hline \\[-9pt]
			Small-5   & \multirow{8}{*}{\shortstack{(vi) \\ \& \\(vii)}} & \ \ 27 (0) & 48 & \multirow{8}{*}{\shortstack{\bm{$0.36$}}} & $2.02 \pm 0.03$ & \multirow{8}{*}{\shortstack{\bm{$0.47$}}} & $1.98 \pm 0.03$ \\
			Large-5   & & \ \ 29 (0) & 52 & & $2.01s \pm 0.03$ & & $1.96 \pm 0.03$ \\ 
			{\bf{Small-10}}  & & \ \ {\bf{27 (0)}} & {\bf{48}} & & {\bm{$2.02 \pm 0.03$}} & & {\bm{$1.98 \pm 0.03$}} \\
			{\bf{Large-10}}  & & \ \ {\bf{29 (0)}} & {\bf{52}} & & {\bm{$2.01 \pm 0.03$}} & & {\bm{$1.96 \pm 0.03$}} \\ 
			Small-15  & & \ \ 29 (0) & 52 & & $2.02 \pm 0.03$ & & $1.98 \pm 0.03$ \\
			Large-15  & & \ \ 27 (0) & 48 & & $2.01 \pm 0.03$ & & $1.95 \pm 0.03$ \\ 
			Small-20  & & \ \ 38 (0) & 68 & & $2.03 \pm 0.02$ & & $1.98 \pm 0.03$ \\
			Large-20  & & \ \ 18 (0) & 32 & & $1.99 \pm 0.04$ & & $1.93 \pm 0.04$ \\ [3pt]
			\hline
			          & &        &    & &                 & & \\
		\end{tabular}}
		\label{tab:fit_TFR}
		
		{\small\raggedright {\bf Notes:} (1) Subsample name (2) Additional selection criteria applied, (3) Number of galaxies in each subsample with outliers in parentheses, (4) Proportion of galaxies in each subsample (after removing outliers), (5) Fixed slope for the TFR using the velocity computed from the mass models, (6) Best-fit zero point (mass models), (7) Fixed slope using the velocity computed from a flat model, (8) Best-fit zero point (flat model). Errors on fit parameters correspond to $1\sigma$ uncertainties.\par}
	\end{table}
	
	We look at the TFR as a function of the environment using the TFR sample. Since there remain too few galaxies in the small structure subsample once all the selection criteria (i to v) are applied, we decided to focus this analysis on two subsamples only. We fit the TFR using different richness thresholds ($N~=~5, 10, 15$ and 20) to separate galaxies into a field/small structure and a large structure subsamples. The best-fit zero points and the slopes values are shown in Table\,\ref{tab:fit_TFR} and in Fig.\,\ref{fig:TFR}. As a comparison, we also show on the bottom panel of Fig\,\ref{fig:TFR} the TFR obtained using a simpler flat model for the rotation curve as defined in \citet{Abril-Melgarejo2021}. This model allows us to measure the galaxies circular velocity without any prior on the baryon distribution and is therefore not affected by our mass modelling.
	
	We find a similar trend between the TFR from the mass models and that from the flat model. Overall, the tightness of the relation using either model makes the zero point values well constrained, with typical uncertainties around $\SI{0.03}{dex}$. When we do not apply any mass or redshift cut, the large structure subsample tends to systematically have a lower zero point between $\SI{0.02}{dex}$ and $\SI{0.04}{dex}$ with respect to the field subsample depending on the richness threshold used\footnote{Note that we fit the circular velocity against stellar mass (independent variable), but we show in Fig.\,\ref{fig:TFR} the inverse relation. Thus the zero point offsets $\beta$ given in the text and in Table\,\ref{tab:fit_TFR} should be read horizontally in the figure.}. This is shown in Table\,\ref{tab:fit_TFR}, as well as in Fig\,\ref{fig:impact_on_TFR}. However, when adding a mass and/or a redshift cut, this offset tends to disappear independently of the model and richness threshold used, as is shown in Fig.\,\ref{fig:TFR}. When using $N=20$, we nevertheless get a small $1\sigma$ significant offset of roughly 0.04\,dex in both TFR. This result suggests that the larger offset values found when applying no cut are certainly the consequence of having different stellar mass distributions between the two subsamples, or might be due to a small impact of the redshift evolution of the TFR. 
	
	Given the disk, bulge and DM halo mass models used to derive the circular velocity (see Sect.\,\ref{sec:kin_modelling} and \ref{Appendix:mass_modelling}), and assuming a constant B/D value of 3\% which is the median value found in the kinematic sample independently of environment, we can write the TFR as a function of the stellar mass $M_{\star, \rm{corr}}$ within $R_{22}$, the stellar fraction $f_{\star} (R_{22})~=~M_{\star, \rm{corr}} / [M_{\star, \rm{corr}} + M_{\rm{DM}} (R_{22})]$, with $M_{\rm{DM}}$ the DM halo mass, both computed at $R_{22}$, and $R_{\rm{d}}$ as
	
	\begin{equation}
		\begin{split}
			\log_{10} \left ( \frac{M_{\star, \rm{corr}}}{\rm{M}_{\odot}} \right ) \approx \ \ & 2 \log_{10} \left ( \frac{V_{22}}{\rm{km~s^{-1}}} \right ) + \log_{10} \left ( \frac{R_{\rm{d}}}{\rm{kpc}} \right ) \\
			                                                              & + \log_{10} \left ( \frac{f_{\star}}{1+0.15 f_{\star}} \right ) + 5.71.
		\end{split}
		\label{eq:TFR_theory}
	\end{equation}
	
	In Eq.\,\ref{eq:TFR_theory}, we see the size-mass relation. Thus, rewriting Eq.\,\ref{eq:TFR_theory} to make the central surface mass density of the disk component appear, and then inserting Eq.\,\ref{eq:central_surface_mass_density_vs_mass_simplified}, we get
	
	\begin{equation}
		\begin{split}
			\log_{10} \left ( \frac{M_{\star, \rm{corr}}}{\rm{M}_{\odot}} \right ) \approx \ \ & 3.03 \log_{10} \left ( \frac{V_{22}}{\rm{km~s^{-1}}} \right ) + 1.52 \log_{10} \left ( \frac{f_{\star}}{1+0.15 f_{\star}} \right ) \\
			& + 3.91 + 1.52~\beta_{\rm{sm}},
		\end{split}
		\label{eq:TFR_function_of_fstar}
	\end{equation}
	where $\beta_{\rm{sm}}$ is the size-mass relation zero point which was found to vary with environment in Sect.\,\ref{sec:mass-size}. In Eq.\,\ref{eq:TFR_function_of_fstar}, we see that only two terms can contribute to an offset on the TFR:
	\begin{enumerate*}[label=(\roman*)]
		\item different zero points on the size-mass relation as a function of environment,
		\item an offset on the stellar fraction measured within $R_{22}$ between the field and the large structure subsamples.
	\end{enumerate*}
	
\begin{figure}[htb!]
		\centering
		\includegraphics[scale=0.7]{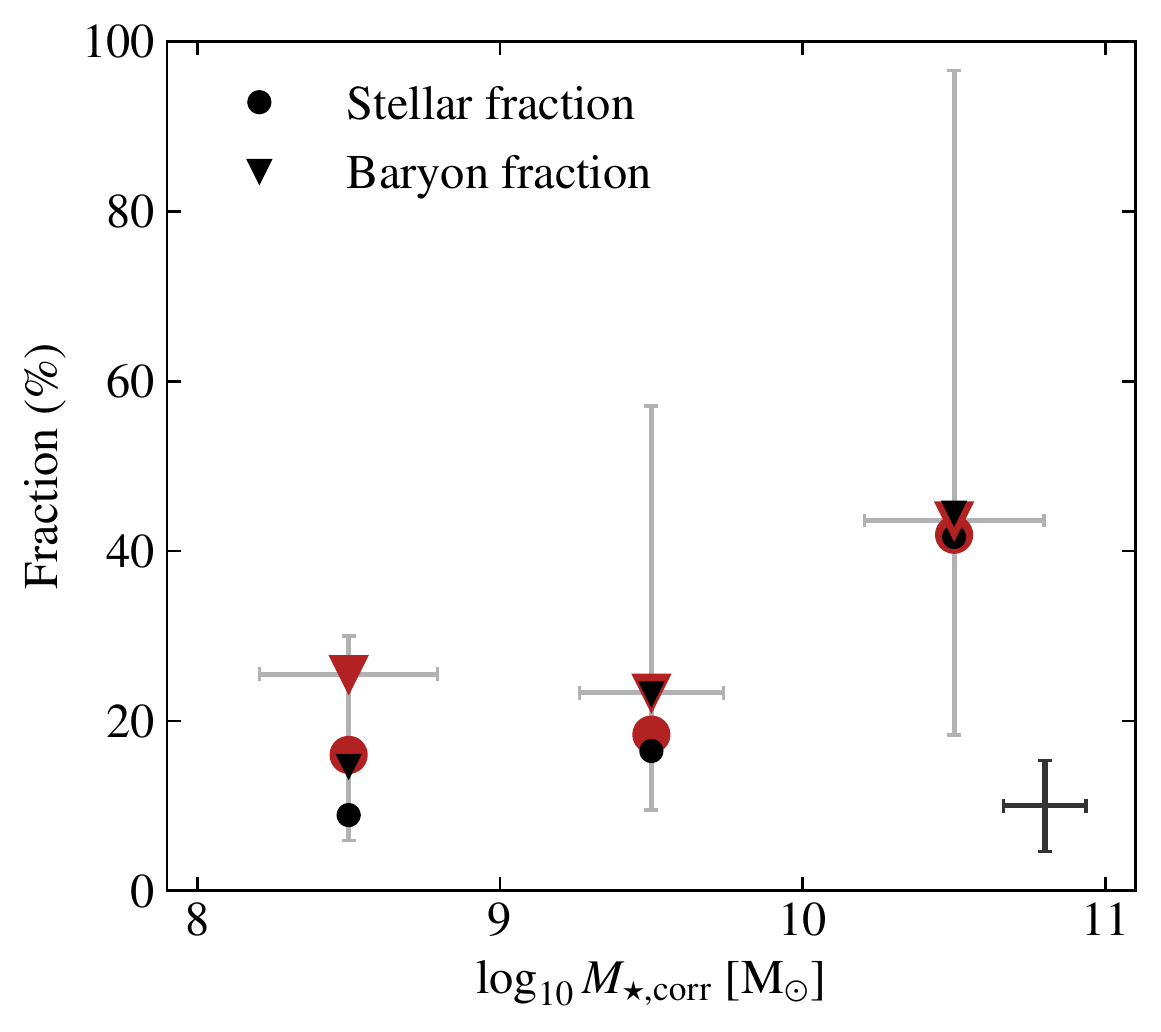}
		\caption{Evolution of the median stellar and baryon fractions for galaxies from the TFR sample in the field (black points) and in large structures (red circles) as a function of stellar mass in mass bins of $\SI{1}{dex}$. Light gray error bars correspond to the 16th and 84th percentiles of the baryon fraction distributions. The typical uncertainty on stellar mass and baryon fraction is shown as a dark grey error bar on the bottom right. Because we removed galaxies whose mass models have large uncertainties, (selection criteria iv and v), the fractions we measure are probably slightly underestimated.}
		\label{fig:stellar_fraction}
	\end{figure}

	If we interpret any offset on the TFR zero point as being an offset on stellar mass at fixed circular velocity, given Eq.\,\ref{eq:TFR_function_of_fstar} we have
	
	\begin{equation}
		\Delta \log_{10} M_{\star, \rm{corr}}~[{\rm{M}_{\odot}}] = 1.52 \left [ \Delta \log_{10} \left ( \frac{f_{\star}}{1+0.15 f_{\star}} \right ) + \Delta \beta_{\rm{sm}} \right ],
	\end{equation}
	where $\Delta \beta_{\rm{sm}}$ is the offset on the zero point of the size-mass relation which is due to the contraction of baryons observed in the most massive structures. With a threshold $N=20$ we have $\Delta \beta_{\rm{sm}} = \SI{0.06}{dex}$, and an offset on the TFR, that is in circular velocity at fixed stellar mass, of 0.04\,dex and 0.05\,dex for the mass and flat models, respectively. The corresponding offset in stellar mass at fixed circular velocity is given by $-\Delta \beta_{\rm{TFR}} / \alpha_{\rm{TFR}}~=~0.11$\,dex for both models. For a typical galaxy in the kinematic sample with a stellar fraction of 20\% this would give a difference between a galaxy in the field and one in the largest structures of roughly 4\%. This result is quite close to the difference in stellar fraction (circles) seen in Fig\,\ref{fig:stellar_fraction} where we have plotted its evolution computed from the mass models in bins of stellar mass between galaxies in the field/small structures (black) and those in large structures (red). We see that the stellar fraction increases as we go towards more massive galaxies, both in the field and in large structures. However, the difference remains small compared to the uncertainty of roughly 10\%. Besides, the distributions tend to be quite spread out, as is shown by the grey error bars, even though there is a significant offset of the stellar fraction distribution and of its dispersion as we go towards larger stellar masses.
	
	Contrary to what was found in \citet{Abril-Melgarejo2021}, we cannot measure an impact of quenching on the TFR since our stellar mass offset is negative, meaning that galaxies in the largest structures would be on average more massive than those in the field. Nevertheless, the difference is quite small ($\sim \SI{0.05}{dex}$) and may not be particularly significant. However, we do measure a significant offset in the MS, which means that quenching does take place somehow within at least some of the galaxies in the largest structures. One way to explain the apparent discrepancy is to look at the timescale over which the SFR we used in the MS is probed. Indeed, we measure the SFR from the \OII{} doublet which mainly probes recent star formation ($\sim \SI{10}{Myr}$). On the other hand, if we consider that the field and large structure subsamples do not have zero points more different than at most their uncertainty ($0.02-\SI{0.03}{dex}$), we can compute an upper bound on the quenching timescale in the large structures using Eq.\,16 of \citet{Abril-Melgarejo2021}. This gives us timescales between roughly $\SI{700}{Myr}$ and $\SI{1.5}{Gyr}$, significantly larger than the $\sim \SI{10}{Myr}$ probed by the SFR from the \OII{} doublet. Hence, the galaxies in the largest structures at $z \sim 0.7$ might have quite recently started being affected by their environment, and thus started being quenched, so that the impact on the TFR might not be visible yet with respect to the field galaxies.
	
	Some authors also implement an asymmetric drift correction to take into account the impact of gas pressure on its dynamics \citep[e.g.][]{meurer_ngc_1996, ubler_evolution_2017, Abril-Melgarejo2021, bouche_muse_2021}. Evaluated at $R_{22}$, the gas pressured corrected circular velocity for a double exponential density profile with a constant thickness writes \citep{meurer_ngc_1996, bouche_muse_2021}
	
	\begin{equation}
		V_{c, 22} = \sqrt{V_{22}^2 + 2.2 \sigma_V^2},
		\label{eq:corrected_velocity}
	\end{equation}
	where $V_{22}$ is the uncorrected circular velocity evaluated at $R_{22}$ and $\sigma_V$ is the velocity dispersion computed as the median value of the beam smearing and LSF corrected velocity dispersion map. Equation\,\ref{eq:corrected_velocity} is only an approximation of the real impact of gas pressure on the measured circular velocity since it only holds for turbulent gas disks with negligible thermal pressure. In the kinematic sample, the median value of the intrinsic velocity dispersion is around $\SI{30}{km~s^{-1}}$ independently of environment. Thus, the impact of the asymmetric drift correction is quite small on the TFR. However, we find that the velocity dispersion is not constant but is correlated with stellar mass such that more massive galaxies are more impacted by the correction than low mass ones. In turn, this tends to align high and low mass galaxies onto a line with roughly the same slope, but with a slightly larger scatter. Indeed, when implementing the asymmetric drift correction, we find virtually the same zero point between the small and large structure subsamples ($\beta_{\rm{TFR}} \approx 2.07$ with the corrected velocity versus $\beta_{\rm{TFR}} \approx 2.02$ with the uncorrected one), independently of the environment or the richness threshold used.
	
	Additionally, we can also include the gas mass into the fit. We compute the gas mass using the Schmidt-Kennicut relation \citep{schmidt_rate_1959, kennicutt_star_1998} assuming the gas is evenly distributed within a disk of radius $R_{22}$:
	
	\begin{equation}
		\begin{split}
			\log_{10} M_{\rm{g}}~[{\rm{M}}_{\odot}] \approx &~8.053 + 0.571 \log_{10} R_{\rm{d}}~[{\rm{kpc}}] \\
			                                                & + 0.714 \log_{10} {\rm{SFR}}~[{\rm{M}}_{\odot}~{\rm{yr}}^{-1}],	
		\end{split}
		\label{eq:gas_mass}
	\end{equation}
	with $M_{\rm{g}}$ the gas mass and SFR the unnormalised SFR (see Sect.\,\ref{sec:gal_properties}). If we replace the size and SFR variables in Eq.\,\ref{eq:gas_mass} by the size-mass and SFR-mass relations found before, we get a correlation between gas and stellar masses such that more massive galaxies also have a higher gas mass. In particular, the offset on the zero point found for the TFR between the field and the large structure subsamples will also lead to a small offset in the gas mass-stellar mass relation. The impact of the gas mass on the mass budget is shown in Fig.\,\ref{fig:stellar_fraction}. We compare the stellar fraction (circles) with the total baryon fraction (triangles) for the field and large structure subsamples. For most galaxies, the gas mass is non-negligible but has a small impact, leading to an offset between stellar and baryon fractions of roughly 5\%. On the other hand, the gas mass has a slightly more significant impact on the lowest mass bin. While the impact is similar to other mass bins for the field sample (roughly 5\%), the impact on the large structure subsample is stronger, reaching about 10\%. This would suggest that the low mass galaxies are more gas rich in the large structures than in the field. However, only a handful of galaxies ($\sim 10$) are located in the lowest mass bin in the large structure subsample. Besides, as is shown by the light gray error bars in Fig.\,\ref{fig:stellar_fraction}, the distribution for the baryon fraction is quite large so that the difference in gas mass is probably not that significant. Another explanation for this slightly larger difference found at low mass might be that these galaxies are experiencing bursts of star formation which would lead to overestimated gas masses, but this effect is not visible in the MS.
	
	When the gas mass is included, we get a tighter TFR, with low mass galaxies which tend to be aligned onto the same line as the high mass ones. In turn, this brings the best-fit slope to a value of $\alpha = 0.31$ when applying the mass and redshift cuts, quite close to the $\alpha = 0.29$ value found when fitting the stellar mass TFR without applying any cut (i.e. driven by massive galaxies). The zero point is almost always similar between the field/small structure and large structure subsamples ($\beta_{\rm{TFR}} \approx 1.99$), independently of the richness threshold used to separate galaxies in the two subsamples. Similarly to the stellar mass TFR, only when using a threshold $N = 20$ do we find a slightly more significant difference in zero point between the field/small structure subsample ($\beta_{\rm{TFR}} = 2.00 \pm 0.02$) and the large structure subsample ($\beta_{\rm{TFR}} = 1.98 \pm 0.02$). However, once we further include the asymmetric drift correction from Eq.\,\ref{eq:corrected_velocity}, the difference vanishes for any richness threshold used ($\beta_{\rm{TFR}} \approx 2.02$).
	
	Thus, if there is an impact on the TFR, it is mostly driven by differences in terms of stellar mass or redshift distributions rather than the environment itself. We note that this result is consistent with \citet{Pelliccia2019}, where they could not find an impact of the environment on the TFR as well, but is contradictory to what was found in \citet{Abril-Melgarejo2021}. By comparing their sample to others such as KMOS3D, KROSS and ORELSE they could find a significant offset in the TFR attributed to probing different environments. This offset was interpreted either as the effect of quenching which reduces the amount of stellar mass in the most massive structures at fixed circular velocity, or as the effect of baryon contraction which leads to an increase of circular velocity at fixed stellar mass. As discussed previously, baryon contraction and quenching are visible in our size-mass and MS relations, but not in the TFR. However, they noted that performing a consistent and reliable comparison between samples using different observing methods, models, tools and selection functions was a difficult task which can lead to multiple sources of uncertainty. These can directly arise from the morphological and kinematic modelling, but can also be driven by uncertainties on the SED-based stellar masses which, depending on the SED fitting code used and the assumptions made on the star formation history, can lead to systematics of the same order of magnitude as the offset found in \citet{Abril-Melgarejo2021}. On the other hand, we argue that our result is quite robust since we have applied the same models, tools, assumptions and selection from the beginning to the end.

	\section{Conclusion}
	\label{sec:conclusion}
	
	We have performed a morpho-kinematic modelling of \NOIISample{} \OII{} emitters from the MAGIC survey using combined HST and MUSE data in the redshift range $0.25 < z < 1.5$. These galaxies are all located in the COSMOS field and have been attributed to structures of various richness (field, small and large structures) using a FoF algorithm. We derived their global properties, such as their stellar mass using the SED fitting code \FAST{} and their SFR using the \OII{} doublet. Their morphological modelling was performed with \Galfit{} on HST F814W images using a bulge and a disk decomposition. The best-fit models were later used to perform a mass modelling to constrain the impact of the baryons on the total rotation curve of the ionised gas. We included a mean prescription for the thickness of stellar disks as a function of redshift to correct for the impact of finite thickness on the mass and rotation curve of the disk component. The kinematic maps (line flux, velocity field, velocity dispersion, etc.) were extracted from the MUSE cubes using the \OII{} doublet as kinematic tracer, and the 2D kinematic modelling was performed by fitting the baryons mass models combined with a NFW profile to describe the DM halo directly onto the observed velocity field, while modelling the impact of the beam smearing to compute the intrinsic velocity dispersion.
	
	Our kinematic sample was divided into subsamples targetting different environments and we decided to focus our analysis on three scaling relations, namely the size-mass relation, MS and the TFR. As a first step, we selected a sample of star forming disk-like galaxies and studied how using different additional selection criteria, in terms of size, S/N and/or redshift would impact the best-fit slope and zero point for each relation. We found that the redshift and mass selection criteria were important in order not to bias the zero point when comparing between environments since their redshift and mass distributions differ. Additionally, the TFR requires additional criteria, especially in terms of inclination to remove galaxies with poorly constrained kinematics.
	
	We find a $1\sigma$ significant difference ($\SI{0.03}{dex}$) in the size-mass relation as a function of environment when using a richness threshold of $N=10$ to separate between small and large structures, and a $2\sigma$ significant difference ($\SI{0.06}{dex}$) using $N=20$. This result suggests that galaxies in the largest structures have, on average, smaller disks ($\sim 14\%$) than their field counterparts at $z~\approx~0.7$, similar to what was found in the literature. Additionally, we get similar results when using the global effective radius rather than the disk effective radius for our disk sizes. Regarding the MS, we find a $2\sigma$ significant impact of the environment on the zero point of the MS ($\SI{0.1}{dex}$) when using $N=10$ and a $3\sigma$ significant difference (0.15\,dex) when using $N=20$. These offsets are consistent with galaxies located in the large structures having reduced SFR by a factor $1.3 - 1.5$ with respect to field galaxies at a similar redshift.
	
	Finally, after applying mass and redshift cuts, we cannot find any difference in the zero point of the TFR between environments, except when using a richness threshold of $N = 20$ to separate between a field/small structure subsample and a large structure subsample. In this instance, we get an offset of 0.04\,dex which is significant to at most $1\sigma$ significant. By interpreting this offset as being an offset in stellar mass at fixed circular velocity, and by including the contribution of the size-mass relation in the interpretation of the TFR, we find that there must be a small difference of roughly 4\% in stellar fraction between field galaxies and those in the largest structures. Because we measure a negative stellar mass offset in the TFR between the field and the large structure subsamples (galaxies in the large structures are more massive than those in the field), we can rule out the effect of quenching as was suggested in \citet{Abril-Melgarejo2021} when using $N=20$ . On the other hand, because there is no measured difference in zero point with $N=5, 10$ and 15 we can compute upper bounds on the quenching timescale of the galaxies in the large structures using the typical uncertainty found on the TFR zero point. If quenching would indeed lead to a deficit in stellar mass in structures at $z \approx 0.7$ with respect to the field, this would suggest that galaxies have been impacted by the largest structures for between at most $\SI{700}{Myr}$ and $\SI{1.5}{Gyr}$. When including the contribution of the gas into the mass budget of the TFR, we find a similarly significant offset of 0.02\,dex between the field and the large structures (using $N=20$). However, as previously discussed, quenching is still ruled out since this leads to a negative mass offset. Nevertheless, we note that these small differences in zero point vanish once we include the contribution of gas pressure into the dynamics (asymmetric drift correction).
	
	The conclusion from our fully self-consistent study differs from that of \citet{Abril-Melgarejo2021}, even though they investigated and took into account as much as possible methodological biases between the samples they compared. Such a difference might be due to uncontrolled biases when they compared the TFR between samples, or from a possible redshift evolution of the TFR since they could not control the redshift distribution of the various samples as much as we did in this analysis.
	
	This outlines the importance of further reducing those biases by using similar datasets, selection functions as well as analysis methods for galaxies in both low- and high-density environment to measure its impact on galaxy evolution.

	\begin{acknowledgements}
		% We thank the referee for providing useful and constructive comments on the submitted version of this paper. 
		We dedicate this article in memory of Hayley Finley. 
		This work was supported by the Programme National Cosmology et Galaxies (PNCG) of CNRS/INSU with INP and IN2P3, co-funded by CEA and CNES. 
		This work has been carried out through the support of the ANR FOGHAR (ANR-13-BS05-0010-02), the OCEVU Labex (ANR-11-LABX-0060), and the A*MIDEX project (ANR-11-IDEX-0001-02), which are funded by the “Investissements d’avenir” French government program managed by the ANR.
		This work has been carried out thanks to the support of the Ministry of Science, Technology and Innovation of Colombia (MINCIENCIAS) PhD fellowship program No. 756-2016.
		JCBP acknowledges financial support of Vicerrectoría de Investigación y Extensión de la Universidad Industrial de Santander under project 2494.
		This research has made use of \matplotlib{} \citep{matplotlib}, \scipy{} \citep{scipy}, \numpy{} \citep{numpy} and \astropy{} \citep{astropy:2013, astropy:2018}.
		We acknowledge David Carton for his investment in the build-up of the project.
	\end{acknowledgements}

	%%%%%%%%%%%%%%%%%%%%%%%%%%
	%      BIBLIOGRAPHY      %
	%%%%%%%%%%%%%%%%%%%%%%%%%%
	
	\bibliographystyle{aa}
	\bibliography{citations}

\begin{thebibliography}{122}
\expandafter\ifx\csname natexlab\endcsname\relax\def\natexlab#1{#1}\fi

\bibitem[{Abril-Melgarejo {et~al.}(2021)Abril-Melgarejo, Epinat, Mercier,
  Contini, Boogaard, Brinchmann, Finley, Michel-Dansac, Ventou, Amram,
  Krajnović, Mahler, Pineda, \& Richard}]{Abril-Melgarejo2021}
Abril-Melgarejo, V., Epinat, B., Mercier, W., {et~al.} 2021, \aap, 647, A152

\bibitem[{{Astropy Collaboration} {et~al.}(2018){Astropy Collaboration},
  {Price-Whelan}, {Sip{\H{o}}cz}, {G{\"u}nther}, {Lim}, {Crawford}, {Conseil},
  {Shupe}, {Craig}, {Dencheva}, {Ginsburg}, {Vand erPlas}, {Bradley},
  {P{\'e}rez-Su{\'a}rez}, {de Val-Borro}, {Aldcroft}, {Cruz}, {Robitaille},
  {Tollerud}, {Ardelean}, {Babej}, {Bach}, {Bachetti}, {Bakanov}, {Bamford},
  {Barentsen}, {Barmby}, {Baumbach}, {Berry}, {Biscani}, {Boquien}, {Bostroem},
  {Bouma}, {Brammer}, {Bray}, {Breytenbach}, {Buddelmeijer}, {Burke},
  {Calderone}, {Cano Rodr{\'\i}guez}, {Cara}, {Cardoso}, {Cheedella}, {Copin},
  {Corrales}, {Crichton}, {D'Avella}, {Deil}, {Depagne}, {Dietrich}, {Donath},
  {Droettboom}, {Earl}, {Erben}, {Fabbro}, {Ferreira}, {Finethy}, {Fox},
  {Garrison}, {Gibbons}, {Goldstein}, {Gommers}, {Greco}, {Greenfield},
  {Groener}, {Grollier}, {Hagen}, {Hirst}, {Homeier}, {Horton}, {Hosseinzadeh},
  {Hu}, {Hunkeler}, {Ivezi{\'c}}, {Jain}, {Jenness}, {Kanarek}, {Kendrew},
  {Kern}, {Kerzendorf}, {Khvalko}, {King}, {Kirkby}, {Kulkarni}, {Kumar},
  {Lee}, {Lenz}, {Littlefair}, {Ma}, {Macleod}, {Mastropietro}, {McCully},
  {Montagnac}, {Morris}, {Mueller}, {Mumford}, {Muna}, {Murphy}, {Nelson},
  {Nguyen}, {Ninan}, {N{\"o}the}, {Ogaz}, {Oh}, {Parejko}, {Parley}, {Pascual},
  {Patil}, {Patil}, {Plunkett}, {Prochaska}, {Rastogi}, {Reddy Janga},
  {Sabater}, {Sakurikar}, {Seifert}, {Sherbert}, {Sherwood-Taylor}, {Shih},
  {Sick}, {Silbiger}, {Singanamalla}, {Singer}, {Sladen}, {Sooley},
  {Sornarajah}, {Streicher}, {Teuben}, {Thomas}, {Tremblay}, {Turner},
  {Terr{\'o}n}, {van Kerkwijk}, {de la Vega}, {Watkins}, {Weaver}, {Whitmore},
  {Woillez}, {Zabalza}, \& {Astropy Contributors}}]{astropy:2018}
{Astropy Collaboration}, {Price-Whelan}, A.~M., {Sip{\H{o}}cz}, B.~M., {et~al.}
  2018, \aj, 156, 123

\bibitem[{{Astropy Collaboration} {et~al.}(2013){Astropy Collaboration},
  {Robitaille}, {Tollerud}, {Greenfield}, {Droettboom}, {Bray}, {Aldcroft},
  {Davis}, {Ginsburg}, {Price-Whelan}, {Kerzendorf}, {Conley}, {Crighton},
  {Barbary}, {Muna}, {Ferguson}, {Grollier}, {Parikh}, {Nair}, {Unther},
  {Deil}, {Woillez}, {Conseil}, {Kramer}, {Turner}, {Singer}, {Fox}, {Weaver},
  {Zabalza}, {Edwards}, {Azalee Bostroem}, {Burke}, {Casey}, {Crawford},
  {Dencheva}, {Ely}, {Jenness}, {Labrie}, {Lim}, {Pierfederici}, {Pontzen},
  {Ptak}, {Refsdal}, {Servillat}, \& {Streicher}}]{astropy:2013}
{Astropy Collaboration}, {Robitaille}, T.~P., {Tollerud}, E.~J., {et~al.} 2013,
  \aap, 558, A33

\bibitem[{Bacon {et~al.}(2017)Bacon, Conseil, Mary, Brinchmann, Shepherd,
  Akhlaghi, Weilbacher, Piqueras, Wisotzki, Lagattuta, Epinat, Guerou, Inami,
  Cantalupo, Courbot, Contini, Richard, Maseda, Bouwens, Bouché, Kollatschny,
  Schaye, Marino, Pello, Herenz, Guiderdoni, \& Carollo}]{bacon_muse_2017}
Bacon, R., Conseil, S., Mary, D., {et~al.} 2017, \aap, 608, A1

\bibitem[{Balogh {et~al.}(2020)Balogh, van~der Burg, Muzzin, Rudnick, Wilson,
  Webb, Biviano, Boak, Cerulo, Chan, Cooper, Gilbank, Gwyn, Lidman, Matharu,
  McGee, Old, Pintos-Castro, Reeves, Shipley, Vulcani, Yee, Alonso, Bellhouse,
  Cooke, Davidson, De~Lucia, Demarco, Drakos, Fillingham, Finoguenov, Forrest,
  Golledge, Jablonka, Lambas~Garcia, McNab, Muriel, Nantais, Noble, Parker,
  Petter, Poggianti, Townsend, Valotto, Webb, \&
  Zaritsky}]{balogh_gogreen_2020}
Balogh, M.~L., van~der Burg, R. F.~J., Muzzin, A., {et~al.} 2020, \mnras, 500,
  358

\bibitem[{Bizyaev {et~al.}(2014)Bizyaev, Kautsch, Mosenkov, Reshetnikov,
  Sotnikova, Yablokova, \& Hillyer}]{bizyaev_catalog_2014}
Bizyaev, D.~V., Kautsch, S.~J., Mosenkov, A.~V., {et~al.} 2014, \apj, 787, 24

\bibitem[{Boogaard {et~al.}(2018)Boogaard, Brinchmann, Bouché, Paalvast,
  Bacon, Bouwens, Contini, Gunawardhana, Inami, Marino, Maseda, Mitchell,
  Nanayakkara, Richard, Schaye, Schreiber, Tacchella, Wisotzki, \&
  Zabl}]{boogaard_muse_2018}
Boogaard, L.~A., Brinchmann, J., Bouché, N., {et~al.} 2018, \aap, 619, A27

\bibitem[{Boselli {et~al.}(2019)Boselli, Epinat, Contini, Abril-Melgarejo,
  Boogaard, Pointecouteau, Ventou, Brinchmann, Carton, Finley, Michel-Dansac,
  Soucail, \& Weilbacher}]{boselli_evidence_2019}
Boselli, A., Epinat, B., Contini, T., {et~al.} 2019, \aap, 631, A114

\bibitem[{Bottinelli {et~al.}(1983)Bottinelli, Gouguenheim, Paturel, \&
  de~Vaucouleurs}]{bottinelli_h_1983}
Bottinelli, L., Gouguenheim, L., Paturel, G., \& de~Vaucouleurs, G. 1983, \aap,
  118, 4

\bibitem[{Bouché {et~al.}(2013)Bouché, Murphy, Kacprzak, Péroux, Contini,
  Martin, \& Dessauges-Zavadsky}]{bouche_signatures_2013}
Bouché, N., Murphy, M.~T., Kacprzak, G.~G., {et~al.} 2013, Science, 341, 50

\bibitem[{Bouché {et~al.}(2021)Bouché, Bera, Krajnovic, Emsellem, Mercier,
  Schaye, Épinat, Richard, Zoutendijk, Abril-Melgarejo, Brinchmann, Bacon,
  Contini, Boogaard, Wisotzki, Maseda, \& Steinmetz}]{bouche_muse_2021}
Bouché, N.~F., Bera, S., Krajnovic, D., {et~al.} 2021, arXiv:2109.07545
  [astro-ph], arXiv: 2109.07545

\bibitem[{{Buchner} {et~al.}(2014){Buchner}, {Georgakakis}, {Nandra}, {Hsu},
  {Rangel}, {Brightman}, {Merloni}, {Salvato}, {Donley}, \&
  {Kocevski}}]{PyMULTINEST}
{Buchner}, J., {Georgakakis}, A., {Nandra}, K., {et~al.} 2014, \aap, 564, A125

\bibitem[{{Calzetti} {et~al.}(2000){Calzetti}, {Armus}, {Bohlin}, {Kinney},
  {Koornneef}, \& {Storchi-Bergmann}}]{Calzetti2000}
{Calzetti}, D., {Armus}, L., {Bohlin}, R.~C., {et~al.} 2000, \apj, 533, 682

\bibitem[{Cappellari {et~al.}(2013)Cappellari, Scott, Alatalo, Blitz, Bois,
  Bournaud, Bureau, Crocker, Davies, Davis, de~Zeeuw, Duc, Emsellem, Khochfar,
  Krajnović, Kuntschner, McDermid, Morganti, Naab, Oosterloo, Sarzi, Serra,
  Weijmans, \& Young}]{ltsfit}
Cappellari, M., Scott, N., Alatalo, K., {et~al.} 2013, \mnras, 432, 1709

\bibitem[{Cardelli {et~al.}(1989)Cardelli, Clayton, \&
  Mathis}]{cardelli_relationship_1989}
Cardelli, J.~A., Clayton, G.~C., \& Mathis, J.~S. 1989, \apj, 345, 245

\bibitem[{{Chabrier}(2003)}]{Chabrier2003}
{Chabrier}, G. 2003, \pasp, 115, 763

\bibitem[{{Conroy} \& {Gunn}(2010)}]{Conroy2010}
{Conroy}, C. \& {Gunn}, J.~E. 2010, \apj, 712, 833

\bibitem[{Contini {et~al.}(2016)Contini, Epinat, Bouché, Brinchmann, Boogaard,
  Ventou, Bacon, Richard, Weilbacher, Wisotzki, Krajnović, Vielfaure,
  Emsellem, Finley, Inami, Schaye, Swinbank, Guérou, Martinsson,
  Michel-Dansac, Schroetter, Shirazi, \& Soucail}]{Contini2016}
Contini, T., Epinat, B., Bouché, N., {et~al.} 2016, \aap, 591, A49

\bibitem[{{Cortese} {et~al.}(2021){Cortese}, {Catinella}, \&
  {Smith}}]{cortese_dawes_2021}
{Cortese}, L., {Catinella}, B., \& {Smith}, R. 2021, \pasa, 38, e035

\bibitem[{Cowie \& McKee(1977)}]{cowie_evaporation_1977}
Cowie, L.~L. \& McKee, C.~F. 1977, \apj, 211, 135

\bibitem[{Cowie \& Songaila(1977)}]{cowie_thermal_1977}
Cowie, L.~L. \& Songaila, A. 1977, Nature, 266, 501

\bibitem[{{Dav{\'e}}(2009)}]{dave_missing_2009}
{Dav{\'e}}, R. 2009, in Astronomical Society of the Pacific Conference Series,
  Vol. 419, Galaxy Evolution: Emerging Insights and Future Challenges, ed.
  S.~{Jogee}, I.~{Marinova}, L.~{Hao}, \& G.~A. {Blanc}, 347

\bibitem[{Dimauro {et~al.}(2018)Dimauro, Huertas-Company, Daddi,
  Pérez-González, Bernardi, Barro, Buitrago, Caro, Cattaneo,
  Dominguez-Sánchez, Faber, Häußler, Kocevski, Koekemoer, Koo, Lee, Mei,
  Margalef-Bentabol, Primack, Rodriguez-Puebla, Salvato, Shankar, \&
  Tuccillo}]{Dimauro+18}
Dimauro, P., Huertas-Company, M., Daddi, E., {et~al.} 2018, \mnras, 478, 5410

\bibitem[{Duncan {et~al.}(2019)Duncan, Conselice, Mundy, Bell, Donley,
  Galametz, Guo, Grogin, Hathi, Kartaltepe, Kocevski, Koekemoer,
  Pérez-González, Mantha, Snyder, \& Stefanon}]{duncan_observational_2019}
Duncan, K., Conselice, C.~J., Mundy, C., {et~al.} 2019, \apj, 876, 110

\bibitem[{Epinat {et~al.}(2010)Epinat, Amram, Balkowski, \&
  Marcelin}]{Epinat2010}
Epinat, B., Amram, P., Balkowski, C., \& Marcelin, M. 2010, \mnras, 401, 2113

\bibitem[{Epinat {et~al.}(2018)Epinat, Contini, Finley, Boogaard, Guérou,
  Brinchmann, Carton, Michel-Dansac, Bacon, Cantalupo, Carollo, Hamer,
  Kollatschny, Krajnović, Marino, Richard, Soucail, Weilbacher, \&
  Wisotzki}]{epinat_ionised_2018}
Epinat, B., Contini, T., Finley, H., {et~al.} 2018, \aap, 609, A40

\bibitem[{{Epinat} {et~al.}(2009){Epinat}, {Contini}, {Le F{\`e}vre},
  {Vergani}, {Garilli}, {Amram}, {Queyrel}, {Tasca}, \& {Tresse}}]{Epinat2009}
{Epinat}, B., {Contini}, T., {Le F{\`e}vre}, O., {et~al.} 2009, \aap, 504, 789

\bibitem[{Epinat {et~al.}(2012)Epinat, Tasca, Amram, Contini, Le~Fèvre,
  Queyrel, Vergani, Garilli, Kissler-Patig, Moultaka, Paioro, Tresse, Bournaud,
  López-Sanjuan, \& Perret}]{Epinat2012}
Epinat, B., Tasca, L., Amram, P., {et~al.} 2012, \aap, 539, A92

\bibitem[{Erb {et~al.}(2006)Erb, Shapley, Pettini, Steidel, Reddy, \&
  Adelberger}]{erb_massmetallicity_2006}
Erb, D.~K., Shapley, A.~E., Pettini, M., {et~al.} 2006, \apj, 644, 813

\bibitem[{Erfanianfar {et~al.}(2016)Erfanianfar, Popesso, Finoguenov, Wilman,
  Wuyts, Biviano, Salvato, Mirkazemi, Morselli, Ziparo, Nandra, Lutz, Elbaz,
  Dickinson, Tanaka, Altieri, Aussel, Bauer, Berta, Bielby, Brandt, Cappelluti,
  Cimatti, Cooper, Fadda, Ilbert, Le~Floch, Magnelli, Mulchaey, Nordon, Newman,
  Poglitsch, \& Pozzi}]{erfanianfar_non-linearity_2016}
Erfanianfar, G., Popesso, P., Finoguenov, A., {et~al.} 2016, \mnras, 455, 2839

\bibitem[{{Feroz} \& {Hobson}(2008)}]{MULTINEST}
{Feroz}, F. \& {Hobson}, M.~P. 2008, \mnras, 384, 449

\bibitem[{Flores {et~al.}(2006)Flores, Hammer, Puech, Amram, \&
  Balkowski}]{flores_3d_2006}
Flores, H., Hammer, F., Puech, M., Amram, P., \& Balkowski, C. 2006, \aap, 455,
  107

\bibitem[{Foster {et~al.}(2017)Foster, van~de Sande, D'Eugenio, Cortese,
  McDermid, Bland-Hawthorn, Brough, Bryant, Croom, Goodwin, Konstantopoulos,
  Lawrence, López-Sánchez, Medling, Owers, Richards, Scott, Taranu, Tonini,
  \& Zafar}]{foster_sami_2017}
Foster, C., van~de Sande, J., D'Eugenio, F., {et~al.} 2017, Monthly Notices of
  the Royal Astronomical Society, 472, 966

\bibitem[{Freeman(1970)}]{freeman_disks_1970}
Freeman, K.~C. 1970, \apj, 160, 811

\bibitem[{Freundlich {et~al.}(2019)Freundlich, Combes, Tacconi, Genzel,
  Garcia-Burillo, Neri, Contini, Bolatto, Lilly, Salomé, Bicalho, Boissier,
  Boone, Bouché, Bournaud, Burkert, Carollo, Cooper, Cox, Feruglio,
  Förster~Schreiber, Juneau, Lippa, Lutz, Naab, Renzini, Saintonge, Sternberg,
  Walter, Weiner, Weiß, \& Wuyts}]{freundlich_phibss2_2019}
Freundlich, J., Combes, F., Tacconi, L.~J., {et~al.} 2019, \aap, 622, A105

\bibitem[{Genzel {et~al.}(2020)Genzel, Price, Übler, Förster~Schreiber,
  Shimizu, Tacconi, Bender, Burkert, Contursi, Coogan, Davies, Davies, Dekel,
  Herrera-Camus, Lee, Lutz, Naab, Neri, Nestor, Renzini, Saglia, Schuster,
  Sternberg, Wisnioski, \& Wuyts}]{genzel_rotation_2020}
Genzel, R., Price, S.~H., Übler, H., {et~al.} 2020, \apj, 902, 98

\bibitem[{Gilbank {et~al.}(2010)Gilbank, Baldry, Balogh, Glazebrook, \&
  Bower}]{gilbank_local_2010}
Gilbank, D.~G., Baldry, I.~K., Balogh, M.~L., Glazebrook, K., \& Bower, R.~G.
  2010, \mnras, no

\bibitem[{Gilbank {et~al.}(2011)Gilbank, Baldry, Balogh, Glazebrook, \&
  Bower}]{gilbank_erratum_2011}
Gilbank, D.~G., Baldry, I.~K., Balogh, M.~L., Glazebrook, K., \& Bower, R.~G.
  2011, \mnras, 412, 2111

\bibitem[{{Graham} {et~al.}(2005){Graham}, {Driver}, {Petrosian}, {Conselice},
  {Bershady}, {Crawford}, \& {Goto}}]{Graham2005}
{Graham}, A.~W., {Driver}, S.~P., {Petrosian}, V., {et~al.} 2005, \aj, 130,
  1535

\bibitem[{Gray {et~al.}(2009)Gray, Wolf, Barden, Peng, Häußler, Bell,
  McIntosh, Guo, Caldwell, Bacon, Balogh, Barazza, Böhm, Heymans, Jahnke,
  Jogee, van Kampen, Lane, Meisenheimer, Sánchez, Taylor, Wisotzki, Zheng,
  Green, Beswick, Saikia, Gilmour, Johnson, \& Papovich}]{gray_stages_2009}
Gray, M.~E., Wolf, C., Barden, M., {et~al.} 2009, \mnras, 393, 1275

\bibitem[{Gunn \& Gott(1972)}]{gunn_infall_1972}
Gunn, J.~E. \& Gott, III, J.~R. 1972, \apj, 176, 1

\bibitem[{Guérou {et~al.}(2017)Guérou, Krajnović, Epinat, Contini, Emsellem,
  Bouché, Bacon, Michel-Dansac, Richard, Weilbacher, Schaye, Marino, den Brok,
  \& Erroz-Ferrer}]{guerou_muse_2017}
Guérou, A., Krajnović, D., Epinat, B., {et~al.} 2017, \aap, 608, A5

\bibitem[{Harris {et~al.}(2020)Harris, Millman, van~der Walt, Gommers,
  Virtanen, Cournapeau, Wieser, Taylor, Berg, Smith, Kern, Picus, Hoyer, van
  Kerkwijk, Brett, Haldane, del R{\'{i}}o, Wiebe, Peterson,
  G{\'{e}}rard-Marchant, Sheppard, Reddy, Weckesser, Abbasi, Gohlke, \&
  Oliphant}]{numpy}
Harris, C.~R., Millman, K.~J., van~der Walt, S.~J., {et~al.} 2020, Nature, 585,
  357

\bibitem[{Heidmann {et~al.}(1972)Heidmann, Heidmann, \&
  de~Vaucouleurs}]{heidmann_inclination_1972}
Heidmann, J., Heidmann, N., \& de~Vaucouleurs, G. 1972, Memoirs of the Royal
  Astronomical Society, 75, 85

\bibitem[{Hernquist(1990)}]{hernquist_analytical_1990}
Hernquist, L. 1990, \apj, 356, 359

\bibitem[{{Hinton}(2016)}]{Hanton2016}
{Hinton}, S. 2016, {MARZ: Redshifting Program}

\bibitem[{Hopkins \& Beacom(2006)}]{hopkins_normalization_2006}
Hopkins, A.~M. \& Beacom, J.~F. 2006, \apj, 651, 142

\bibitem[{Hopkins {et~al.}(2012)Hopkins, Quataert, \&
  Murray}]{hopkins_stellar_2012}
Hopkins, P.~F., Quataert, E., \& Murray, N. 2012, \mnras, 421, 3522

\bibitem[{Hunter(2007)}]{matplotlib}
Hunter, J.~D. 2007, Computing in Science \& Engineering, 9, 90

\bibitem[{Ilbert {et~al.}(2010)Ilbert, Salvato, Le~Floc'h, Aussel, Capak,
  McCracken, Mobasher, Kartaltepe, Scoville, Sanders, Arnouts, Bundy, Cassata,
  Kneib, Koekemoer, Le~Fèvre, Lilly, Surace, Taniguchi, Tasca, Thompson,
  Tresse, Zamojski, Zamorani, \& Zucca}]{ilbert_galaxy_2010}
Ilbert, O., Salvato, M., Le~Floc'h, E., {et~al.} 2010, \apj, 709, 644

\bibitem[{Inami {et~al.}(2017)Inami, Bacon, Brinchmann, Richard, Contini,
  Conseil, Hamer, Akhlaghi, Bouché, Clément, Desprez, Drake, Hashimoto,
  Leclercq, Maseda, Michel-Dansac, Paalvast, Tresse, Ventou, Kollatschny,
  Boogaard, Finley, Marino, Schaye, \& Wisotzki}]{inami2017}
Inami, H., Bacon, R., Brinchmann, J., {et~al.} 2017, \aap, 608, A2

\bibitem[{Iovino {et~al.}(2016)Iovino, Petropoulou, Scodeggio, Bolzonella,
  Zamorani, Bardelli, Cucciati, Pozzetti, Tasca, Vergani, Zucca, Finoguenov,
  Ilbert, Tanaka, Salvato, Kovač, \& Cassata}]{iovino_high_2016}
Iovino, A., Petropoulou, V., Scodeggio, M., {et~al.} 2016, \aap, 592, A78

\bibitem[{Jaffe(1983)}]{jaffe_simple_1983}
Jaffe, W. 1983, \mnras, 202, 995

\bibitem[{Kelkar {et~al.}(2015)Kelkar, Aragón-Salamanca, Gray, Maltby,
  Vulcani, De~Lucia, Poggianti, \& Zaritsky}]{kelkar_galaxy_2015}
Kelkar, K., Aragón-Salamanca, A., Gray, M.~E., {et~al.} 2015, \mnras, 450,
  1246

\bibitem[{Kennicutt(1992)}]{kennicutt_integrated_1992}
Kennicutt, Jr., R.~C. 1992, \apj, 388, 310

\bibitem[{Kennicutt(1998{\natexlab{a}})}]{kennicutt_star_1998}
Kennicutt, Jr., R.~C. 1998{\natexlab{a}}, \apj, 498, 541

\bibitem[{Kennicutt(1998{\natexlab{b}})}]{kennicutt_jr_global_1998}
Kennicutt, Jr., R.~C. 1998{\natexlab{b}}, \araa, 36, 189

\bibitem[{{Kere{\v{s}}} {et~al.}(2005){Kere{\v{s}}}, {Katz}, {Weinberg}, \&
  {Dav{\'e}}}]{kere_how_2005}
{Kere{\v{s}}}, D., {Katz}, N., {Weinberg}, D.~H., \& {Dav{\'e}}, R. 2005,
  \mnras, 363, 2

\bibitem[{{Knobel} {et~al.}(2012){Knobel}, {Lilly}, {Iovino}, {Kova{\v{c}}},
  {Bschorr}, {Presotto}, {Oesch}, {Kampczyk}, {Carollo}, {Contini}, {Kneib},
  {Le Fevre}, {Mainieri}, {Renzini}, {Scodeggio}, {Zamorani}, {Bardelli},
  {Bolzonella}, {Bongiorno}, {Caputi}, {Cucciati}, {de la Torre}, {de Ravel},
  {Franzetti}, {Garilli}, {Lamareille}, {Le Borgne}, {Le Brun}, {Maier},
  {Mignoli}, {Pello}, {Peng}, {Perez Montero}, {Silverman}, {Tanaka}, {Tasca},
  {Tresse}, {Vergani}, {Zucca}, {Barnes}, {Bordoloi}, {Cappi}, {Cimatti},
  {Coppa}, {Koekemoer}, {L{\'o}pez-Sanjuan}, {McCracken}, {Moresco}, {Nair},
  {Pozzetti}, \& {Welikala}}]{Knobel2012}
{Knobel}, C., {Lilly}, S.~J., {Iovino}, A., {et~al.} 2012, \apj, 753, 121

\bibitem[{{Knobel} {et~al.}(2009){Knobel}, {Lilly}, {Iovino}, {Porciani},
  {Kova{\v{c}}}, {Cucciati}, {Finoguenov}, {Kitzbichler}, {Carollo}, {Contini},
  {Kneib}, {Le F{\`e}vre}, {Mainieri}, {Renzini}, {Scodeggio}, {Zamorani},
  {Bardelli}, {Bolzonella}, {Bongiorno}, {Caputi}, {Coppa}, {de la Torre}, {de
  Ravel}, {Franzetti}, {Garilli}, {Kampczyk}, {Lamareille}, {Le Borgne}, {Le
  Brun}, {Maier}, {Mignoli}, {Pello}, {Peng}, {Perez Montero}, {Ricciardelli},
  {Silverman}, {Tanaka}, {Tasca}, {Tresse}, {Vergani}, {Zucca}, {Abbas},
  {Bottini}, {Cappi}, {Cassata}, {Cimatti}, {Fumana}, {Guzzo}, {Koekemoer},
  {Leauthaud}, {Maccagni}, {Marinoni}, {McCracken}, {Memeo}, {Meneux}, {Oesch},
  {Pozzetti}, \& {Scaramella}}]{Knobel2009}
{Knobel}, C., {Lilly}, S.~J., {Iovino}, A., {et~al.} 2009, \apj, 697, 1842

\bibitem[{Koekemoer {et~al.}(2007)Koekemoer, Aussel, Calzetti, Capak,
  Giavalisco, Kneib, Leauthaud, Le~Fevre, McCracken, Massey, Mobasher, Rhodes,
  Scoville, \& Shopbell}]{Koekemoer2007}
Koekemoer, A.~M., Aussel, H., Calzetti, D., {et~al.} 2007, \apjs, 172, 196

\bibitem[{Korsaga {et~al.}(2019)Korsaga, Epinat, Amram, Carignan, Adamczyk, \&
  Sorgho}]{korsaga_ghasp_2019}
Korsaga, M., Epinat, B., Amram, P., {et~al.} 2019, \mnras, 490, 2977

\bibitem[{{Kriek} {et~al.}(2009){Kriek}, {van Dokkum}, {Labb{\'e}}, {Franx},
  {Illingworth}, {Marchesini}, \& {Quadri}}]{FAST}
{Kriek}, M., {van Dokkum}, P.~G., {Labb{\'e}}, I., {et~al.} 2009, \apj, 700,
  221

\bibitem[{Kuchner {et~al.}(2017)Kuchner, Ziegler, Verdugo, Bamford, \&
  Häußler}]{kuchner_effects_2017}
Kuchner, U., Ziegler, B., Verdugo, M., Bamford, S., \& Häußler, B. 2017,
  \aap, 604, A54

\bibitem[{{Laigle} {et~al.}(2016){Laigle}, {McCracken}, {Ilbert}, {Hsieh},
  {Davidzon}, {Capak}, {Hasinger}, {Silverman}, {Pichon}, {Coupon}, {Aussel},
  {Le Borgne}, {Caputi}, {Cassata}, {Chang}, {Civano}, {Dunlop}, {Fynbo},
  {Kartaltepe}, {Koekemoer}, {Le F{\`e}vre}, {Le Floc'h}, {Leauthaud}, {Lilly},
  {Lin}, {Marchesi}, {Milvang-Jensen}, {Salvato}, {Sanders}, {Scoville},
  {Smolcic}, {Stockmann}, {Taniguchi}, {Tasca}, {Toft}, {Vaccari}, \&
  {Zabl}}]{Laigle2016}
{Laigle}, C., {McCracken}, H.~J., {Ilbert}, O., {et~al.} 2016, \apjs, 224, 24

\bibitem[{Laigle {et~al.}(2016)Laigle, McCracken, Ilbert, Hsieh, Davidzon,
  Capak, Hasinger, Silverman, Pichon, Coupon, Aussel, Le~Borgne, Caputi,
  Cassata, Chang, Civano, Dunlop, Fynbo, Kartaltepe, Koekemoer, Le~Fèvre,
  Le~Floc’h, Leauthaud, Lilly, Lin, Marchesi, Milvang-Jensen, Salvato,
  Sanders, Scoville, Smolcic, Stockmann, Taniguchi, Tasca, Toft, Vaccari, \&
  Zabl}]{cosmos2015}
Laigle, C., McCracken, H.~J., Ilbert, O., {et~al.} 2016, \apjs, 224, 24

\bibitem[{Leja {et~al.}(2018)Leja, Johnson, Conroy, \& Dokkum}]{leja_hot_2018}
Leja, J., Johnson, B.~D., Conroy, C., \& Dokkum, P.~v. 2018, \apjl, 854, 62

\bibitem[{Lubin {et~al.}(2009)Lubin, Gal, Lemaux, Kocevski, \&
  Squires}]{ORELSE}
Lubin, L.~M., Gal, R.~R., Lemaux, B.~C., Kocevski, D.~D., \& Squires, G.~K.
  2009, \apj, 137, 4867

\bibitem[{López-Sanjuan {et~al.}(2012)López-Sanjuan, Le~Fèvre, Ilbert,
  Tasca, Bridge, Cucciati, Kampczyk, Pozzetti, Xu, Carollo, Contini, Kneib,
  Lilly, Mainieri, Renzini, Sanders, Scodeggio, Scoville, Taniguchi, Zamorani,
  Aussel, Bardelli, Bolzonella, Bongiorno, Capak, Caputi, de~la Torre,
  de~Ravel, Franzetti, Garilli, Iovino, Knobel, Kovač, Lamareille, Le~Borgne,
  Le~Brun, Le~Floc’h, Maier, McCracken, Mignoli, Pelló, Peng,
  Pérez-Montero, Presotto, Ricciardelli, Salvato, Silverman, Tanaka, Tresse,
  Vergani, Zucca, Barnes, Bordoloi, Cappi, Cimatti, Coppa, Koekemoer, Liu,
  Moresco, Nair, Oesch, Schawinski, \& Welikala}]{lopez-sanjuan_dominant_2012}
López-Sanjuan, C., Le~Fèvre, O., Ilbert, O., {et~al.} 2012, \aap, 548, A7

\bibitem[{Maltby {et~al.}(2010)Maltby, Aragón-Salamanca, Gray, Barden,
  Häußler, Wolf, Peng, Jahnke, McIntosh, Böhm, \& van
  Kampen}]{maltby_environmental_2010}
Maltby, D.~T., Aragón-Salamanca, A., Gray, M.~E., {et~al.} 2010, \mnras, 402,
  282

\bibitem[{Mantha {et~al.}(2018)Mantha, McIntosh, Brennan, Ferguson, Kodra,
  Newman, Rafelski, Somerville, Conselice, Cook, Hathi, Koo, Lotz, Simmons,
  Straughn, Snyder, Wuyts, Bell, Dekel, Kartaltepe, Kocevski, Koekemoer, Lee,
  Lucas, Pacifici, Peth, Barro, Dahlen, Finkelstein, Fontana, Galametz, Grogin,
  Guo, Mobasher, Nayyeri, Pérez-González, Pforr, Santini, Stefanon, \&
  Wiklind}]{mantha_major_2018}
Mantha, K.~B., McIntosh, D.~H., Brennan, R., {et~al.} 2018, \mnras, 475, 1549

\bibitem[{{Markwardt}(2009)}]{MPFIT}
{Markwardt}, C.~B. 2009, in Astronomical Society of the Pacific Conference
  Series, Vol. 411, Astronomical Data Analysis Software and Systems XVIII, ed.
  D.~A. {Bohlender}, D.~{Durand}, \& P.~{Dowler}, 251

\bibitem[{Massey {et~al.}(2010)Massey, Stoughton, Leauthaud, Rhodes, Koekemoer,
  Ellis, \& Shaghoulian}]{Massey2010}
Massey, R., Stoughton, C., Leauthaud, A., {et~al.} 2010, \mnras, 401, 371

\bibitem[{Matharu {et~al.}(2019)Matharu, Muzzin, Brammer, van~der Burg, Auger,
  Hewett, van~der Wel, van Dokkum, Balogh, Chan, Demarco, Marchesini, Nelson,
  Noble, Wilson, \& Yee}]{matharu_hst_2019}
Matharu, J., Muzzin, A., Brammer, G.~B., {et~al.} 2019, \mnras, 484, 595

\bibitem[{Meurer {et~al.}(1996)Meurer, Carignan, Beaulieu, \&
  Freeman}]{meurer_ngc_1996}
Meurer, G.~R., Carignan, C., Beaulieu, S.~F., \& Freeman, K.~C. 1996, \apj,
  111, 1551

\bibitem[{{Mowla} {et~al.}(2019){Mowla}, {van Dokkum}, {Brammer}, {Momcheva},
  {van der Wel}, {Whitaker}, {Nelson}, {Bezanson}, {Muzzin}, {Franx},
  {MacKenty}, {Leja}, {Kriek}, \& {Marchesini}}]{mowla_cosmos-dash:_2018}
{Mowla}, L.~A., {van Dokkum}, P., {Brammer}, G.~B., {et~al.} 2019, \apj, 880,
  57

\bibitem[{Muzzin {et~al.}(2013)Muzzin, Marchesini, Stefanon, Franx, McCracken,
  Milvang-Jensen, Dunlop, Fynbo, Brammer, Labbé, \& van
  Dokkum}]{muzzin_evolution_2013}
Muzzin, A., Marchesini, D., Stefanon, M., {et~al.} 2013, \apj, 777, 18

\bibitem[{Muzzin {et~al.}(2009)Muzzin, Wilson, Yee, Hoekstra, Gilbank, Surace,
  Lacy, Blindert, Majumdar, Demarco, Gardner, Gladders, \&
  Lonsdale}]{muzzin_spectroscopic_2009}
Muzzin, A., Wilson, G., Yee, H. K.~C., {et~al.} 2009, \apj, 698, 1934

\bibitem[{Nantais {et~al.}(2020)Nantais, Wilson, Muzzin, Old, Demarco, Cerulo,
  Balogh, Rudnick, Chan, Cooper, Forrest, Hayden, Lidman, Noble, Perlmutter,
  Rhea, Surace, van~der Burg, \& van Kampen}]{nantais_h_2020}
Nantais, J., Wilson, G., Muzzin, A., {et~al.} 2020, \mnras, 499, 3061

\bibitem[{Navarro {et~al.}(1995)Navarro, Frenk, \&
  White}]{navarro_assembly_1995}
Navarro, J.~F., Frenk, C.~S., \& White, S. D.~M. 1995, \mnras, 275, 56

\bibitem[{Navarro {et~al.}(1996)Navarro, Frenk, \&
  White}]{navarro_structure_1996}
Navarro, J.~F., Frenk, C.~S., \& White, S. D.~M. 1996, \apj, 462, 563

\bibitem[{Noeske {et~al.}(2007)Noeske, Weiner, Faber, Papovich, Koo,
  Somerville, Bundy, Conselice, Newman, Schiminovich, Le~Floc'h, Coil, Rieke,
  Lotz, Primack, Barmby, Cooper, Davis, Ellis, Fazio, Guhathakurta, Huang,
  Kassin, Martin, Phillips, Rich, Small, Willmer, \& Wilson}]{noeske_star_2007}
Noeske, K.~G., Weiner, B.~J., Faber, S.~M., {et~al.} 2007, \apj, 660, L43

\bibitem[{Ocvirk {et~al.}(2008)Ocvirk, Pichon, \&
  Teyssier}]{ocvirk_bimodal_2008}
Ocvirk, P., Pichon, C., \& Teyssier, R. 2008, \mnras

\bibitem[{Old {et~al.}(2020{\natexlab{a}})Old, Balogh, van~der Burg, Biviano,
  Yee, Pintos-Castro, Webb, Muzzin, Rudnick, Vulcani, Poggianti, Cooper,
  Zaritsky, Cerulo, Wilson, Chan, Lidman, McGee, Demarco, Forrest, De~Lucia,
  Gilbank, Kukstas, McCarthy, Jablonka, Nantais, Noble, Reeves, \&
  Shipley}]{old_erratum_2020}
Old, L.~J., Balogh, M.~L., van~der Burg, R. F.~J., {et~al.} 2020{\natexlab{a}},
  \mnras, 500, 355

\bibitem[{Old {et~al.}(2020{\natexlab{b}})Old, Balogh, van~der Burg, Biviano,
  Yee, Pintos-Castro, Webb, Muzzin, Rudnick, Vulcani, Poggianti, Cooper,
  Zaritsky, Cerulo, Wilson, Chan, Lidman, McGee, Demarco, Forrest, De~Lucia,
  Gilbank, Kukstas, McCarthy, Jablonka, Nantais, Noble, Reeves, \&
  Shipley}]{old_gogreen_2020}
Old, L.~J., Balogh, M.~L., van~der Burg, R. F.~J., {et~al.} 2020{\natexlab{b}},
  \mnras, 493, 5987

\bibitem[{{Osterbrock} \& {Ferland}(2006)}]{Osterbrock2006}
{Osterbrock}, D.~E. \& {Ferland}, G.~J. 2006, {Astrophysics of gaseous nebulae
  and active galactic nuclei}

\bibitem[{{Padilla} {et~al.}(2009){Padilla}, {Lagos}, \&
  {Strauss}}]{padilla_stellar_2009}
{Padilla}, N.~D., {Lagos}, C., \& {Strauss}, M. 2009, in American Institute of
  Physics Conference Series, Vol. 1201, The Monster's Fiery Breath: Feedback in
  Galaxies, Groups, and Clusters, ed. S.~{Heinz} \& E.~{Wilcots}, 100--103

\bibitem[{Pelliccia {et~al.}(2019)Pelliccia, Lemaux, Tomczak, Lubin, Shen,
  Epinat, Wu, Gal, Rumbaugh, Kocevski, Tresse, \& Squires}]{Pelliccia2019}
Pelliccia, D., Lemaux, B.~C., Tomczak, A.~R., {et~al.} 2019, \mnras, 482, 3514

\bibitem[{Pelliccia {et~al.}(2017)Pelliccia, Tresse, Epinat, Ilbert, Scoville,
  Amram, Lemaux, \& Zamorani}]{Pelliccia2017}
Pelliccia, D., Tresse, L., Epinat, B., {et~al.} 2017, \aap, 599, A25

\bibitem[{{Peng} {et~al.}(2002{\natexlab{a}}){Peng}, {Ho}, {Impey}, \&
  {Rix}}]{GALFIT}
{Peng}, C.~Y., {Ho}, L.~C., {Impey}, C.~D., \& {Rix}, H.-W. 2002{\natexlab{a}},
  \aj, 124, 266

\bibitem[{{Peng} {et~al.}(2002{\natexlab{b}}){Peng}, {Peng}, {Chou}, \&
  {Lin}}]{Peng2002}
{Peng}, Q.-H., {Peng}, F., {Chou}, C.-K., \& {Lin}, Y.-H. 2002{\natexlab{b}},
  \apss, 282, 499

\bibitem[{Peng {et~al.}(2010)Peng, Lilly, Kovač, Bolzonella, Pozzetti,
  Renzini, Zamorani, Ilbert, Knobel, Iovino, Maier, Cucciati, Tasca, Carollo,
  Silverman, Kampczyk, de~Ravel, Sanders, Scoville, Contini, Mainieri,
  Scodeggio, Kneib, Le~Fèvre, Bardelli, Bongiorno, Caputi, Coppa, de~la Torre,
  Franzetti, Garilli, Lamareille, Le~Borgne, Le~Brun, Mignoli, Montero, Pello,
  Ricciardelli, Tanaka, Tresse, Vergani, Welikala, Zucca, Oesch, Abbas, Barnes,
  Bordoloi, Bottini, Cappi, Cassata, Cimatti, Fumana, Hasinger, Koekemoer,
  Leauthaud, Maccagni, Marinoni, McCracken, Memeo, Meneux, Nair, Porciani,
  Presotto, \& Scaramella}]{peng_mass_2010}
Peng, Y.-j., Lilly, S.~J., Kovač, K., {et~al.} 2010, \apj, 721, 193

\bibitem[{Plummer(1911)}]{plummer_problem_1911}
Plummer, H.~C. 1911, \mnras, 71, 460

\bibitem[{Rousseeuw \& Van~Driessen(2006)}]{rousseeuw_computing_2006}
Rousseeuw, P.~J. \& Van~Driessen, K. 2006, Data Mining and Knowledge Discovery,
  12, 29

\bibitem[{Schmidt(1959)}]{schmidt_rate_1959}
Schmidt, M. 1959, \apj, 129, 243

\bibitem[{{Schroetter} {et~al.}(2019){Schroetter}, {Bouch{\'e}}, {Zabl},
  {Contini}, {Wendt}, {Schaye}, {Mitchell}, {Muzahid}, {Marino}, {Bacon},
  {Lilly}, {Richard}, \& {Wisotzki}}]{schroetter_muse_2019}
{Schroetter}, I., {Bouch{\'e}}, N.~F., {Zabl}, J., {et~al.} 2019, \mnras, 490,
  4368

\bibitem[{Scoville {et~al.}(2007)Scoville, Aussel, Benson, Blain, Calzetti,
  Capak, Ellis, El‐Zant, Finoguenov, Giavalisco, Guzzo, Hasinger, Koda,
  Le~Fevre, Massey, McCracken, Mobasher, Renzini, Rhodes, Salvato, Sanders,
  Sasaki, Schinnerer, Sheth, Shopbell, Taniguchi, Taylor, \&
  Thompson}]{scoville_large_2007}
Scoville, N., Aussel, H., Benson, A., {et~al.} 2007, \apjs, 172, 150

\bibitem[{{Scoville} {et~al.}(2007){Scoville}, {Aussel}, {Brusa}, {Capak},
  {Carollo}, {Elvis}, {Giavalisco}, {Guzzo}, {Hasinger}, {Impey}, {Kneib},
  {LeFevre}, {Lilly}, {Mobasher}, {Renzini}, {Rich}, {Sanders}, {Schinnerer},
  {Schminovich}, {Shopbell}, {Taniguchi}, \& {Tyson}}]{Scoville2007}
{Scoville}, N., {Aussel}, H., {Brusa}, M., {et~al.} 2007, \apjs, 172, 1

\bibitem[{Shen {et~al.}(2003)Shen, Mo, White, Blanton, Kauffmann, Voges,
  Brinkmann, \& Csabai}]{shen_size_2003}
Shen, S., Mo, H.~J., White, S. D.~M., {et~al.} 2003, \mnras, 343, 978

\bibitem[{{Soto} {et~al.}(2016){Soto}, {Lilly}, {Bacon}, {Richard}, \&
  {Conseil}}]{ZAP}
{Soto}, K.~T., {Lilly}, S.~J., {Bacon}, R., {Richard}, J., \& {Conseil}, S.
  2016, \mnras, 458, 3210

\bibitem[{Speagle {et~al.}(2014)Speagle, Steinhardt, Capak, \&
  Silverman}]{speagle_highly_2014}
Speagle, J.~S., Steinhardt, C.~L., Capak, P.~L., \& Silverman, J.~D. 2014,
  \apjs, 214, 15

\bibitem[{Tacconi {et~al.}(2018)Tacconi, Genzel, Saintonge, Combes,
  García-Burillo, Neri, Bolatto, Contini, Schreiber, Lilly, Lutz, Wuyts,
  Accurso, Boissier, Boone, Bouché, Bournaud, Burkert, Carollo, Cooper, Cox,
  Feruglio, Freundlich, Herrera-Camus, Juneau, Lippa, Naab, Renzini, Salome,
  Sternberg, Tadaki, Übler, Walter, Weiner, \& Weiss}]{tacconi_phibss_2018}
Tacconi, L.~J., Genzel, R., Saintonge, A., {et~al.} 2018, \apjs, 853, 179

\bibitem[{Tiley {et~al.}(2019)Tiley, Bureau, Cortese, Harrison, Johnson, Stott,
  Swinbank, Smail, Sobral, Bunker, Glazebrook, Bower, Obreschkow, Bryant,
  Jarvis, Bland-Hawthorn, Magdis, Medling, Sweet, Tonini, Turner, Sharples,
  Croom, Goodwin, Konstantopoulos, Lorente, Lawrence, Mould, Owers, \&
  Richards}]{tiley_krosssami_2019}
Tiley, A.~L., Bureau, M., Cortese, L., {et~al.} 2019, \mnras, 482, 2166

\bibitem[{Toomre(1963)}]{toomre_distribution_1963}
Toomre, A. 1963, \apj, 138, 385

\bibitem[{Tremonti {et~al.}(2004)Tremonti, Heckman, Kauffmann, Brinchmann,
  Charlot, White, Seibert, Peng, Schlegel, Uomoto, Fukugita, \&
  Brinkmann}]{tremonti_origin_2004}
Tremonti, C.~A., Heckman, T.~M., Kauffmann, G., {et~al.} 2004, \apj, 613, 898

\bibitem[{Trujillo {et~al.}(2007)Trujillo, Conselice, Bundy, Cooper,
  Eisenhardt, \& Ellis}]{trujillo_strong_2007}
Trujillo, I., Conselice, C.~J., Bundy, K., {et~al.} 2007, \mnras, 382, 109

\bibitem[{Tully \& Fisher(1977)}]{tully_new_1977}
Tully, R.~B. \& Fisher, J.~R. 1977, \aap, 54, 661

\bibitem[{van~der Wel {et~al.}(2014{\natexlab{a}})van~der Wel, Chang, Bell,
  Holden, Ferguson, Giavalisco, Rix, Skelton, Whitaker, Momcheva, Brammer,
  Kassin, Martig, Dekel, Ceverino, Koo, Mozena, van Dokkum, Franx, Faber, \&
  Primack}]{van_der_wel_geometry_2014}
van~der Wel, A., Chang, Y.-Y., Bell, E.~F., {et~al.} 2014{\natexlab{a}}, \apj,
  792, L6

\bibitem[{van~der Wel {et~al.}(2014{\natexlab{b}})van~der Wel, Franx, van
  Dokkum, Skelton, Momcheva, Whitaker, Brammer, Bell, Rix, Wuyts, Ferguson,
  Holden, Barro, Koekemoer, Chang, McGrath, Häussler, Dekel, Behroozi,
  Fumagalli, Leja, Lundgren, Maseda, Nelson, Wake, Patel, Labbé, Faber,
  Grogin, \& Kocevski}]{van_der_wel_3d-hstcandels_2014}
van~der Wel, A., Franx, M., van Dokkum, P.~G., {et~al.} 2014{\natexlab{b}},
  \apj, 788, 28

\bibitem[{Ventou {et~al.}(2017)Ventou, Contini, Bouché, Epinat, Brinchmann,
  Bacon, Inami, Lam, Drake, Garel, Michel-Dansac, Pello, Steinmetz, Weilbacher,
  Wisotzki, \& Carollo}]{ventou_muse_2017}
Ventou, E., Contini, T., Bouché, N., {et~al.} 2017, \aap, 608, A9

\bibitem[{Ventou {et~al.}(2019)Ventou, Contini, Bouché, Epinat, Brinchmann,
  Inami, Richard, Schroetter, Soucail, Steinmetz, \&
  Weilbacher}]{ventou_new_2019}
Ventou, E., Contini, T., Bouché, N., {et~al.} 2019, \aap, 631, A87

\bibitem[{Vergani {et~al.}(2012)Vergani, Epinat, Contini, Tasca, Tresse, Amram,
  Garilli, Kissler-Patig, Le~Fèvre, Moultaka, Paioro, Queyrel, \&
  López-Sanjuan}]{Vergani2012}
Vergani, D., Epinat, B., Contini, T., {et~al.} 2012, \aap, 546, A118

\bibitem[{Virtanen {et~al.}(2020)Virtanen, Gommers, Oliphant, Haberland, Reddy,
  Cournapeau, Burovski, Peterson, Weckesser, Bright, {van der Walt}, Brett,
  Wilson, Millman, Mayorov, Nelson, Jones, Kern, Larson, Carey, Polat, Feng,
  Moore, {VanderPlas}, Laxalde, Perktold, Cimrman, Henriksen, Quintero, Harris,
  Archibald, Ribeiro, Pedregosa, {van Mulbregt}, \& {SciPy 1.0
  Contributors}}]{scipy}
Virtanen, P., Gommers, R., Oliphant, T.~E., {et~al.} 2020, Nature Methods, 17,
  261

\bibitem[{Walter {et~al.}(2020)Walter, Carilli, Neeleman, Decarli, Popping,
  Somerville, Aravena, Bertoldi, Boogaard, Cox, da~Cunha, Magnelli, Obreschkow,
  Riechers, Rix, Smail, Weiss, Assef, Bauer, Bouwens, Contini, Cortes, Daddi,
  Diaz-Santos, González-López, Hennawi, Hodge, Inami, Ivison, Oesch, Sargent,
  van~der Werf, Wagg, \& Yung}]{walter_evolution_2020}
Walter, F., Carilli, C., Neeleman, M., {et~al.} 2020, \apj, 902, 111

\bibitem[{{Weilbacher} {et~al.}(2020){Weilbacher}, {Palsa}, {Streicher},
  {Bacon}, {Urrutia}, {Wisotzki}, {Conseil}, {Husemann}, {Jarno}, {Kelz},
  {P{\'e}contal-Rousset}, {Richard}, {Roth}, {Selman}, \&
  {Vernet}}]{Weilbacher2020}
{Weilbacher}, P.~M., {Palsa}, R., {Streicher}, O., {et~al.} 2020, \aap, 641,
  A28

\bibitem[{Whitaker {et~al.}(2014)Whitaker, Franx, Leja, van Dokkum, Henry,
  Skelton, Fumagalli, Momcheva, Brammer, Labbé, Nelson, \&
  Rigby}]{whitaker_constraining_2014}
Whitaker, K.~E., Franx, M., Leja, J., {et~al.} 2014, \apj, 795, 104

\bibitem[{Williams {et~al.}(2010)Williams, Bureau, \& Cappellari}]{MPFITEXY}
Williams, M.~J., Bureau, M., \& Cappellari, M. 2010, \mnras, 409, 1330

\bibitem[{Wuyts {et~al.}(2011)Wuyts, Förster~Schreiber, Lutz, Nordon, Berta,
  Altieri, Andreani, Aussel, Bongiovanni, Cepa, Cimatti, Daddi, Elbaz, Genzel,
  Koekemoer, Magnelli, Maiolino, McGrath, García, Poglitsch, Popesso, Pozzi,
  Sanchez-Portal, Sturm, Tacconi, \& Valtchanov}]{wuyts_star_2011}
Wuyts, S., Förster~Schreiber, N.~M., Lutz, D., {et~al.} 2011, \apj, 738, 106

\bibitem[{Yang {et~al.}(2008)Yang, Flores, Hammer, Neichel, Puech, Nesvadba,
  Rawat, Cesarsky, Lehnert, Pozzetti, Fuentes-Carrera, Amram, Balkowski,
  Dannerbauer, di~Serego~Alighieri, Guiderdoni, Kembhavi, Liang, Östlin,
  Ravikumar, Vergani, Vernet, \& Wozniak}]{yang_images._2008}
Yang, Y., Flores, H., Hammer, F., {et~al.} 2008, \aap, 477, 789

\bibitem[{Zabl {et~al.}(2019)Zabl, Bouché, Schroetter, Wendt, Finley, Schaye,
  Conseil, Contini, Marino, Mitchell, Muzahid, Pezzulli, \&
  Wisotzki}]{zabl_muse_2019}
Zabl, J., Bouché, N.~F., Schroetter, I., {et~al.} 2019, \mnras, 485, 1961

\bibitem[{Zhang {et~al.}(2019)Zhang, Primack, Faber, Koo, Dekel, Chen,
  Ceverino, Chang, Fang, Guo, Lin, \& Wel}]{zhang_evolution_2019}
Zhang, H., Primack, J.~R., Faber, S.~M., {et~al.} 2019, \mnras, 484, 5170

\bibitem[{Übler {et~al.}(2017)Übler, Förster~Schreiber, Genzel, Wisnioski,
  Wuyts, Lang, Naab, Burkert, van Dokkum, Tacconi, Wilman, Fossati, Mendel,
  Beifiori, Belli, Bender, Brammer, Chan, Davies, Fabricius, Galametz, Lutz,
  Momcheva, Nelson, Saglia, Seitz, \& Tadaki}]{ubler_evolution_2017}
Übler, H., Förster~Schreiber, N.~M., Genzel, R., {et~al.} 2017, \apj, 842,
  121

\end{thebibliography}
	
	%%%%%%%%%%%%%%%%%%%%%%%%
	%      APPENDICES      %
	%%%%%%%%%%%%%%%%%%%%%%%%
	
	\clearpage
	\begin{appendix}
		\section{Additional plots and tables}
		\label{Appendix:rotation_comparison}		
		
		\begin{figure}[hbt!]
			\includegraphics[scale=0.7]{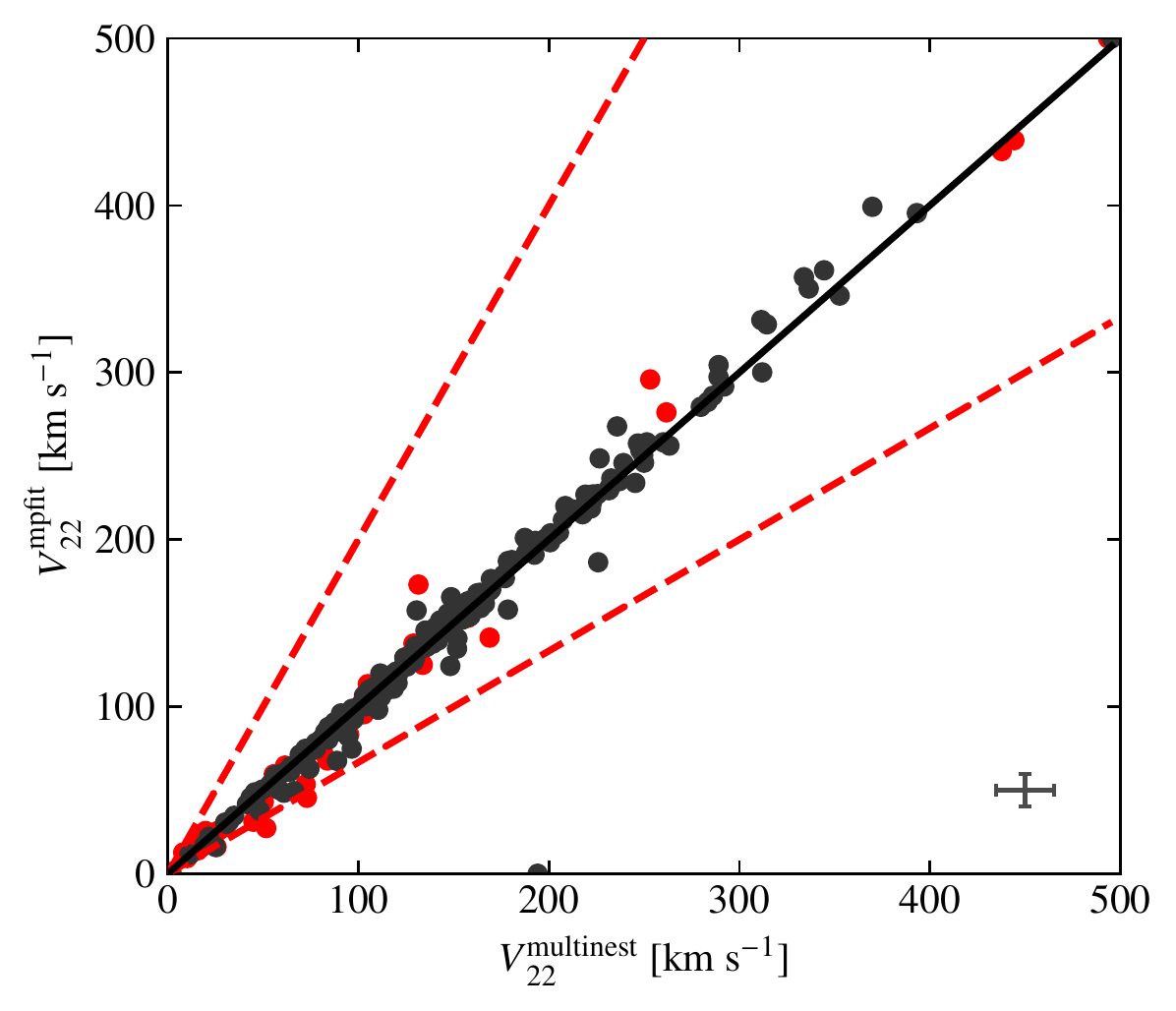}
			\caption{Comparison of the circular velocity $V_{22} = V(R_{22})$ using \Mocking{} between \Multinest{} and \MPFIT{} for galaxies from the MS sample. The rotation curve used was a flat model, and we removed galaxies whose circular velocity could not be reliably constrained ($R_{22}$ falls outside the range where there is sufficient S/N in the MUSE data cube to derive the kinematic). Red points correspond to galaxies visually classified as having no apparent velocity field in their kinematic maps, and red dashed lines correspond to a 50\% difference between the two methods. The typical uncertainty is shown on the bottom right part of the plot. Overall, values are consistent within their error bars.}
			\label{fig:rotation_comparison}
		\end{figure}
		
		\begin{figure}[hbt!]
			\includegraphics[scale=0.7]{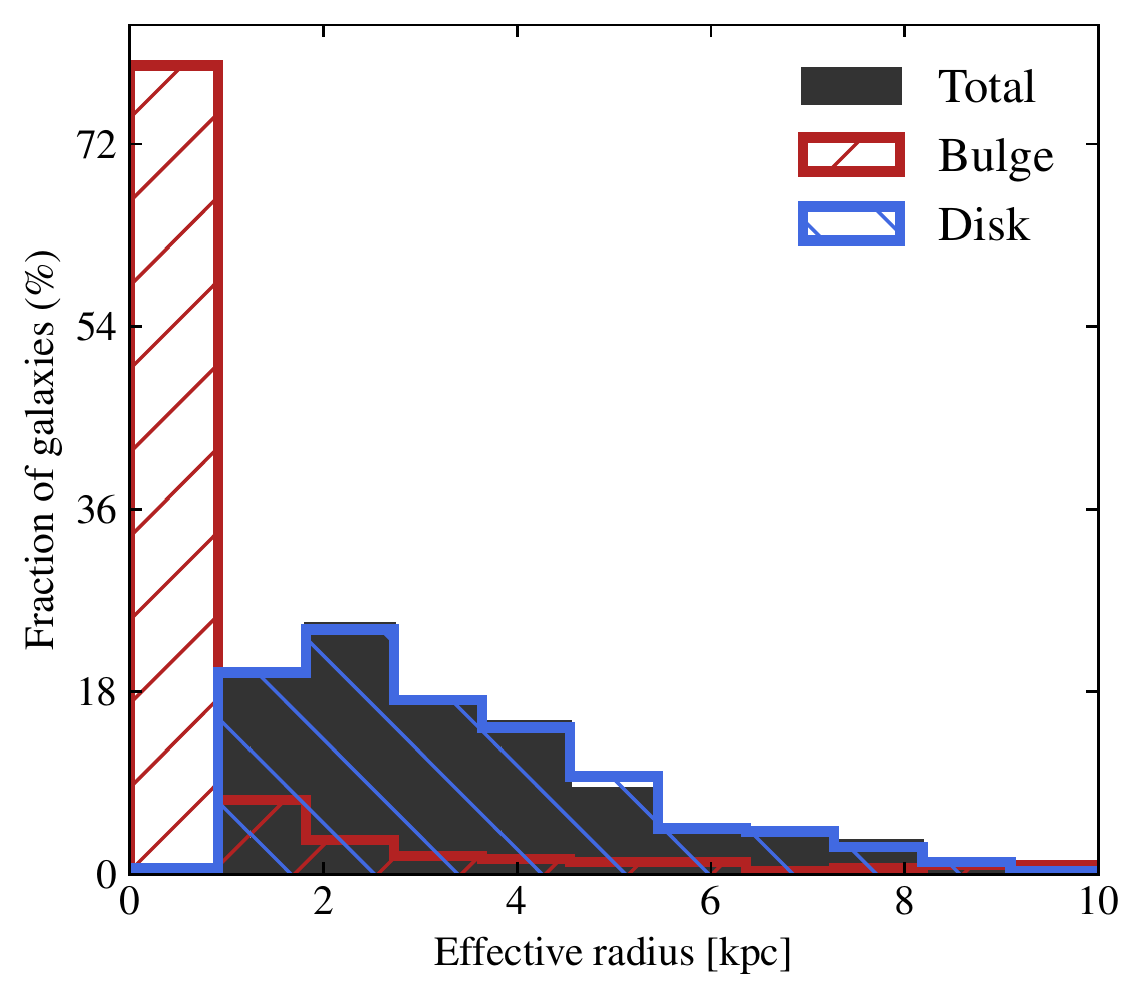}
			\caption{Distribution of effective radii for galaxies in the morphological sample. In grey (filled) is shown the total size, in red (hatched) the bulge size and in blue (hatched) the disk size. Disks are mostly found between roughly $\SI{1}{kpc}$ and $\SI{6}{kpc}$, with very few galaxies with disk sizes beyond $\SI{10}{kpc}$. The lack of disks below $\SI{1}{kpc}$ is due to the size selection criterion from Sect.\,\ref{sec:samples_selection}. On the other hand, the majority of bulges are found below $\SI{2}{kpc}$. The total size of galaxies is mainly driven by the disk component.}
			\label{fig:bulge_Reff_distribution}
		\end{figure}
		
		\begin{figure}
			\centering
			\includegraphics[scale=0.7]{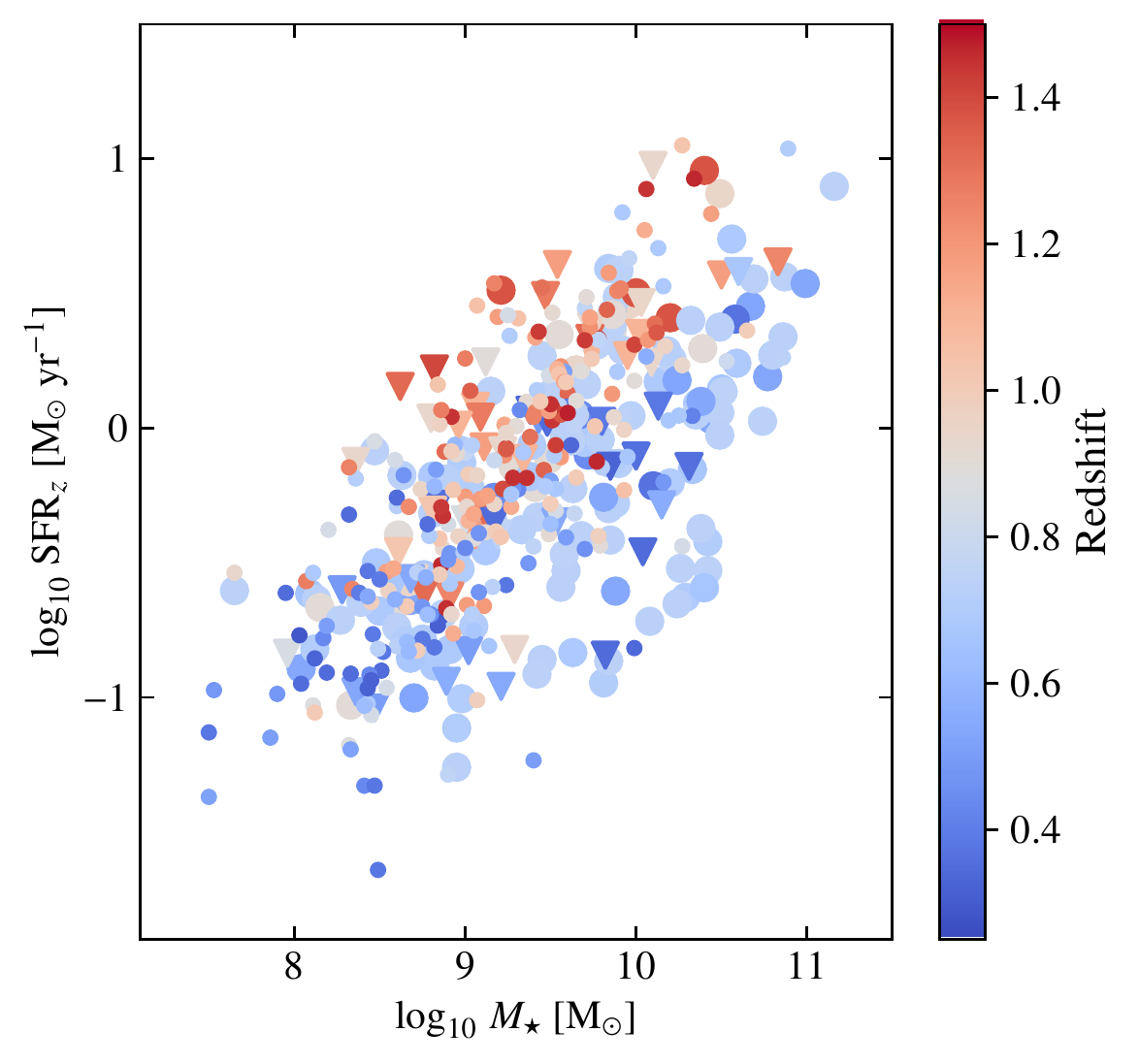}
			\caption{SFR-stellar mass relation for galaxies from the MS sample, colour coded as a function of redshift. Despite the applied normalisation to a redshift $z_0 = 0.7$, we still see a dichotomy. High redshift galaxies tend to align along lines with the largest specific Star Formation Rate (sSFR) while low redshift galaxies tend to align along lines with the lowest sSFR because of the survey design.}
			\label{fig:MS_redshift}
		\end{figure}
		
		\clearpage
		\section{Impact of selection}
		\label{Appendix:impact_of_selection}
		
		\begin{figure}[hbt!]
			\centering
			\includegraphics[scale=0.51]{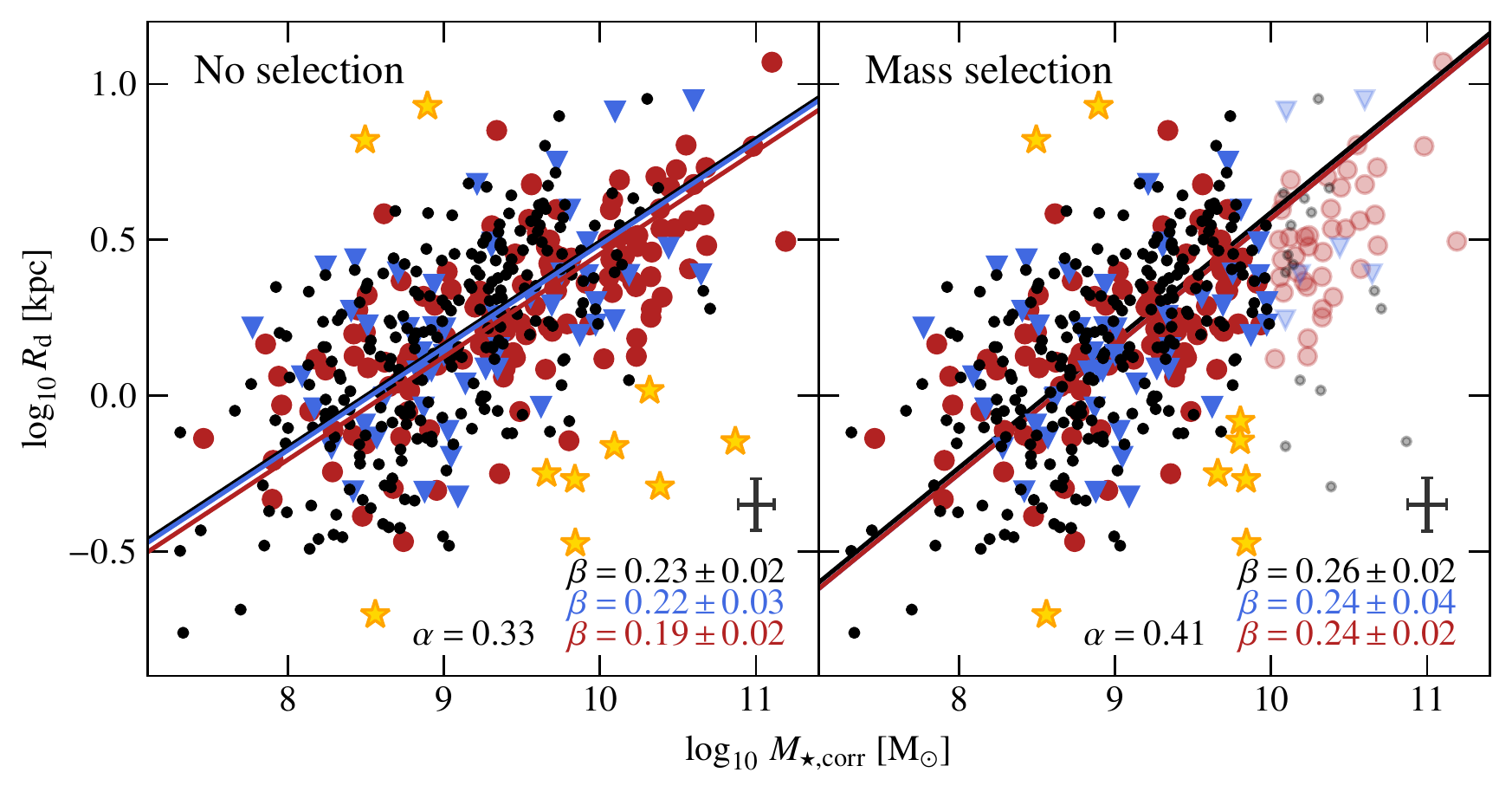}
			\caption{Size-mass relation with and without applying the mass selection criterion (vi) on galaxies from the MS sample. The data points and best-fit lines are similar to Fig.\,\ref{fig:fit_size-mass}. As an indication, we also show as semi-transparent symbols galaxies removed by the mass cut in the right panel. The typical uncertainty on stellar mass and disk size is shown on both panels as a grey errorbar.}
			\label{fig:impact_on_mass-size_relation}
		\end{figure}
		
		\begin{figure}[hbt!]
			\centering
			\includegraphics[scale=0.51]{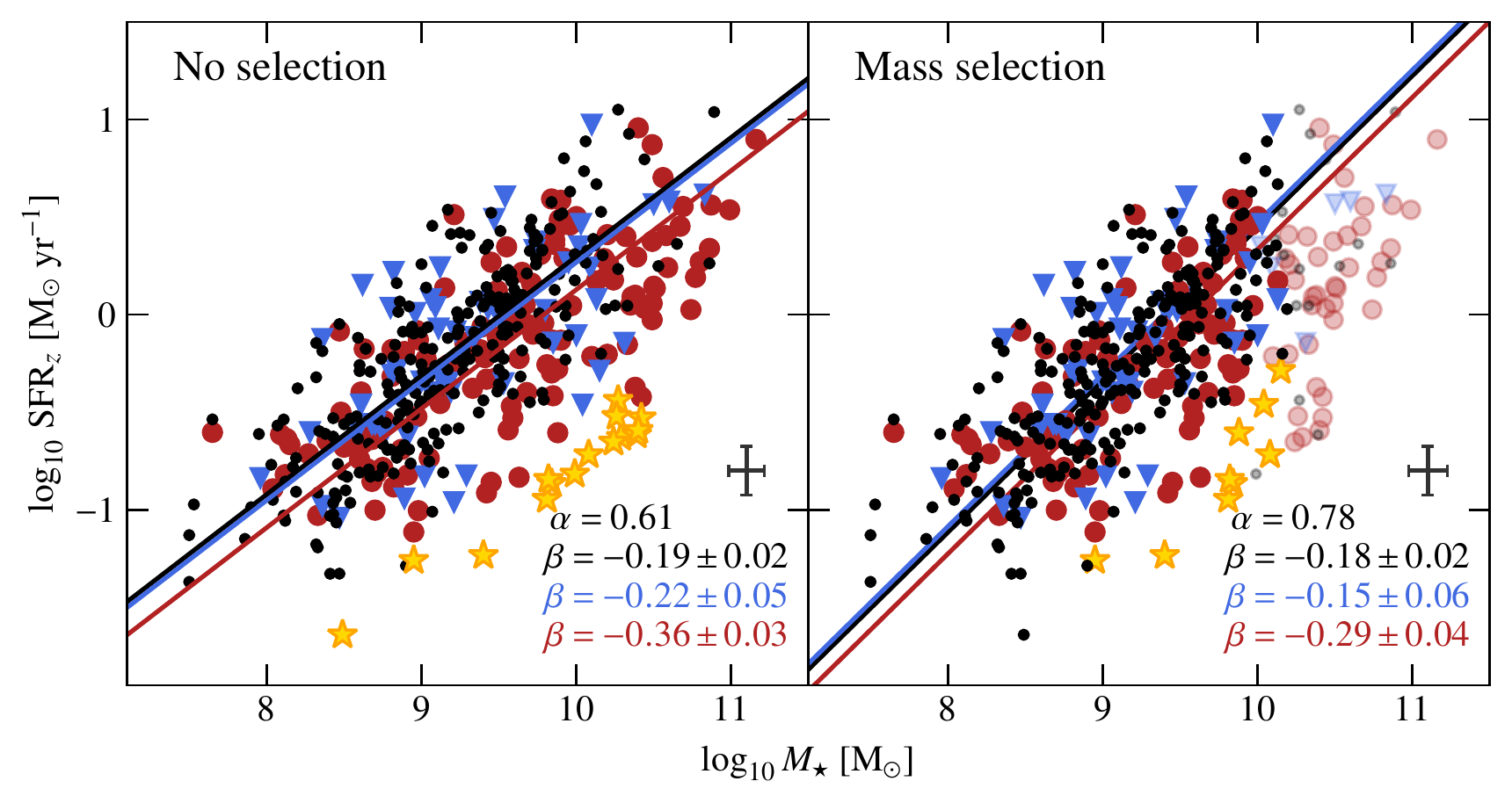}
			\caption{SFR-mass relation with and without applying the mass selection criterion (vi) on galaxies from the kinematic sample. The data points and best-fit lines are similar to Fig.\,\ref{fig:fit_MS}. As an indication, we also show as semi-transparent symbols galaxies removed by the mass cut in the right panel. The typical uncertainty on stellar mass and SFR is shown on both panels as a grey errorbar.}
			\label{fig:impact_on_MS}
		\end{figure}
		
		\begin{figure}
			\centering
			\includegraphics[scale=0.51]{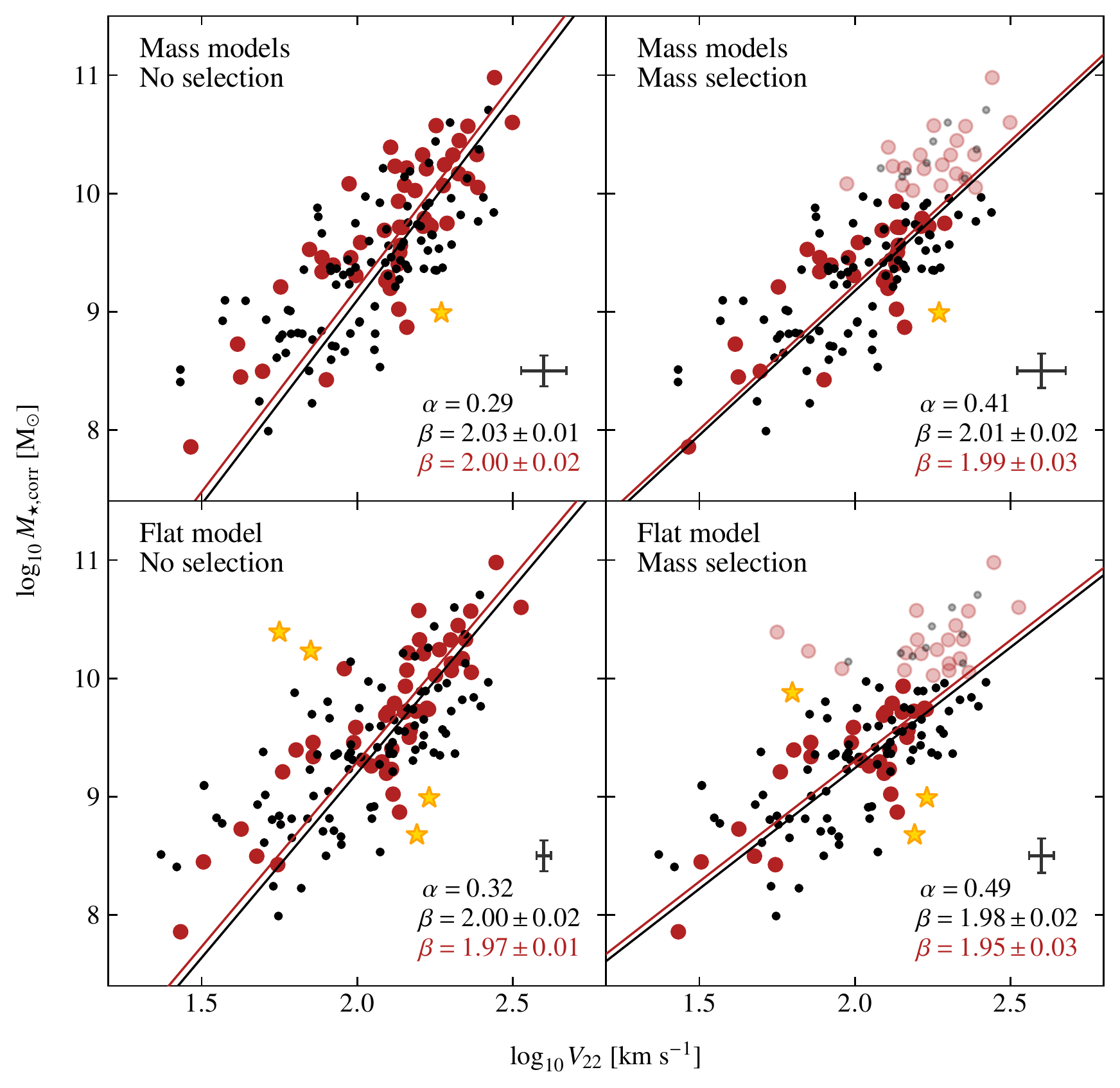}
			\caption{TFR with and without applying the mass selection criterion (vi) on galaxies from the TFR sample. The data points and best-fit lines are similar to Fig.\,\ref{fig:TFR}. The first row shows the TFR using the velocity derived from the best-fit mass models, and the second row the TFR using the flat model. As an indication, we also show as semi-transparent symbols galaxies removed by the mass cut in the rightmost panels. The typical uncertainty on stellar mass and velocity is shown on each panel as a grey errorbar.}
			\label{fig:impact_on_TFR}
		\end{figure}

		\clearpage
		\begin{figure*}[hbt!]
			\centering
			\includegraphics[scale=0.9]{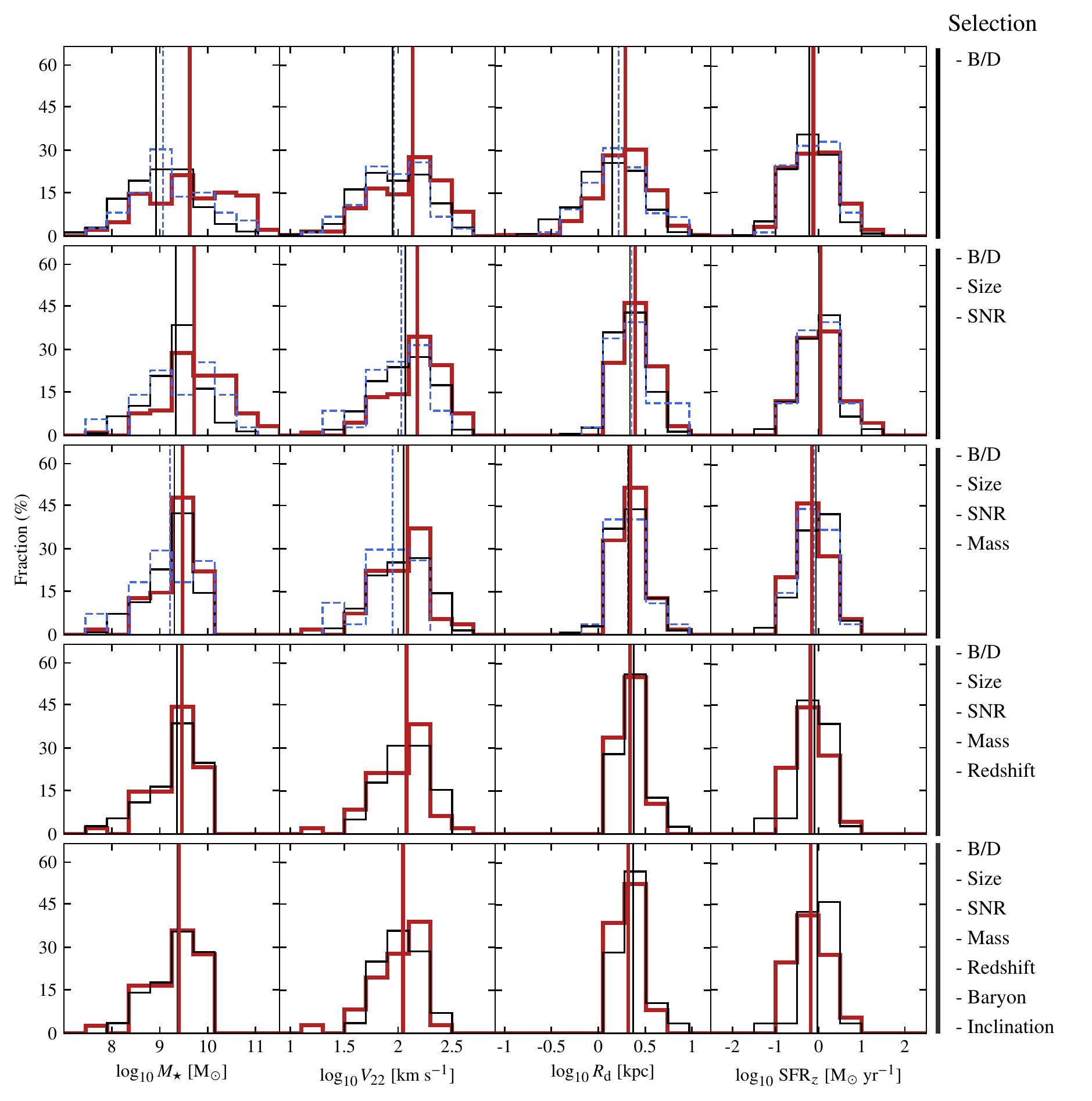}
			\caption{Impact of selection criteria on the main parameters distributions for galaxies from the kinematic sample. Each row represents a different selection. The black full line corresponds to the field galaxies subsample, the blue dashed line to the small structures and the red thick line to the large structures (using a threshold of $N = 10$ to separate between structures). We also show the median values for each subsample as vertical lines. We do not show the small structure subsample in the last two rows since there remain too few galaxies.}
			\label{fig:selection_impact_distributions}
		\end{figure*}

		\clearpage
		\section{Bulge-disk decomposition}%
		\label{Appendix:bulge_disk_decomposition}
		
		\begin{figure}[hbt!]
			\includegraphics[scale=0.7]{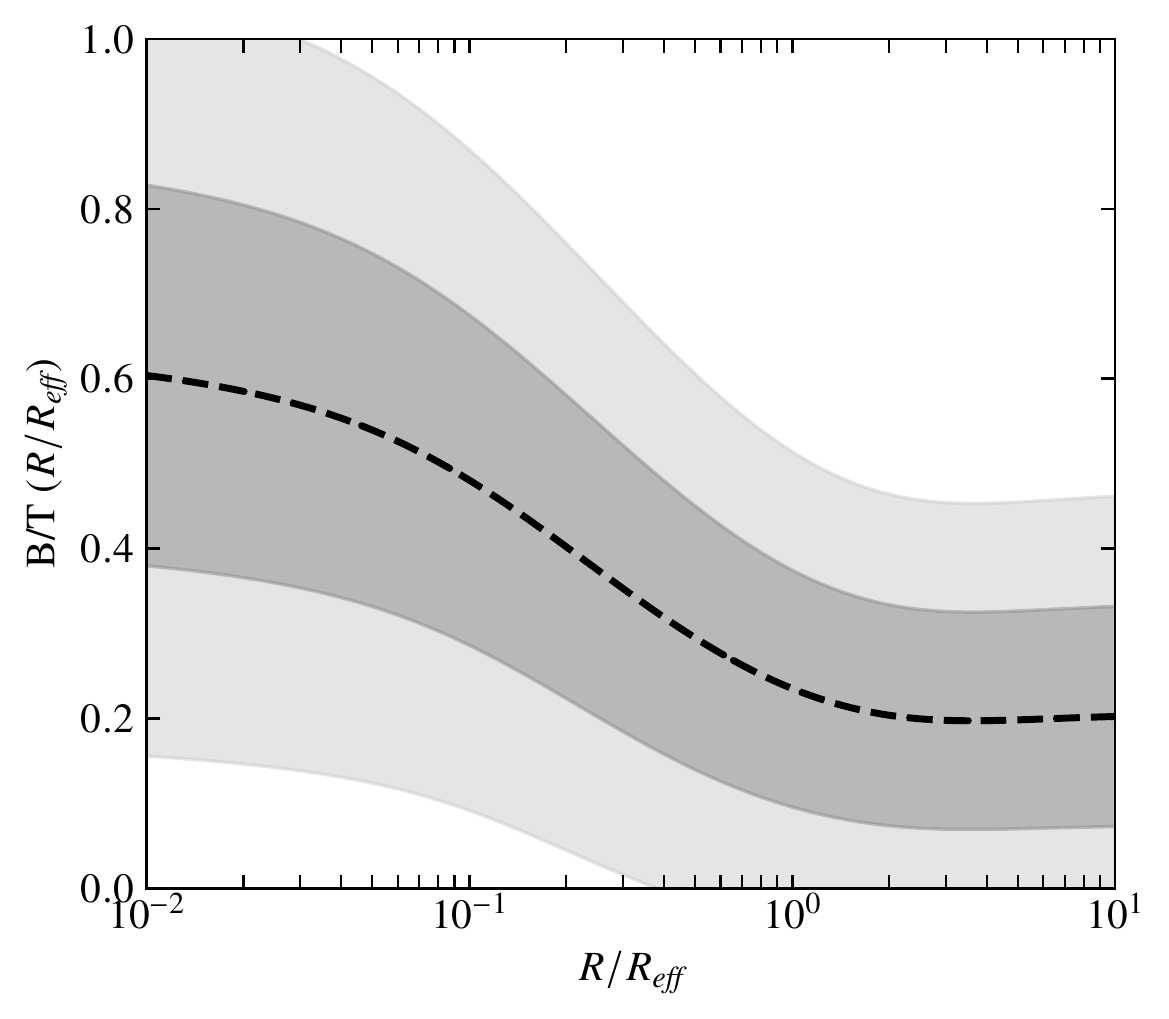}
			\caption{Mean B/T for galaxies in the morphological sample computed at various radii in units of $R_{\textit{eff}}$. The areas correspond to the $1\sigma$ (dark gray) and $2\sigma$ (light gray) dispersions. The bulge component dominates the central parts of the galaxies whereas the disk takes over completely after roughly one effective radius. Even as far as $10 R_{\textit{eff}}$, we find a nearly constant non-zero $\rm{B/T} \approx 0.2$ indicative of a non-negligible bulge contribution to the overall flux budget.}
			\label{fig:med_B/T}
		\end{figure}
		
		Figure\,\ref{fig:med_B/T} represents the median value of the bulge-to-total flux ratio (B/T) for the morphological sample as a function of radius. We see that beyond one effective radius the disk dominates the flux budget. When computed near the centre, B/T is close to one, consistent with the bulge dominating the inner parts. Even though the disk dominates at large distances, B/T does not reach zero. This is a consequence of the chosen bulge-disk decomposition. Indeed, for a Sérsic profile with parameters $(n, \Sigma_{\textit{eff}}, R_{\textit{eff}})$, the integrated flux up to radius $r$ is given by
	
	\begin{equation}
		F (<r) = 2\pi n~\Sigma_{\textit{eff}}~R_{\textit{eff}}^2~e^{b_n}~\gamma \left (2n, b_n \left (r/R_{\textit{eff}} \right )^{1/n} \right ) / b_n^{2n},
		\label{eq:flux_Sersic}
	\end{equation}
	where $\gamma$ is the lower incomplete gamma function and where $b_n$ is the solution of the equation $\Gamma \left (2n \right )~=~2\gamma\left (2n, b_n \right )$ \citep{Graham2005}, with $\Gamma$ the complete gamma function. Therefore, for a bulge-disk decomposition the total flux ratio between the two components is given by
	
	\begin{equation}
		{\rm{B/T}} (r \rightarrow \infty) \approx \Sigma_{\textit{eff}, \rm{b}}~R_{\textit{eff}, \rm{b}}^2~/ \left ( \Sigma_{\textit{eff}, \rm{b}}~R_{\textit{eff}, \rm{b}}^2 + 0.527~\Sigma_{\textit{eff}, \rm{d}}~R_{\textit{eff}, \rm{d}}^2 \right ),
		\label{eq:flux_ratio}
	\end{equation}
	where $(\Sigma_{\textit{eff}, \rm{b}}, R_{\textit{eff}, \rm{b}})$ and $(\Sigma_{\textit{eff}, \rm{d}}, R_{\textit{eff}, \rm{d}})$ are the bulge and disk parameters, respectively. The only case for which Eq.\,\ref{eq:flux_ratio} vanishes is when the bulge contribution can be neglected with respect to the disk. Otherwise, when B/T$ (\infty)$ is sufficiently larger than 0, this reflects a non-negligible contribution of the bulge to the overall flux budget. The fact that the median value for the morphological sample is around 0.2 is therefore a good indication of the relevance of performing a bulge-disk decomposition with respect to using a single disk model.
	
	The half-light radius of a multi-component decomposition involving only Sérsic models does not necessarily have to be computed through numerical integration but can also be derived by finding the single zero of a given function. Indeed, for a bulge-disk decomposition, from the definition of the global half-light radius (that is the radius which encloses half of the total flux), we have
	
	\begin{equation}
		F_{\rm{d}} (R_{\textit{eff}}) + F_{\rm{b}} (R_{\textit{eff}}) = \left ( F_{\rm{tot, d}} + F_{\rm{tot, b}} \right ) / 2,
		\label{eq:global_effective_radius_general_def}
	\end{equation}
	where $F_{\rm{d}} (R_{\textit{eff}})$ and $F_{\rm{b}} (R_{\textit{eff}})$ are the disk and bulge fluxes at the global effective radius $R_{\textit{eff}}$, and $F_{\rm{tot, d}}$, $F_{\rm{tot, b}}$ are the disk and bulge total fluxes, respectively. Given Eq.\,\ref{eq:flux_Sersic}, one can rewrite Eq.\,\ref{eq:global_effective_radius_general_def} as
	
	\begin{equation}
		\label{eq:global_effective_radius_simplified}
		\begin{split}
			& F_{\rm{tot, d}} \left [ \gamma \left ( 2, b_1 \frac{R_{\textit{eff}}}{R_{\textit{eff}, \rm{d}}} \right ) - 0.5 \right ] + \\
			& \frac{F_{\rm{tot, b}}}{\Gamma(8)} \left [ \gamma \left (8, b_4 \left ( \frac{R_{\textit{eff}}}{R_{\textit{eff}, \rm{b}}} \right )^{1/4} \right ) - \Gamma(8) / 2 \right ] = 0.
		\end{split}
	\end{equation}
	
	Furthermore, if one defines the total magnitude of a component $i$ as ${\rm{mag}}_i = -2.5 \log_{10} F_{{\rm{tot}}, i} + \rm{zpt}$, where zpt is a zero point which is the same for all the components, and normalises by the total flux, then Eq.\,\ref{eq:global_effective_radius_simplified} simplifies to
	
	\begin{equation}
		\label{eq:global_effective_radius_even_more_simplified}
		f(R_{\textit{eff}}) / f(\infty) = 0,
	\end{equation}
	with the function $f$ defined as
	\begin{equation}
		\begin{split}
			f(x) = & 10^{-{\rm{mag_d}} /2.5} \left [ \gamma \left (2, b_1 \frac{x}{R_{\textit{eff} \rm{d}}}  \right ) - 0.5 \right ]  + \\
			& 10^{-{\rm{mag_b}} /2.5} \left [ \gamma \left (8, b_4 \left (\frac{x}{R_{\textit{eff}, \rm{b}}} \right )^{1/4} \right ) / \Gamma(8) - 0.5 \right ].
		\end{split}
		\label{eq:global_effective_radius_function}
	\end{equation}
	
	Equation \ref{eq:global_effective_radius_even_more_simplified} can be solved by searching for a zero in the range $\left [\min \left ( {R_{\textit{eff}, \rm{d}}, R_{\textit{eff}, \rm{b}}} \right), \max \left ( {R_{\textit{eff}, \rm{d}}, R_{\textit{eff}, \rm{b}}} \right) \right ]$. Indeed, if $R_{\textit{eff}}~>~\max \left ( {R_{\textit{eff}, \rm{d}}, R_{\textit{eff}, \rm{b}}} \right)$, the flux at $R_{\textit{eff}}$ would be the sum of $F_{\rm{d}} (R_{\textit{eff}})~>~F_{\rm{d, tot}} / 2$ and $F_{\rm{b}} (R_{\textit{eff}})~>~F_{\rm{b, tot}} / 2$ such that it would be larger than the expected $F_{\rm{tot}} / 2$ value. Thus $R_{\textit{eff}}$ cannot be greater than $\max \left ( {R_{\textit{eff}, \rm{d}}, R_{\textit{eff}, \rm{b}}} \right)$, and the same argument can be given for the case $R_{\textit{eff}}~<~\min \left ( {R_{\textit{eff}, \rm{d}}, R_{\textit{eff}, \rm{b}}} \right)$.
	
	Finally, there is only one zero which is solution of Eq.\,\ref{eq:global_effective_radius_even_more_simplified}, and this can be shown by noticing that $f$ is a monotonously increasing function of $x$ whose normalised form $f(x) / f(\infty)$ is bounded between -1 for $x = 0$ and 1 for $x = \infty$.
		
	\section{Mass modelling}%
	\label{Appendix:mass_modelling}
	
	The methodology used in \citet{Abril-Melgarejo2021} to derive the galaxies dynamics is only an approximation of the intrinsic ionised gas kinematics. The flat model used for the rotation curve is ad-hoc, based on observations of local and intermediate redshift rotation curves of dark matter dominated galaxies. While being a good approximation for dark matter dominated systems, this kinematic modelling does not take into account information from the morphological modelling. In theory, one should derive the ionised gas kinematics, assuming in our case that the gas is distributed within an infinitely thin disk, from the 3D mass distribution of the different galaxy components. Even though we do not have access to such distributions, we can nevertheless constrain the gas kinematics under a few assumptions. In Sect.\,\ref{sec:morpho modelling}, we have assumed that the sky projected surface density of the stars can be described by a bulge-disk decomposition, where the surface density of stellar disk is represented by an exponential profile and the stellar bulge is assumed to be spherically symmetric with a surface density described by a de Vaucouleurs profile. If one can find 3D flux densities which, when projected onto the line of sight, become the corresponding surface densities, then one has found the corresponding mass densities up to a multiplicative factor which is the mass to light ratio $\Upsilon = ({\rm{M/L}})_\star$.
	
	\subsection{Theoretical background}
	
	For any mass density $\rho_{\rm{M}} (\vec r)$, we can derive the corresponding potential $\Phi$ from Poisson equation
	
	\begin{equation}
		\nabla^2 \Phi (\vec r) = 4\pi G \rho_{\rm{M}} (\vec r).
		\label{eq:Poisson}
	\end{equation}
	
	The observed velocity maps are derived from the ionised gas kinematics, which is assumed to be located within an infinitely thin disk, therefore we are only interested in the velocity of the gas within the plane of the galaxy disk. If we further assume that the mass distribution $\rho_{\rm{M}}$ is in equilibrium within its gravitational potential, then the centrifugal acceleration caused by its rotation must balance the radial gradient of the potential $\Phi$ in the galaxy plane, that is
	
	\begin{equation}
		\frac{V_{\rm{circ}}^2}{R}(R) = - \frac{\partial \Phi}{\partial R}(R, z=0),
		\label{eq:circular_speed}
	\end{equation}
	with $V_{\rm{circ}}$ the circular velocity, $R$ the radial distance in the plane of the galaxy, and where we have assumed that the potential and circular velocity are independent of the azimuth because of the symmetry of the mass distributions used in the following. Since the mass distributions and therefore the potentials add up, the circular velocity can be simply written as
	
	\begin{equation}
		V_{\rm{circ}}^2 (R) = \sum_i V^2_{\rm{circ}, i} (R),
		\label{eq:sum_circular_speed}
	\end{equation}
	where $V_{\rm{circ}, i}$ is the circular velocity of the component $i$ obeying Eq.\,\ref{eq:circular_speed} for the corresponding potential well. In our case, the components which will contribute the most to the rotation curve are the stellar disk, stellar bulge and the dark matter halo to account for constant or slowly declining observed rotation curves at large radii. We do not model the contribution of the gas, which will therefore slightly contribute to the dark matter halo profile. In the case of the stellar components, we transform from stellar light distributions $\rho_i$ to mass distributions $\rho_{\rm{M}, i}$ using
	
	\begin{equation}
		\rho_{\rm{M}, i} (\vec r) = \Upsilon \rho_i (\vec r),
		\label{eq:Upsilon}
	\end{equation}
	where we have further assumed that the mass to light ratio $\Upsilon$ is constant throughout the galaxy, and we compute it using the SED-based estimator of the stellar mass as 
	
	\begin{equation}
		\Upsilon = \rm{M}_{\star}/\rm{F_{SP}} (\SI{1.5}{\arcsec}),
		\label{eq:mass-to-light}
	\end{equation}
	where $M_{\star}$ is the SED-based mass computed in a circular aperture of diameter $\SI{3}{\arcsec}$, and $F_{\rm{SP}} (\SI{1.5}{\arcsec})$ is the flux integrated on the plane of the sky in the same aperture. In this analysis, we assume a similar $\Upsilon$ for both disk and bulge because it would require at least two HST bands to constrain efficiently the M/L for both components individually as done for instance in \citet{Dimauro+18}.
	
	\subsection{Razor thin stellar disk}
	\label{Appendix:razor_thin_disk}
	
	To begin with, we assume the stellar disk to be infinitely thin, so that the stellar light density can be written as 	
	
	\begin{align}			
		\label{eq:razor_thin_disk}	
		\rho (\vec r) & = \Sigma_{\rm{RT}} (R)~\delta (z) \\
		\Sigma_{\rm{RT}} (R) & = \Sigma_{\rm{RT}} (0)~e^{- b_1 [ R/R_{\textit{eff}, \rm{d}} ]}
		\label{eq:razor_thin_surface_density}
	\end{align}	
	where $\Sigma_{\rm{RT}}$ represents the light distribution in the plane of the disk, with $\Sigma_{\rm{RT}} (0)$ the central surface density, $b_1 \approx 1.6783$, $R_{\textit{eff}, \rm{d}}$ the disk effective radius, and $\delta$ is the Dirac distribution. The rotation curve for such a distribution was computed for the first time by \citet{freeman_disks_1970} using the method described in \citet{toomre_distribution_1963}:
	
	\begin{equation}
		V_{\rm{RT}}(R) = V_{\rm{RT, max}} \times \frac{y f(y)}{1.075~f(1.075)},
		\label{eq:rotation_curve_razor_thin}
	\end{equation}
	with $f(y) = \sqrt{I_0(y) K_0(y) - I_1(y) K_1(y)}$ and $y = R / (2 R_{\rm{d}})$. The effective radius of the disk is related to the disk scale length appearing in Eq.\,\ref{eq:rotation_curve_razor_thin} through $R_{\textit{eff}, \rm{d}} = b_1 R_{\rm{d}}$. The maximum circular velocity is reached at a radius $R = 2.15 R_{\rm{d}}$ and is equal to 
	
	\begin{equation}
		V_{\rm{RT, max}}~=~ 2.15 f(1.075) \sqrt{\pi G {R_{\rm{d}}} \Upsilon~\Sigma_{\rm{RT}}(0)},
		\label{eq:max_curve_razor_thin}
	\end{equation}
	where $G$ is the gravitational constant.
	
	\subsection{Thin stellar disk}
	\label{Appendix:thin_disk}
	
	\begin{figure}
		\includegraphics[scale=0.7]{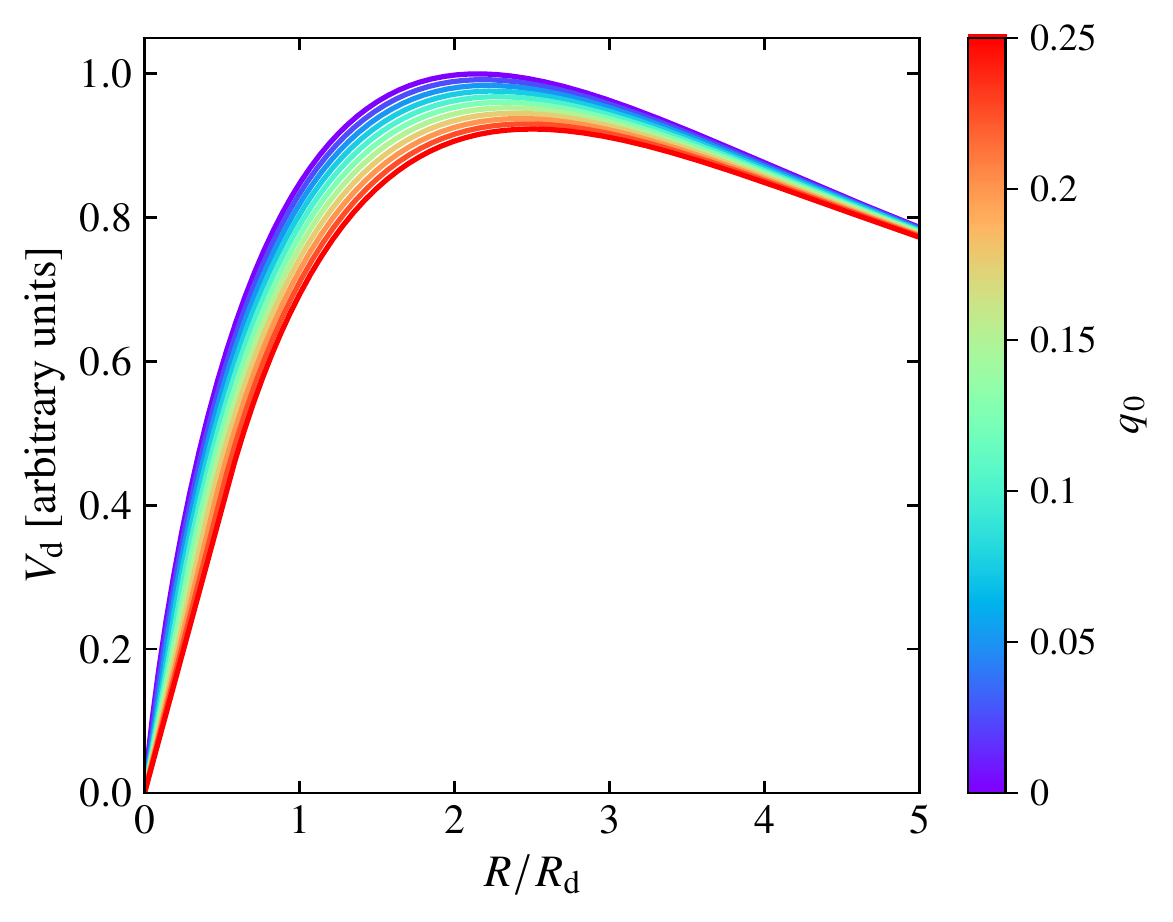}
		\caption{Impact of the thickness on the shape of the rotation curve for a thin disk. The finite thickness only impacts the inner parts, and changes both the amplitude and the radius where the maximum is reached.}
		\label{fig:example_thin_disk_rot_curve}
	\end{figure}
	
	To refine the mass modelling of the stellar disk, we consider a disk model with a finite thickness. Assuming the light distribution can be correctly represented by a double exponential profile, we have
	
	\begin{equation}
		\rho (\vec r) =  \Sigma_{\rm{RT}} (R)~e^{-|z|/h_z} / (2 h_z),
		\label{eq:thin_disk}
	\end{equation}	
	where $h_z$ is the disk scale height. It can be shown \citep{Peng2002} that the potential in the plane of the galaxy for such a density can be written as
	
	\begin{equation}
		\Phi (R) = - (2\pi~G /h_z) \int_{\mathbb{R}_{+}}  dk~(1/h_z + k)^{-1} J_0 (kR)~S_0(k),
		\label{eq:potential_thin_disk}
	\end{equation}
	where $S_0 (k)$ is the Hankel transform of order 0 of the surface density $\Sigma_{\rm{d}} (R)$. For thin disks with small $h_z$, an approximation of the circular velocity in the plane of the galaxy is given by\footnote{For a derivation of this approximation see Eq.\,8.73 in Chapter 8 of Bovy J. Dynamics and Astrophysics of Galaxies. Princeton University Press, Princeton, NJ (in preparation) whose online version can be found at \url{https://galaxiesbook.org/}.}
	
	\begin{equation}
		V_{\rm{d}}^2 (R) = V_{\rm{RT}}^2 (R) - V_{\rm{corr, max}}^2 \times \frac{R~e^{1-R/{R_{\rm{d}}}}}{{R_{\rm{d}}}},
		\label{eq:approximation_curve_thin_disk}
	\end{equation}
	where $V_{\rm{RT}}$ is the razor-thin circular velocity defined in Eq.\,\ref{eq:rotation_curve_razor_thin} and $R_{\rm{d}}$ is the disk scale length. For typical values of $h_z / R_{\rm{d}}~\approx~0.2-0.3$, this approximation gives a circular velocity which is different from numerical integration by less than 2\% for most of the radial range, except near the central parts where the relative difference rises, though the absolute difference remains negligible in practice as the circular velocity quickly drops to zero near the centre. The maximum of the correction is reached at $R_{\rm{d}}$ (see Fig.\,\ref{fig:example_thin_disk_rot_curve}) and is given by
	
	\begin{equation}
		V_{\rm{corr, max}} = \sqrt{2\pi G h_z \Upsilon~\Sigma_{\rm{RT}}(0) / e}.
	\end{equation}
	
	\subsection{Impact of thickness on inclination and central surface density}		
	
	In the case of a razor-thin disk projected at an inclination $i$ with respect to the line of sight, the apparent central surface density $\Sigma_{\rm{RT, obs}}(0)$ and axis ratio $q = b/a$, with $a$ and $b$ the semi-major and semi-minor axes, respectively, scale with the inclination as
	
	\begin{align}
		\label{eq:central_density_razor}
		\Sigma_{\rm{RT, obs}}(0) & = \Sigma_{\rm{RT}}(0) / \cos i, \\
		q & = \cos i.
		\label{eq:q_razor}
	\end{align}
	
	Writing Eq.\ref{eq:central_density_razor} is equivalent to saying that the total flux of the disk must be independent of its inclination on the sky, and Eq.\ref{eq:q_razor} comes from the fact that the isophotes of a projected razor-thin disk are ellipses. However, in the case of a disk with non-zero thickness the surface density profile gets more complicated, and must be computed as the integral of the inclined density distribution along the line of sight. We give in Appendix \ref{Appendix:double_exponential_integral} a derivation of this integral in the general case. For the apparent central density, it simplifies to
	
	\begin{equation}
		\label{eq:central_density_disk}
		\begin{split}
			\Sigma_{\rm{d, obs}} (0) & = \frac{\Sigma_{\rm{RT}}(0)~R_{\rm{d}}}{2 h_z \sin i_0} \int_{\mathbb{R}} dv~e^{-|v|~(1 + \beta)} \\
			                    & = \frac{\Sigma_{\rm{RT}} (0)}{q_0 \sin i_0 + \cos i_0},
		\end{split}
	\end{equation}
	with $q_0 = h_z / R_{\rm{d}}$ the real axis ratio, $R_{\rm{d}}$ the disk scale length, $\Sigma_{\rm{RT}}(0)$ the central surface density if the galaxy was seen face-on, and $i_0$ the real inclination of the galaxy. We see that when the disk is infinitely thin (i.e. $h_z = 0$) we recover Eq.\,\ref{eq:central_density_razor}, as should be expected. For a perfectly edge-on galaxy, that is $i = \SI{90}{\degree}$, Eq.\,\ref{eq:central_density_razor} diverges, which is due to the fact that a razor-thin disk seen edge-on does not have its flux distributed onto a surface any more, but onto a line. For a disk with non-zero thickness, this is not the case, and therefore Eq.\,\ref{eq:central_density_disk} remains finite for an edge-on galaxy.
	
	For a disk with finite thickness, there is no trivial way to relate the observed axis ratio $q$ to the real one $q_0$. In practice, the isophotes of a projected disk can be approximated by ellipses but with an ellipticity which depends on position, disk scale length, scale height and inclination. Still, we expect the observed axis ratio to be 1 for a face-on galaxy, and equal to $q_0$ for a perfectly edge-on galaxy. For an oblate system, we can relate the observed axis ratio to the intrinsic one and the galaxy inclination $i_0$ as \citep{bottinelli_h_1983}:
	
	\begin{equation}
		\label{Eq:correction_inclination}
		\cos^2 i_0 = (q^2 - q_0^2) / (1 - q_0^2).
	\end{equation}
	
	Technically, the isodensity surfaces of a double exponential profile are not oblate but have a biconical shape, which means that Eq.\,\ref{Eq:correction_inclination} is only an approximation of the real dependence of the observed axis ratio on $q_0$ and inclination. In Sect.\,\ref{sec:morpho modelling}, we have fitted 2D profiles of galaxies using a bulge-disk decomposition, assuming that the disk is exponential with zero thickness. Its apparent central surface density is therefore given by Eq.\,\ref{eq:central_density_razor} with $i$ the apparent inclination related to the observed axis ratio through Eq.\,\ref{eq:q_razor}. If the stellar disk 3D distribution is actually described by a double exponential profile, then its apparent central surface density given by Eq.\,\ref{eq:central_density_disk} must match that of the fitted single exponential profile. Using Eq.\,\ref{Eq:correction_inclination} to express the apparent inclination in terms of the real inclination $i_0$ and intrinsic axis ratio $q_0$, we can derive the ratio $r_0$ of the central surface density computed using a double exponential profile against that computed from a single exponential fit as
	
	\begin{equation}
		\label{eq:ratio_central_density}
		r_0 = \frac{q_0 \sin i_0 + \cos i_0}{\sqrt{q_0^2 \sin^2 i_0 + \cos^2 i_0}}.
	\end{equation}
	
\begin{figure}
		\includegraphics[scale=0.7]{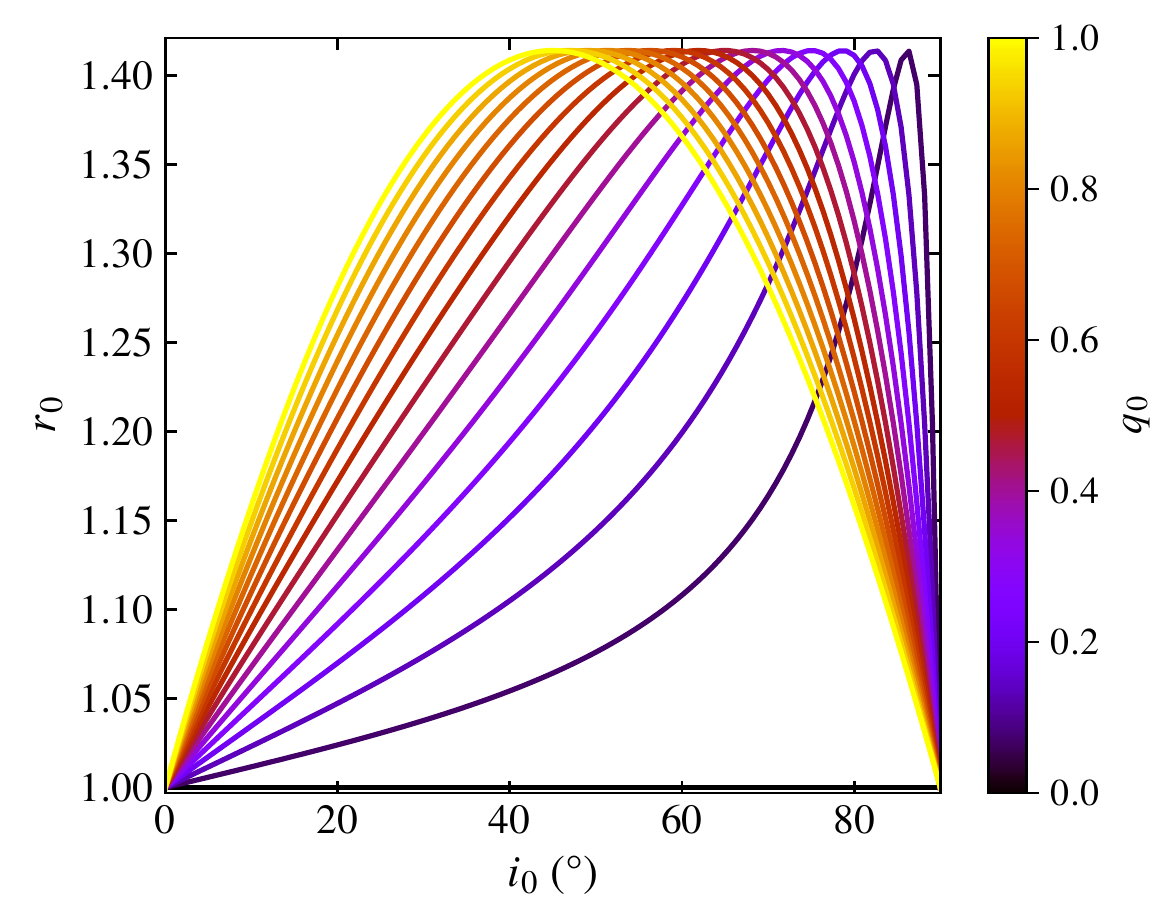}
		\caption{Ratio of the central density assuming a double exponential profile with that derived assuming a razor thin disk exponential fit as a function of the galaxy real inclination $i_0$ and intrinsic axis ratio $q_0 = h_z / R_{\rm{d}}$, with $R_{\rm{d}}$ the disk scale length. The maximum value is equal to $\sqrt{2}$ and is reached at $i_0 = \arctan(1/q_0)$.}
		\label{fig:ratio_central_density}
	\end{figure}

	The ratio of the central surface densities is plotted in Fig.\,\ref{fig:ratio_central_density} as a function of the intrinsic axis ratio and real inclination. The central surface density derived in the case of a disk with non-zero thickness is always larger than its infinitely thin disk counterpart, the ratio reaching a maximum 
	
	\begin{equation}
		\max_{i_0} r_0 = \sqrt{2},
	\end{equation}
	at $i_0 = \arctan (1 / q_0)$. As is expected, when the disk becomes more and more flattened the ratio reaches unity. Similarly, when the galaxy is viewed face-on, the central surface densities for both models are equal.
	
	\subsection{Correction in the inner parts}

	The Bovy approximation to the rotation curve of a double exponential profile given by Eq.\,\ref{eq:approximation_curve_thin_disk} has the disadvantage to reach a null velocity as soon as the correction term on the right hand side becomes larger than the velocity of the razor-thin disk appearing in the equation, that is at $R > 0$. However, the real rotation curve would reach a null velocity at $R=0$ if one integrates it numerically. The impact of using Eq.\,\ref{eq:approximation_curve_thin_disk} would be small since we lack the resolution in our MUSE data to model precisely the velocity in the inner parts and because beam-smearing strongly affects the velocity field near the centre. Nevertheless, it can be useful to slightly modify it in order to have a rotation curve that behaves more physically in the inner parts.
	
	To do so, we decided to replace the rotation curve for the double exponential profile near the centre with the tangential line Bovy approximation which passes through $R=0$. This means that the rotation curve will behave linearly in the inner parts until it reaches the tangential point where Bovy approximation will take over. Let us call $R_0$ the radius at which the corresponding tangential line passes through the point $R=0$, then the tangent must obey the following equation
	
	\begin{equation}
		\frac{dV_{\rm{d}}}{dR} (R_0) \times R = V_{\rm{d}} (R_0) \times R / R_0.
	\end{equation}
	
	Defining $y = R / (2 R_{\rm{d}})$ and $y_0 = R_0 / (2 R_{\rm{d}})$, this simplifies to
	
	\begin{equation}
		y_0 \times \frac{dV_{\rm{d}}^2}{dy}(y_0) = 2 V_{\rm{d}}^2 (y_0),
	\end{equation}
	with the derivative of $V_{\rm{d}}^2$ given by
	
	\begin{equation}
		\begin{split}
			\frac{dV^2_{\rm{d}}}{dy} (y_0) = & V^2_{\rm{d}} (y_0) / y_0 + \\
			& \alpha y_0 \left [ f^2(y_0) + y_0 \frac{df^2}{dy} (y_0) + 2 q_0 e^{-2y_0} \right ],
		\end{split}
	\end{equation}
	where $f$ is defined in Appendix\,\ref{Appendix:razor_thin_disk} and $\alpha = 4\pi G R_{\rm{d}} \Upsilon \Sigma_{\rm{RT}} (0)$. Furthermore, the derivative of $f^2$ is given by
	
	\begin{equation}
		\frac{df^2}{dy} (y_0) = 2 I_1 (y_0) K_0(y_0) + 2 I_1(y_0) K_1 (y_0) / y_0 - 2 I_0(y) K_1(y_0).
	\end{equation}
	
	Thus, combining everything together, the equation one needs to solve to find $y_0 = R / R_{\rm{d}}$ as a function of the disk thickness $q_0$ is
	
	\begin{equation}
		\begin{split}
			y_0^2 \left [ I_1 (y_0) K_0 (y_0) - I_0 (y_0) K_1 (y_0) \right ] + & y_0 I_1 (y_0) K_1 (y_0) + \\
			& q_0 \left (y_0 + 0.5 \right ) e^{-2 y_0} = 0.
		\end{split}
		\label{eq:solve_y0}
	\end{equation}
	
\begin{figure}[hbt!]
		\centering
		\includegraphics[scale=0.7]{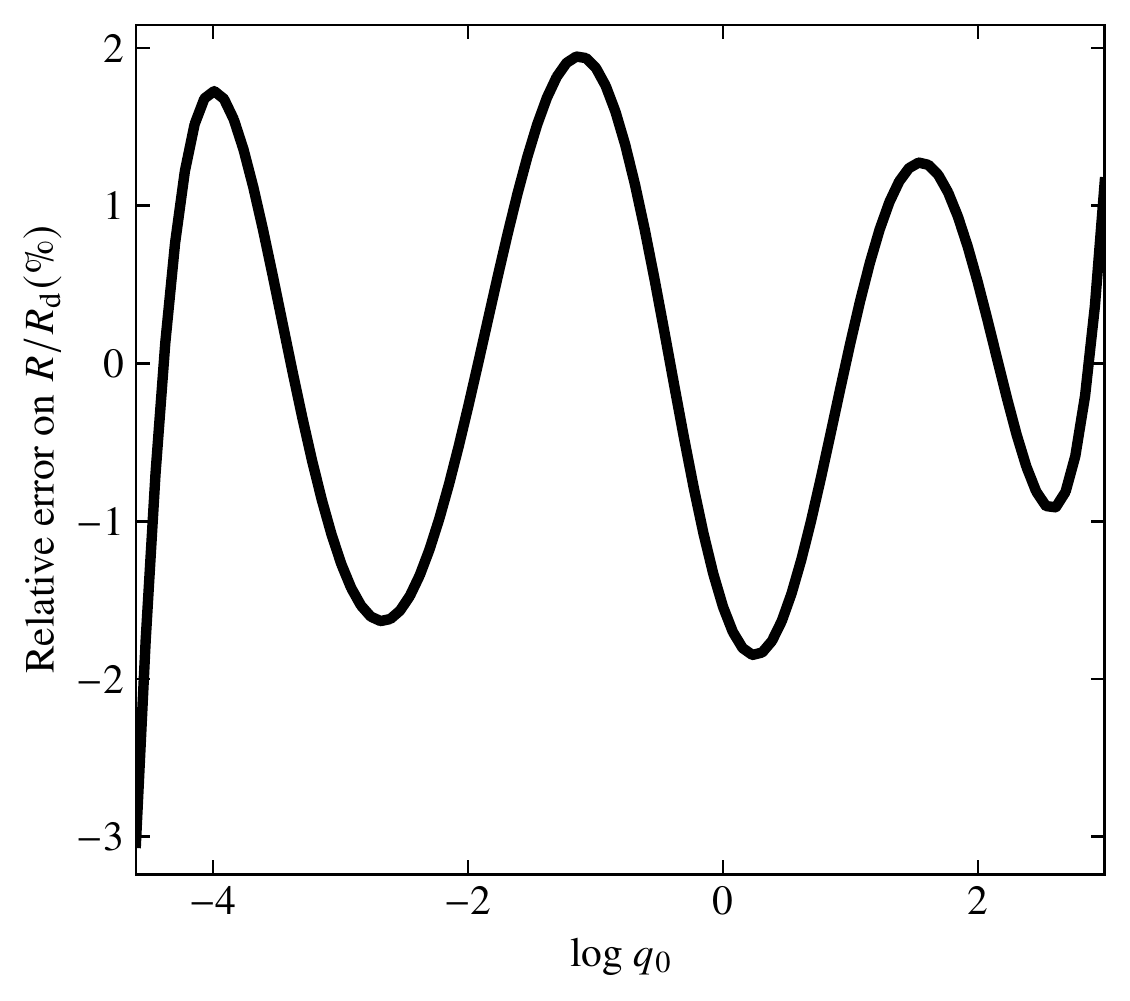}
		\caption{Relative error on $R/R_{\rm{d}}$ between the numerical solution of Eq.\,\ref{eq:solve_y0} and the analytical approximation given by Eq.\,\ref{eq:analytical_approximation_y0} as a function of the disk thickness $q_0$. In the range of disk thicknesses we are interested in the error does not exceed 2\%.}
		\label{fig:error_on_y0}
	\end{figure}

	Equation\,\ref{eq:solve_y0} was solved numerically for a range of $q_0$ values and was then fitted by a polynomial function of degree five in order to get an analytical approximation of $y_0$ as function of $q_0$. We found that the best polynomial fit is given by
	
	\begin{equation}
		\begin{split}
			y_0 =\, & 0.76679 + 0.86230 q_0 -0.13703 q_0^2 -0.02308 q_0^3 + \\
			& 0.00452 q_0^4 + 0.00102 q_0^5,
		\end{split}
		\label{eq:analytical_approximation_y0}
	\end{equation}
	and we show in Fig.\,\ref{fig:error_on_y0} the relative error on $y_0 = R / R_{\rm{d}}$ between the analytical approximation given by Eq.\,\ref{eq:analytical_approximation_y0} and the numerical solution as a function of the disk thickness.
	
	\subsection{Stellar bulge}
	\label{Appendix:bulge_model}
	
	Galaxy bulges can be described by various 3D distributions such as Plummer or Jaffe profiles \citep{plummer_problem_1911, jaffe_simple_1983}, but the most interesting one remains the Hernquist profile \citep{hernquist_analytical_1990}
		
	\begin{equation}
		\rho_{\rm{M}} (r) = \frac{M_{\rm{b}}}{2\pi} \frac{a}{r} \left ( r + a \right )^{-3},
		\label{eq:hernquist}
	\end{equation}
	with $M_{\rm{b}}$ the total bulge mass and $a$ a scale radius related to the half-mass size $r_{1/2, \rm{M}}$ through the relation $a = r_{1/2, \rm{M}} / \left (1 + \sqrt{2} \right )$. In the case of a light distribution, the total bulge mass $M_{\rm{b}}$ is replaced by the total bulge flux $F_{\rm{b}} = M_{\rm{b}} / \Upsilon$. This profile has the advantage of being spherically symmetric, with analytical forms of its gravitational potential and circular velocity, while having a line of sight projected surface density close to a de Vaucouleurs profile, except towards the inner parts. Therefore describing the bulge 3D mass distribution as an Hernquist profile seems to be the most relevant choice. The circular velocity can be written as
	
	\begin{equation}
		V_{\rm{b}}(r) = 2 V_{\rm{b, max}} \sqrt{a r} \left (a + r \right )^{-1},
		\label{eq:rotation_curve_bulge}
	\end{equation}
	where $V_{\rm{b, max}} = 0.5 \times \sqrt{G \Upsilon F_{\rm{b}} / a}$ is the maximum circular velocity reached at a radius $r = a$.
	
	\subsection{Hernquist - de Vaucouleurs mapping}
	
	\begin{figure}[hbt!]
		\includegraphics[scale=0.7]{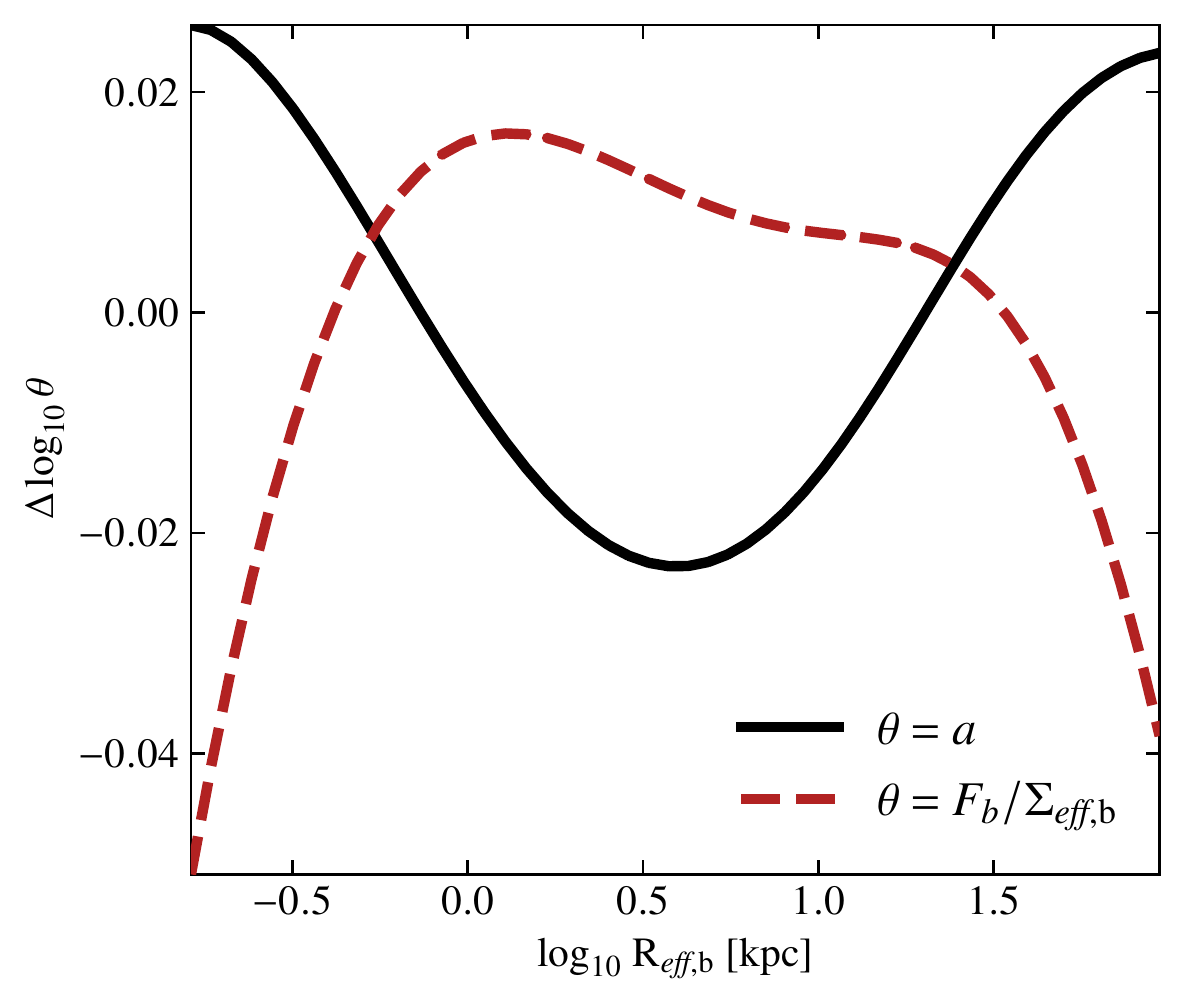}
		\caption{Log difference of the best-fit scaling relations from Eq.\,\ref{eq:a_functionof_Re} and \ref{eq:Lb/Sigma_e_functionof_Re} with the derived parameters $\theta \in \lbrace a, F_{\rm{b}}	 / \Sigma_{\textit{eff}, \rm{b}} \rbrace$ as a function of the bulge effective radius. The variation of the Hernquist parameters $a$ and $F_{\rm{b}}$ with $R_{\textit{eff}, \rm{b}}$ was derived by generating a grid of Hernquist models, by projecting each model along the line of sight and fitting them with de Vaucouleurs profiles. In the range of bulge sizes we are interested in, the error on the parameters is around 5\%.}
		\label{fig:sersic_hernquist}
	\end{figure}
	
	\begin{figure}[h]
		\includegraphics[scale=0.7]{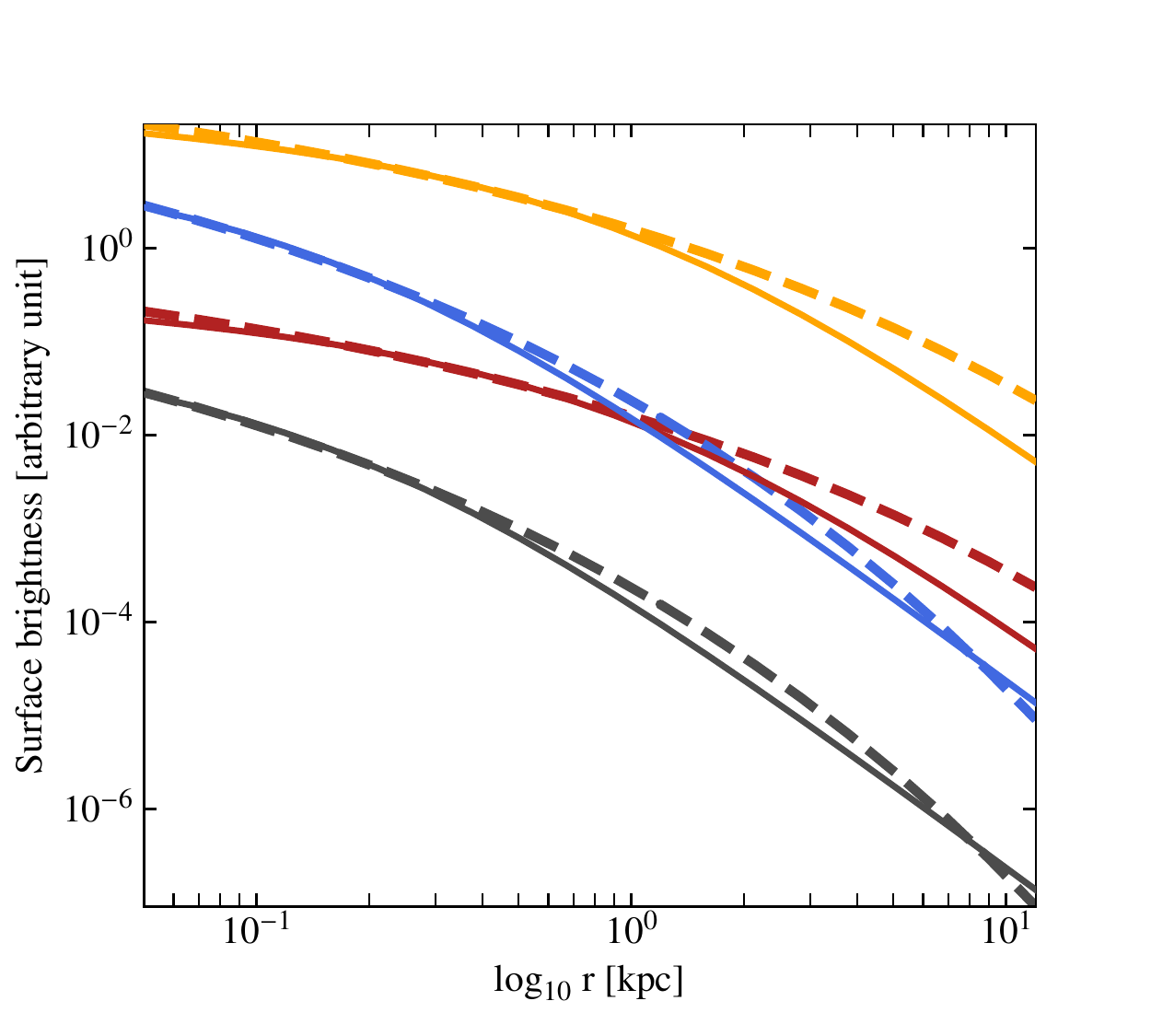}
		\caption{Examples of de Vaucouleurs profiles (dashed lines) and their corresponding sky projected Hernquist profiles (continuous lines) using the scaling relations in Eq.\,\ref{eq:a_functionof_Re} and \ref{eq:Lb/Sigma_e_functionof_Re}. From top to bottom, the Sérsic parameters are $(\Sigma_{\textit{eff}}, R_{\textit{eff}}) = (10^{-3}, 0.5)$ (orange), $(10^{-3}, 6)$ (blue), $(0.1, 0.5)$ (red) and $(0.1, 6)$ (grey). Because the deviation of the projected Hernquist profile to the Sérsic one occurs mainly at large distances, where the surface brightness quickly drops, the overall fluxes are actually in quite good agreement.}
		\label{fig:comparison_sersic_Hernquist}
	\end{figure}
	
	To compute the rotation curve of the bulge component, one needs to map the de Vaucouleurs parameters $\left (\Sigma_{\textit{eff}, \rm{b}}, R_{\textit{eff}, \rm{b}} \right)$ from \Galfit{} with the parameters $\left (F_{\rm{b}}, a \right )$ of the Hernquist model whose line of sight projected surface brightness matches best that of the Sérsic model. We generated 2\,500 line of sight projected Hernquist models on a $\log_{10} a~-~\log_{10} F_{\rm{b}}$ grid in the ranges $-1 \leq \log_{10} a/\rm{kpc} \leq 1$ and $-4 \leq \log_{10} F_{\rm{b}} /\SI{e-20}{erg s^{-1} \angstrom^{-1}} \leq 6$, and for each model, a de Vaucouleurs profile was fitted by minimising the root mean square error using the Levenberg-Maquart algorithm. The bounds for both parameters were chosen based on previous tests which showed that these values correspond to typical sizes and surface brightness we have in our HST data. After inspection, it seems that the Hernquist parameters can be mapped to the Sérsic ones through the two following scaling relations
	
	\begin{align}
		\label{eq:a_functionof_Re}
		\log_{10} a \,[\rm{kpc}] &= \alpha_a + \beta_a \log_{10} R_{\textit{eff}, \rm{b}} [\rm{kpc}] \\
		\log_{10} F_{\rm{b}} / \Sigma_{\textit{eff}, \rm{b}} \,[\rm{cm}^2] &=  \alpha_F + \beta_F \log_{10} R_{\textit{eff}, \rm{b}} [\rm{kpc}]
		\label{eq:Lb/Sigma_e_functionof_Re}
	\end{align}
	
	The error on these two scaling relations is shown in Fig.\,\ref{fig:sersic_hernquist}. While not being perfect, for typical bulge sizes around $\SI{2}{kpc}$ the error is around 5\%. We find the following best-fit scaling parameters: $\alpha_a = -0.454$, $\beta_a = 0.725$, $\alpha_F = 1.194$ and $\beta_F = 1.75$. Examples of Sérsic profiles and their associated projected Hernquist profiles using Eq.\,\ref{eq:a_functionof_Re} and \ref{eq:Lb/Sigma_e_functionof_Re} are shown in Fig.\,\ref{fig:comparison_sersic_Hernquist}. The two profiles start diverging towards large radii where the Sérsic profile drops more rapidly than the Hernquist one.
	
	By construction, the Hernquist amplitude parameter $F_{\rm{b}}$ should be equal to the total de Vaucouleurs flux, but because the total flux is proportional to $R_{\textit{eff}, \rm{b}}^2$ while $F_{\rm{b}}$ is proportional to $R_{\textit{eff}, \rm{b}}^{1.75}$, in practice, this means that our parametrisation, while recovering the shape of a de Vaucouleurs profile for a broad part of the radial range, will underestimate or overestimate the real flux contribution, and therefore the maximum circular velocity of the bulge component. Using Eq.\,\ref{eq:flux_Sersic}, \ref{eq:Lb/Sigma_e_functionof_Re} and \ref{eq:rotation_curve_bulge}, we can derive the error on $V_{\rm{b, max}}$ as a function of the de Vaucouleurs parameters
	
	\begin{equation}
		\Delta V_{\rm{b, max}} / V_{\rm{b, max}} (F_{\rm{tot}}) = 0.5 \left [ 1.174 \left (R_{\textit{eff}, \rm{b}} / \rm{kpc} \right)^{-0.125}  - 1 \right ],
	\end{equation}
	where $\Delta V_{\rm{b, max}} = V_{\rm{b, max}} (F_{\rm{b}}) - V_{\rm{b, max}} (F_{\rm{tot}})$, with $V_{\rm{b, max}} (F_{\rm{b}})$ and $V_{\rm{b, max}} (F_{\rm{tot}})$ the maximum circular velocities from Eq.\,\ref{eq:rotation_curve_bulge} using the Hernquist amplitude parameter and the total de Vaucouleurs flux respectively. Therefore, our parametrisation overestimates the bulge circular velocity for bulge sizes $R_{\textit{eff}, \rm{b}}~\lesssim~\SI{3.6}{kpc}$, and underestimates it beyond, with a maximum relative difference of 50\% when $R_{\textit{eff}, \rm{b}} \rightarrow \infty$. Nevertheless, these differences need to be weighted out by two facts
	\begin{enumerate*}[label=(\roman*)]
		\item as can be seen in Fig.\,\ref{fig:bulge_Reff_distribution}, bulges mainly have radii below $1.5-\SI{2}{kpc}$ where the difference is mostly negligible given the uncertainties on the other parameters and the assumption of a constant mass to light ratio,
		\item very small bulge sizes where we may expect the largest differences to arise are in practice associated with really weak bulge contribution, that is $\Sigma_{\textit{eff}, \rm{b}} \sim 0$, and therefore to a negligible rotation.
	\end{enumerate*}

	\subsection{Dark matter halo}
	\label{Appendix:DM_halo}
	
	Apart from the baryonic disk and bulge components, we also model the galaxies dark matter halo with a NFW profile \citep{navarro_assembly_1995}
	
	\begin{equation}
		\rho(r) = \delta_{\rm{c}} \rho_{\rm{crit}} (r / r_{\rm{s}})^{-1} (1 + r/r_{\rm{s}})^{-2},
	\end{equation}
	where $r_s = r_{200} / c$ is the halo scale radius, with $r_{200}$ the virial radius of the halo where the mean overdensity is equal to 200 and $c$ the halo concentration, $\rho_{\rm{crit}} = 3H_0^2 / (8\pi G)$ the Universe closure density and $\delta_c$ the halo characteristic overdensity \citep{navarro_structure_1996}. The associated circular velocity is given by
	
	\begin{equation}
		V_{\rm{h}}(r) = \frac{V_{\rm{h, max}}}{0.46499} \left [ \frac{\ln\left (1 + r/r_{\rm{s}} \right )}{r/r_{\rm{s}}} - \frac{1}{1 + r/r_{\rm{s}}} \right ]^{1/2},
	\end{equation}
	where $V_{\rm{h, max}}$ is the maximum rotation velocity reached at a radius $r \approx 2.163 r_{\rm{s}}$.
		
	\section{Sky projection of a double exponential profile}%
	\label{Appendix:double_exponential_integral}
		
	\begin{figure}[hbt!]
		\centering
		\includegraphics[scale=0.35]{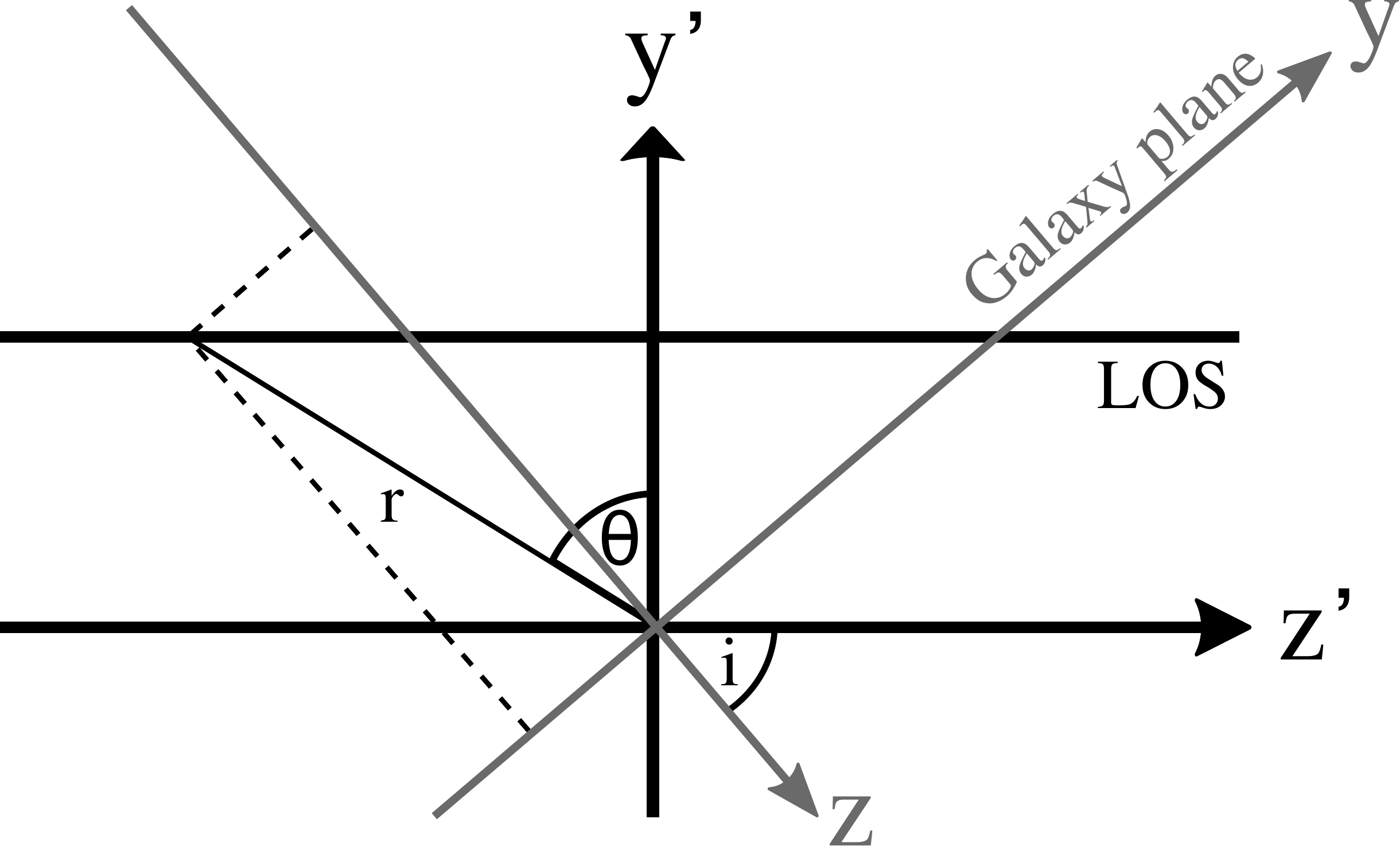}
		\caption{Geometry of the line of sight integration problem. For each point $(x', y')$ in the plane of the sky, the 3D density disk distribution $\rho_{\rm{d}}(x, y, z)$ must be integrated along a line of constant $R'~=~(x'^{2}~+~y'^2)^{1/2}$. The angle $\theta$ is oriented such that it is positive for $z' > 0$ and negative otherwise.}
		\label{fig:change_of_coord}
	\end{figure}
		
	We consider the double exponential disk model of the form $\rho_{\rm{d}}(\vec r) = \rho_{\rm{d}}(R, z)$ with $R$ the radius in the plane of the disk and $z$ the direction orthogonal to the disk. Let us define three new coordinates $(x', y', z')$ such that $(x', y', 0)$ corresponds to the plane of the sky (see Fig.\,\ref{fig:change_of_coord}). Furthermore, the axis defined by $x=x'$ corresponds to the intersection between the plane of the disk and the plane of the sky. Computing the surface density of the inclined 3D distribution at position $(x', y')$ on the plane of the sky amounts to solving the following integral
		
		\begin{equation}
			\label{eq:sigma_original_integral}
			\Sigma_{\rm{d}}(x', y') = \int_{\mathbb{R}} dz' \rho_{\rm{d}}(R, z).
		\end{equation}
		
		Therefore, one must write $R$ and $z$ as functions of $x'$, $y'$ and $z'$. To do so, let us define $r$, the distance of a point in the $(y', z')$ plane and $\theta$, the angle between the $r$ axis and $y'$, where $\theta$ is an oriented angle varying between $-\pi/2$ and $\pi/2$. We have
		
		\begin{align}
			y' & = r \cos \theta, &  z' & = r \sin \theta \label{eq:y'}, \\
			y  & = r \cos (\theta - i), & z  & = r \sin (\theta - i).\label{eq:y}
		\end{align}
		
		Since the integral is computed along a line of constant $y'$, we can plug Eq.\,\ref{eq:y'} into Eq.\,\ref{eq:y} after developing the cosine and sine terms to get
		
		\begin{align}
			y & = z' \sin i + y' \cos i, \\
			z & = z' \cos i - y' \sin i.
		\end{align}
		
		Using the expression for the double exponential profile (see Eqs.\,\ref{eq:razor_thin_surface_density} and \ref{eq:thin_disk}) we get
		
		\begin{equation}
			\label{eq:sigma_full_integral}
			\begin{split}
				\Sigma_{\rm{d}}(x', y') = \frac{\Sigma_{\rm{d}} (0,0)}{2h_z} \int_{\mathbb{R}} dz' \exp \left \lbrace -\frac{\sqrt{x'^2 + \left ( z' \sin i + y' \cos i \right )^2}}{R_{{\rm{d}}}} \right. \\
				\left. - \frac{| z' \cos i - y' \sin i |}{h_z} \right \rbrace
			\end{split}.
		\end{equation}
		
		We can simplify this integral by making the change of variable $v = y / R_{\rm{d}}$ and by defining the following parameters
		
		\begin{align}
			\alpha & = x / R_{\rm{d}}, \\
			\beta  & = \left ( q_0 \tan i \right )^{-1}, \\
			\gamma & = \frac{y'}{h_z} \left ( \sin i + \cos^2 i / \sin i \right ),
		\end{align}
		with $q_0 = h_z / R_{\rm{d}}$ the intrinsic axis ratio of the galaxy. The integral becomes
		
		\begin{equation}
			\label{eq:sigma_simple_integral}
			\Sigma_{\rm{d}}(x', y') = \frac{\Sigma_{\rm{d}}(0, 0)}{2 q_0 \sin i} \int_{\mathbb{R}} dv \exp \left \lbrace - \sqrt{\alpha^2 + v^2} - |\beta v - \gamma | \right \rbrace.
		\end{equation}
		
		The original problem of solving Eq.\,\ref{eq:sigma_original_integral} for the double exponential profile required 6 free parameters, namely $x'$, $y'$, $\Sigma_{\rm{d}}(0, 0)$, $R_{\rm{d}}$, $h_z$ and $i$, with $\Sigma_{\rm{d}}(0, 0)$ only acting as an amplitude parameter, but Eq.\,\ref{eq:sigma_simple_integral} reduces the dimensionality of the problem to 3 free parameters only to compute the integral. In the general case, there is no straightforward analytical solution or numerical approximation to the integral above, though a solution can be derived along the $y'$-axis when $x = 0$
		
		\begin{equation}
			\begin{split}
				\Sigma_{\rm{d}}(0, y') & = \frac{\Sigma_{\rm{d}}(0, 0)}{2 q_0 \sin i} \int_{\mathbb{R}} dv \exp \left \lbrace - | v | - |\beta v - \gamma | \right \rbrace \\
				& = \frac{\Sigma_{\rm{d}}(0,0)}{q_0 \sin i} \frac{e^{-\gamma} - \beta e^{-\gamma/\beta}}{{1 - \beta^2}}.
			\end{split}
		\end{equation}
		
	\section{MAGIC catalogue}%
	
	\begin{table*}[t]
		\caption{Column description of the MAGIC catalogue containing morpho-kinematics and physical parameters for the MS sample of \NMSSample{} galaxies.}

		\begin{tabular}{lll}
			\hline
			\hline
		    No. & Title      & Description \\
		    1   & ID         & MUSE galaxy ID in the form \textsc{X-CGrY}, where X refers to the galaxy identification number within the field \\
		        &            & targeting COSMOS group \textsc{CGrY} \\
		    2   & z          & Spectroscopic systemic redshift derived from kinematics modelling                                                                     \\
		    3   & RA         & J2000 Right Ascension of morphological centre in decimal degrees                                                                     \\
		    4   & Dec        & J2000 Declination of morphological centre in decimal degrees                                                                         \\
		    5   & N          & Number of galaxies in structures with more than three members                                                                        \\
		    6   & Reffd      & Disk effective radius in kpc ($R_{\textit{eff},\rm{d}}$)                                                                             \\
		    7   & Reffb      & Bulge effective radius in kpc ($R_{\textit{eff},\rm{b}}$)                                                                            \\
		    8   & Reff       & Global effective radius in kpc ($R_{\textit{eff}}$)                                                                                  \\
		    9   & logBD      & Logarithm of the bulge-to-disk ratio at $R_{\textit{eff}}$                                                                           \\
		   10   & q          & Axis ratio of the disk ($q$)                                                                                                         \\
		   11   & PAm        & Morphological position angle of the major axis in degrees                                                                            \\
		   12   & FWHM       & Median PSF FWHM, corresponding to narrow band \OII\ MUSE observations in arcsecond                                                   \\
		   13   & OIIflux(R22) & \OII\ flux derived from MUSE flux maps at $R_{22} = 1.311 \times R_{\textit{eff},\rm{d}}$ in $10^{-21}$~erg~s$^{-1}$~cm$^{-2}$       \\
		   14   & OIIflux    & \OII\ flux derived from MUSE flux maps at $\SI{3}{\arcsec}$ in $10^{-21}$~erg~s$^{-1}$~cm$^{-2}$                                             \\
		   15   & SNR        & Total \OII\ signal-to-noise ratio ((S/N)$_{\rm{tot}}$)                                                                               \\
		   16   & i          & Disk inclination corrected for thickness in degrees ($i$)                                                                            \\
		   17   & PAk        & Kinematic position angle of the major axis in degrees                                                                                \\
		   18   & rs         & NFW halo scale radius in kpc ($r_s$)                                                                                                 \\
		   19   & Vhmax      & Maximum rotation velocity of the NFW rotation curve in km~s$^{-1}$ ($V_{\rm{h,max}}$)                                                 \\
		   20   & Vr22       & Rotation velocity at $R_{22}$ in km~s$^{-1}$ ($V_{22}$)                                                                              \\
		   21   & sigma      & Median velocity dispersion in km~s$^{-1}$ ($\sigma_V$)                                                                                   \\
		   22   & Vc22       & Corrected rotation velocity at $R_{22}$ in km~s$^{-1}$ ($V_{c, 22}$)                                                                 \\
		   23   & logM*      & Logarithm of the stellar mass ($M_{\star}$ / M$_{\odot}$) within an aperture of $\SI{3}{\arcsec}$                                           \\
		   24   & logM*(R22) & Logarithm of the corrected stellar ($M_{\star,\rm{corr}}$ / M$_{\odot}$) inside $R_{22}$                                              \\
		   25   & logSFR     & Logarithm of the SFR (SFR / [M$_{\odot}$~yr$^{-1}$]) at $\SI{3}{\arcsec}$ using \citet{gilbank_local_2010, gilbank_erratum_2011} prescription \\
		   26   & logMg      & Logarithm of the gas mass ($M_{\rm{g}}$ / M$_{\odot}$) computed from the Schmidt-Kennicutt law and \OII\ flux \\
		        &            & measured at $R_{22}$    \\
		   27   & logMdyn    & Logarithm of the dynamical mass ($M_{\rm{dyn}}$ / M$_{\odot}$) computed at $R_{22}$ from the mass model                               \\
		% (   12 &        kpc        rt       radius at which the plateau velocity is reached)
		% (   13 &        km/s       Vt       velocity of the plateau)
		  \hline
		\end{tabular}
		\label{tab:catalog_CDS}
	\end{table*}
	
	\clearpage
	\section{Example of morpho-kinematics maps}
	\label{Sec:example_morpho_kin_maps}
	
	We show below an example of a morpho-kinematic map. The maps for all the galaxies in the MS sample are sorted according to their (RA2000, DEC2000) coordinates and can be found online.
	
	\begin{figure}[hbt!]
      \centering
      \includegraphics[scale=0.5]{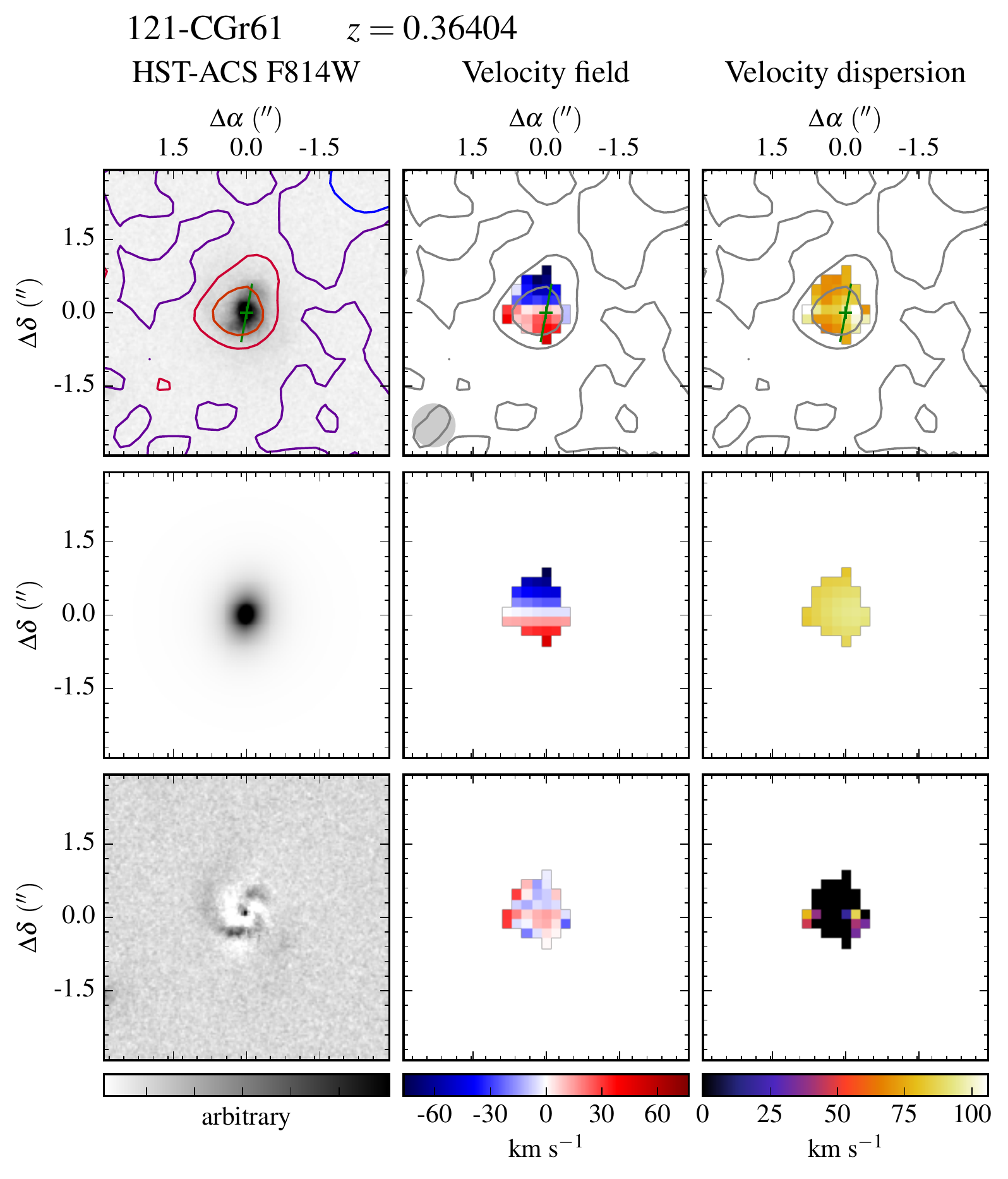}
      \caption{Morpho-kinematics map for galaxy \textsc{121-CGr61}. In each panel, from top to bottom and left to right: HST-ACS image, \textsc{Galfit} model, HST residuals, \textsc{Camel} velocity field, \textsc{Mocking} velocity field model, Velocity field residuals, \textsc{Camel} velocity dispersion map, \textsc{Mocking} beam smearing model including spectral resolution broadening, beam smearing and LSF corrected velocity dispersion map. The morpho-kinematic centre and the morphological position angle are shown in the HST image and the \textsc{Camel} maps as a green cross and a green line whose length corresponds to $R_{22}$, respectively. The PSF FWHM is indicated as the grey disk in the velocity field. The \textsc{[Oii]} surface brightness distribution is overlaid on top of the HST and MUSE \textsc{[Oii]} flux maps with contours at levels $\Sigma_{\textsc{[Oii]}}~=~2.5, 5, 10, 20, 40~{\rm{and}}~\SI{80e-18}{erg~s^{-1}~cm^{-2}~arcsec^{-2}}$.}
      \label{Fig:example_kin_map_appendix}
   \end{figure}
		
	\end{appendix}
	
\end{document}